\documentclass[letterpaper,english,notitlepage,superscriptaddress,nofootinbib]{revtex4-2}
\usepackage[T1]{fontenc}
\usepackage{textcomp}
\usepackage[latin9]{inputenc}
\usepackage{color}
\usepackage{babel}
\usepackage{array}
\usepackage{multirow}
\usepackage{amsmath}
\usepackage{amssymb}
\usepackage{graphicx}
\usepackage[pdfusetitle,
 bookmarks=true,bookmarksnumbered=false,bookmarksopen=false,
 breaklinks=false,pdfborder={0 0 1},backref=false,colorlinks=false]
 {hyperref}

\makeatletter

\pdfpageheight\paperheight
\pdfpagewidth\paperwidth

\providecommand{\tabularnewline}{\\}

\count\footins = 1000  
\usepackage{xcolor}
\usepackage{babel}
\usepackage{tikz}

\pdfpageheight\paperheight
\pdfpagewidth\paperwidth

\providecommand{\tabularnewline}{\\}

\makeatother

\begin{document}
\title{The exclusive photoproduction of $\chi_{c}\gamma$ pairs in the small-$x$
kinematics}
\author{M. Siddikov}
\affiliation{Departamento de Física, Universidad Técnica Federico Santa María,  Casilla 110-V, Valparaíso,
Chile~~~~~~~~~~\\
 }
\affiliation{Centro Científico - Tecnológico de Valparaíso, Casilla 110-V, Valparaíso,
Chile}
\author{I. Zemlyakov}
\affiliation{Departamento de Física, Universidad Técnica Federico Santa María,  Casilla 110-V, Valparaíso,
Chile~~~~~~~~~~\\
 }
\affiliation{Instituto de Física, Pontificia Universidad Católica de Valparaíso,
Av. Brasil 2950, Valparaíso, Chile}
\author{M. Roa}
\affiliation{Departamento de Física, Universidad Técnica Federico Santa María,  Casilla 110-V, Valparaíso,
Chile~~~~~~~~~~\\
}
\affiliation{Facultad de Ingeniería, Laboratorio DataScience, Universidad de Playa
Ancha, ~~~~\\
 Leopoldo Carvallo 270, Valparaíso, Chile,~~~~~~ ~~~}
\begin{abstract}

In this manuscript we analyze the exclusive photoproduction of the
$\chi_{c}\gamma$ pairs. We focus on the small-$x$ kinematics and
evaluate the cross-sections in the Color Glass Condensate framework.
We found that in the leading order in the strong coupling $\alpha_{s}$,
this process is sensitive only to the forward color dipole scattering
amplitude. We estimated numerically the cross-sections for different
polarizations of $\chi_{c}$ mesons in the kinematics of the ultraperipheral
collisions at LHC and the future Electron Ion Collider. We also analyzed
the role of this process as a potential background to exclusive $\chi_{c}$
photoproduction, which has been recently suggested as an alternative
channel for studies of odderons. According to our estimates, the cross-section
of $\chi_{c}\gamma$ with undetected photons is comparable to that
of odderons at small momentum transfer $|t|\lesssim1$ GeV$^{2}$,
but becomes less relevant at larger $|t|$. We also found that a dominant
background to odderon-mediated production of $\chi_{c}$ comes from
the radiative decays of $\psi(2S)$ mesons, which potentially can
present a challenge for odderon studies via $\chi_{c}$ photoproduction.
\end{abstract}
\maketitle

\section{Introduction}

The Color Glass Condensate (CGC) approach~ \cite{McLerran:1993ni,McLerran:1993ka,McLerran:1994vd,Gelis:2010nm,Iancu:2003uh}
has been established as a theoretical framework for systematic analysis
of the hadronic processes in the small-$x$ kinematics. It naturally
incorporates saturation effects and allows us to obtain a self-consistent
and phenomenologically acceptable description of various lepton-hadron
and hadron-hadron processes~\cite{Aidala:2020mzt,Ma:2014mri,Cheung:2024qvw,Mantysaari:2020lhf,Kang:2023doo,Tuchin:2004,Blaizot:2004,ALICE:2012,ATLAS:2016,IPSat,RESH,watt:bcgc,DaSilveira:2018haa,Albacete:2012td,LHCb:2022ahs,Mantysaari:2023xcu}.
The interactions of partons with the target in this approach are encoded
in the nonperturbative $n$-point correlators of Wilson lines (forward
multipole scattering amplitudes), whose dependence on rapidity obeys
the evolution equations of Balitsky-Kovchegov Jalilian-Marian, Iancu,
McLerran, Weigert, Leonidov and Kovner (BK-JIMWLK)~\cite{Kovchegov:1999yj,Kovchegov:2006vj,Balitsky:2008zza,Balitsky:1996,Balitsky:2001re,JIMWLK:1997,JIMWLK:1998,JIMWLK:2,JIMWLK:2a,JIMWLK:3,Iancu:2011ns}.
 In the presence of appropriate hard scale which controls the transverse
size of the system, it is expected that the dominant contribution
stems from the quark-antiquark component of the wave function, convoluted
with forward dipole scattering amplitude. On this premise, a number
of processes has been used in phenomenological analyses in order to
fix the forward dipole scattering amplitude. However, the higher 
order multipoles may contribute in loop corrections even in the simplest
and well-known processes, like Deep Inelastic Scattering (DIS)~\cite{Beuf:2022ndu,Beuf:2021qqa}.
A recent numerical analysis of the next-to-leading order (NLO) corrections
to various processes demonstrated that they could be numerically sizable~\cite{Banu:2019uei,Mantysaari:2022kdm,Mantysaari:2022bsp,Iancu:2022gpw},
thus shedding doubts on phenomenological extractions based on leading
order picture. Unfortunately, such estimates remain ambiguous due
to dependence on poorly known parametrization of these multipoint
correlators. Conversely, a reasonable phenomenological description
of the existing experimental data by completely different parametrizations,
disregarding NLO corrections altogether, suggests that the existing
data don't uniquely fix  the dipole amplitude~\cite{IPSat,watt:bcgc,RESH}.

While it is impossible to switch off completely the contributions
of the higher twist correlators, it is possible to reveal their presence
testing the universality of the forward dipole scattering amplitude
in various processes. The exclusive $2\to3$ processes from this
point of view are well-suited for this analysis, since they have a
different structure of the amplitude, and thus can be used as independent
tools for phenomenological studies ~\cite{Roy:2019cux,Roy:2019hwr,Qiu:2024mny,Qiu:2023mrm,Deja:2023ahc,Siddikov:2022bku,Siddikov:2023qbd,GPD2x3:10,GPD2x3:11}.
 Due to high luminosity of the ongoing photoproduction experiments
at LHC (in ultraperipheral kinematics)~\cite{Mangano:2017tke} and
the future experiments at the Electron Ion Collider (EIC)~\cite{AbdulKhalek:2021gbh,Burkert:2022hjz},
measurement of such processes is feasible in the nearest future. A
special interest in this program deserves the production of quarkonia-photon
pairs. The heavy quark mass serves as a natural hard scale in processes
involving their production and justifies perturbative description
of the $Q\bar{Q}$ pair formation, as well as the use of NRQCD framework
for their hadronization~\cite{Bodwin:1994jh,Maltoni:1997pt,Feng:2015cba,Brambilla:2010cs}.
 Due to strong suppression of $C$-parity exchanges in the $t$-channel,
the largest cross-section have the quarkonia-photon pairs which include
$C$-even quarkonia, like $\eta_{c}$ and $\chi_{c}$. The production
of $\eta_{c}\gamma$ pairs has been studied recently in~\cite{Siddikov:2024bre},
and it was discussed how the expected kinematic distributions of the
final-state particles are related to the dipole amplitude. However,
that channel suffers from significant backgrounds, most notably from
the radiative decays of other quarkonia, and for this reason requires
additional kinematic cutoffs which reduce the expected yields. In
this paper we extend that analysis and suggest to study the production
of the $\chi_{c}\gamma$ meson pairs. In what follows we will focus
on the high energy (small-$x$) photon-proton collisions, which can
be studied as a subprocess in ultraperipheral $pA$ and $AA$ collisions
at LHC, as well as in the future high energy runs at Electron Ion
Collider, the Large Hadron electron Collider (LHeC)~\cite{Agostini:2020fmq},
and the Future Circular Collider (FCC-he)~\cite{Abada:2019lih}.

In addition to studies of the dipole scattering amplitudes, the $\chi_{c}\gamma$
photoproduction can also present interest for two other problems.
Historically, the triplet of the $P$-wave quarkonia $\chi_{c}$ with
different spins  and nearly the same masses has been considered in
the literature as a possible tool for precision tests of NRQCD. In
the infinitely heavy quark mass limit, it is expected that the long-distance
matrix elements (LDMEs) of all $\chi_{c}$ mesons are related to each
other due to heavy quark spin symmetry (HQSS). However, the experimental
results for the ratio $\sigma_{\chi_{c2}}/\sigma_{\chi_{c1}}$in hadroproduction
experiments disagrees significantly with HQSS-based expectations,
and the phenomenologically successful descriptions of that channel
either assume that HQSS is heavily broken for charm, or introduce
sizable contributions of the color octet LDMEs~\cite{Ma:2010vd,Likhoded:2014kfa,Zhang:2014coi,Baranov:2019lhm,Baranov:2023ckv}.
 The study of the same ratio in exclusive $\chi_{c}\gamma$ photoproduction
experiments can decisively clarify the validity of HQSS because this
channel does not get contributions from the color octet LDMEs.  Furthermore,
the exclusive $\chi_{c}$ photoproduction (via $\gamma p\to\chi_{c}p$
channel) recently has been suggested in~~\cite{Jia:2022oyl,Benic:2024}
as a sensitive tool for precision studies of odderons. The photoproduction
of $\chi_{c}\gamma$ pairs in this context deserves interest as a
potential background to odderon searches. 

The paper is structured as follows. In the next Section~\ref{sec:Formalism}
we discuss in detail the kinematics of the process, and derive the
theoretical results for the cross-section in the Color Glass Condensate
framework. In Section~\ref{sec:Numer} we provide tentative estimates
for the cross-sections and expected counting rates obtained with a
publicly available parametrization of the forward dipole scattering
amplitude. We also estimate the contribution of this process and the
$\psi(2S)\to\chi_{c}\gamma$ radiative decay to odderon-mediated $\gamma p\to\chi_{c}p$
suggested in~\cite{Benic:2024}. Finally, in Section~\ref{sec:Conclusions}
we draw conclusions.

\section{The theoretical framework}

\label{sec:Formalism} As we explained in the introduction, we will analyze the $\gamma p\to \chi_c \gamma p$ process in the high energy (small-$x$) kinematics, using for description the CGC framework. We will focus on the kinematics of quasireal incoming photons, which dominate the spectrum of equivalent photons in ultraperipheral
collisions and electroproduction experiments. In the subsections~\ref{subsec:Kinematics}~-~\ref{subsec:Formalism}
below we will discuss the kinematics of the process and the light-cone wave functions of $\chi_c$ mesons with different helicities, introduce briefly the CGC framework and derive the amplitude of the process in this approach. The final theoretical expressions for different helicity components are given in Section~\ref{subsec:Formalism}, in Eqns.~(\ref{eq:Amplitude_PP-1}~-~\ref{eq:Amplitude_PP--1},~\ref{eq:Amplitude_PP-1-1}~-~\ref{eq:Amplitude_PP-1--1}). The latter equations will be used for numerical estimates in the next section.

\subsection{Kinematics of the process}

\label{subsec:Kinematics} 
\begin{figure}
\includegraphics[width=10cm]{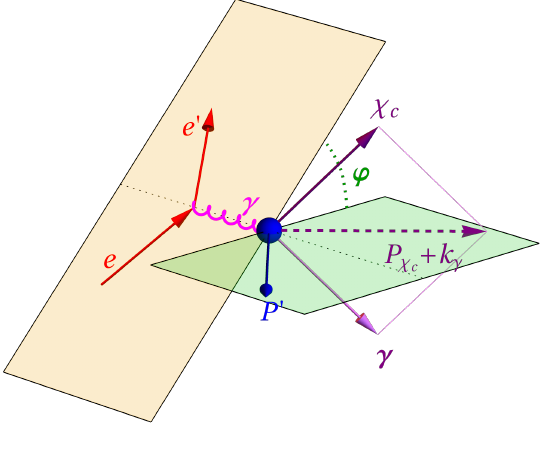}

\caption{(Color online) The definition of the angle $\varphi$ between leptonic
and hadronic planes for the $ep\to e'p'\chi_{c}\gamma$ electroproduction
channel, as seen from the target rest frame. The lepton scattering
plane is formed by the three-momenta of incoming and scattered electrons
(marked with labels $e,e'$), and the hadron scattering plane may
be defined as a plane that includes the three-momenta of the recoiled
proton $P_{{\rm out}}$ and the total three-momentum $p_{\chi_{c}}+k_{\gamma}$
of the produced $\gamma\chi_{c}$ pair. Due to momentum conservation,
this plane also includes the 3-momentum of the virtual photon $\gamma^{*}$.
In the photon-proton collision frame, the incoming proton propagates
in direction anticollinear to the incoming virtual photon $\gamma^{*}$,
so the momentum of the recoil proton $P_{{\rm out}}$ is almost anticollinear
to the momentum of the virtual photon $\gamma^{*}$.}
\label{fig:Planes} \label{fig:CGCBasic-2}
\end{figure}

The evaluation of the invariant cross-section
can be realized in the so-called photon-proton collision frame, where
the 3-momenta of the incoming proton and photon are anticollinear
to each other. For the sake of definiteness we'll assume that the
collision axis coincides with the axis $z$ of our reference frame,
and that the photon and proton propagate in its positive and negative
directions respectively. The mass of the heavy quarkonium $M_{\chi_{c}}$
and the invariant mass of the photon quarkonium pair $M_{\gamma\chi_c}$will
be considered as hard scales, in agreement with~\cite{Siddikov:2024blb}.
The evaluation of the physical amplitude requires to define the light-cone
decomposition of various momenta. In what follows we will utilize
the convention of Kogut-Soper~\cite{Brodsky:1997de}, assuming that
any 4-vector $v^{\mu}$ can be represented as 
\begin{equation}
v^{\mu}=\left(v^{+},\quad v^{-},\quad\boldsymbol{v}_{\perp}\right),\qquad v^{\pm}=\frac{v^{0}\pm v^{3}}{\sqrt{2}},\qquad\boldsymbol{v}_{\perp}=v_{x}\hat{\boldsymbol{x}}+v_{y}\hat{\boldsymbol{y}},
\end{equation}
so its square and convolution with Dirac matrix are given by 
\begin{equation}
v^{2}\equiv v^{\mu}v_{\mu}=2v^{+}v^{-}-\boldsymbol{v}_{\perp}^{2},\qquad\hat{v}\equiv\gamma^{\mu}v_{\mu}=\gamma^{+}v^{-}+\gamma^{-}v^{+}-\boldsymbol{\gamma}_{\perp}\cdot\boldsymbol{v}_{\perp}.\label{eq:gammaSlash}
\end{equation}
The light-cone decomposition of particle's momenta will be chosen
in agreement with earlier studies on meson-photon production~\cite{GPD2x3:8,GPD2x3:9},
\begin{align}
q^{\mu} & =\left(\sqrt{\frac{s}{2}},\,0,\,\boldsymbol{0}_{\perp}\right),\qquad k^{\mu}=\left(\bar{\alpha}_{\chi_{c}}\sqrt{\frac{s}{2}},\,\frac{\left(\boldsymbol{k}_{\perp}^{\gamma}\right)^{2}}{\bar{\alpha}_{\chi_{c}}\sqrt{2s}},\,\boldsymbol{k}_{\perp}^{\gamma}\right)\label{eq:q}\\
P_{{\rm in}}^{\mu} & =\left(\frac{m_{N}^{2}}{\sqrt{2s}\left(1+\xi\right)},\,\left(1+\xi\right)\sqrt{\frac{s}{2}},\,\boldsymbol{0}_{\perp}\right),\quad P_{{\rm out}}^{\mu}=\left(\frac{m_{N}^{2}+\boldsymbol{\Delta}_{\perp}^{2}}{\sqrt{2s}\left(1-\xi\right)},\,\left(1-\xi\right)\sqrt{\frac{s}{2}},\,\boldsymbol{\Delta}_{\perp}\right),\\
\Delta^{\mu} & =P_{{\rm out}}^{\mu}-P_{{\rm in}}^{\mu}=\left(\frac{2\xi m_{N}^{2}+\left(1+\xi\right)\boldsymbol{\Delta}_{\perp}^{2}}{\sqrt{2s}\left(1-\xi^{2}\right)},\,-2\xi\,\sqrt{\frac{s}{2}},\,\boldsymbol{\Delta}_{\perp}\right),\\
\end{align}

\begin{align}
p_{\chi_{c}}^{\mu} & =\left(\alpha_{\chi_{c}}\sqrt{\frac{s}{2}},\,\frac{\left(\boldsymbol{p}_{\perp}^{\chi_{c}}\right)^{2}+M_{\chi_{c}}^{2}}{\alpha_{\chi_{c}}\sqrt{2s}},\,\boldsymbol{p}_{\perp}^{\chi_{c}}\right),\\
 & \boldsymbol{k}_{\perp}^{\gamma}:=\boldsymbol{p}_{\perp}-\frac{\boldsymbol{\Delta}_{\perp}}{2},\quad\boldsymbol{p}_{\perp}^{\chi_{c}}:=-\boldsymbol{p}_{\perp}-\frac{\boldsymbol{\Delta}_{\perp}}{2},\qquad\boldsymbol{p}_{\perp}\equiv\left(\boldsymbol{k}_{\perp}^{\gamma}-\boldsymbol{p}_{\perp}^{\chi_{c}}\right)/2\label{eq:k}
\end{align}
where $q$ and $k$ are the momenta of the incoming and outgoing (emitted)
photons, $P_{{\rm in}},P_{{\rm out}}$ are the momenta of the proton
before and after the interaction, and $p_{\chi_{c}}$ is the momentum
of produced $\chi_{c}$ meson. The parameter $\alpha_{\chi_{c}}$
stands for the fraction of the photon's light-cone momentum carried
by $\chi_{c}$, and a shorthand notation $\bar{\alpha}_{\chi_{c}}\equiv1-\alpha_{\chi_{c}}$
corresponds to a similar light-cone fraction carried by the scattered
photon. For the proton which moves in the opposite direction, the
parameter $\xi$ controls the longitudinal momentum transfer to the
proton during the process. The hard scale $s$ can be related to the
photon-proton invariant~energy $W$ of the process as 
\begin{align}
S_{\gamma N} & \equiv W^{2}=\left(q+P_{{\rm in}}\right)^{2}=s\left(1+\xi\right)+m_{N}^{2},\qquad\sqrt{s}=\sqrt{\frac{W^{2}-m_{N}^{2}}{1+\xi}}.\label{eq:W}
\end{align}
As will be demonstrated below, for high energy collisions $(W\gg m_{N})$,
the variable $\xi\ll1$, so~(\ref{eq:W}) can be approximated as
$\sqrt{s}\approx W.$ The light-cone decomposition in the target rest
frame (r.f.) can be obtained from~(\ref{eq:q}-\ref{eq:k}) making
a longitudinal boost 
\[
v^{+}\to\Lambda v^{+},\quad v^{-}\to\Lambda^{-1}v^{-},\quad\boldsymbol{v}_{\perp}\to\boldsymbol{v}_{\perp}
\]
with $\Lambda=\sqrt{s}\left(1+\xi\right)/m_{N}$, yielding
\begin{align}
q_{{\rm r.f.}}^{\mu} & =\left(\frac{s\left(1+\xi\right)}{m_{N}\sqrt{2}},\,0,\,\boldsymbol{0}_{\perp}\right),\quad k_{{\rm r.f.}}^{\mu}=\left(\frac{\bar{\alpha}_{\chi_{c}}s\left(1+\xi\right)}{\sqrt{2}m_{N}},\,\frac{\left(\boldsymbol{k}_{\perp}^{\gamma}\right)^{2}}{\bar{\alpha}_{\chi_{c}}s\sqrt{2}}\,\frac{m_{N}}{\left(1+\xi\right)},\,\boldsymbol{k}_{\perp}^{\gamma}\right)\label{eq:q-2}\\
P_{{\rm in},{\rm r.f.}}^{\mu} & =\left(\frac{m_{N}}{\sqrt{2}},\,\frac{m_{N}}{\sqrt{2}},\,\boldsymbol{0}_{\perp}\right),\quad P_{{\rm out}}^{\mu}=\left(\frac{m_{N}^{2}+\boldsymbol{\Delta}_{\perp}^{2}}{\sqrt{2}\left(1-\xi\right)}\frac{\left(1+\xi\right)}{m_{N}},\,\frac{m_{N}}{\sqrt{2}}\frac{1-\xi}{1+\xi},\,\boldsymbol{\Delta}_{\perp}\right),\\
\Delta_{{\rm r.f.}}^{\mu} & =P_{{\rm out}}^{\mu}-P_{{\rm in}}^{\mu}=\left(\frac{2\xi m_{N}^{2}+\left(1+\xi\right)\boldsymbol{\Delta}_{\perp}^{2}}{\sqrt{2}m_{N}}\frac{1+\xi}{1-\xi},\,\frac{-\sqrt{2}\xi\,m_{N}}{\left(1+\xi\right)},\,\boldsymbol{\Delta}_{\perp}\right)\\
p_{\chi_{c},{\rm r.f.}}^{\mu} & =\left(\frac{\alpha_{\chi_{c}}s\left(1+\xi\right)}{\sqrt{2}m_{N}},\,\frac{\left(\boldsymbol{p}_{\perp}^{\chi_{c}}\right)^{2}+M_{\chi_{c}}^{2}}{\alpha_{\chi_{c}}s\sqrt{2}}\frac{m_{N}}{\left(1+\xi\right)},\,\boldsymbol{p}_{\perp}^{\chi_{c}}\right).\label{eq:k-2}
\end{align}
The light-cone decomposition~~(\ref{eq:q}-\ref{eq:k},~\ref{eq:q-2}-\ref{eq:k-2})
satisfies the energy-momentum conservation and guarantees that all
particles are on shell. As we discussed in~\cite{Siddikov:2024bre,Siddikov:2024blb},
it is possible to express the variables $\xi,\,p_{\perp},\,\Delta_{\perp},\,\alpha_{\chi_{c}}$,
which define the kinematics, in terms of the invariant mass $M_{\gamma\chi_{c}}^{2}$,
momentum transfers to the proton ($t$) and to the photon ($t'$).
\begin{align}
t & =\Delta^{2}=\left(P_{{\rm out}}-P_{{\rm in}}\right)^{2}=-\frac{1+\xi}{1-\xi}\boldsymbol{\Delta}_{\perp}^{2}-\frac{4\xi^{2}m_{N}^{2}}{1-\xi^{2}}\le-\frac{4\xi^{2}m_{N}^{2}}{1-\xi^{2}}.\label{eq:tDep}
\end{align}
\begin{equation}
t'=\left(k-q\right)^{2}=-2q\cdot k=-\frac{\left(\boldsymbol{p}_{\perp}-\boldsymbol{\Delta}_{\perp}/2\right)^{2}}{\left(1-\alpha_{\chi_{c}}\right)}\le0\label{eq:tPrimeDef}
\end{equation}
\begin{align}
 & M_{\gamma\chi_{c}}^{2}=\left(k+p_{\chi_{c}}\right)^{2}=\left(q+P_{{\rm in}}-P_{{\rm out}}\right)^{2}=t-2q\cdot\Delta=t+2s\xi\,\label{eq:uPrimetPrime}
\end{align}
Resolving these equations, it is possible to obtain exact identities~\cite{Siddikov:2024bre,Siddikov:2024blb}
\begin{equation}
\xi=\frac{M_{\gamma\chi_{c}}^{2}-t}{2\left(W^{2}-m_{N}^{2}\right)-M_{\gamma\chi_{c}}^{2}+t}>0,\quad s=W^{2}-m_{N}^{2}+\frac{t-M_{\gamma\chi_{c}}^{2}}{2},\quad2\xi s=M_{\gamma\chi_{c}}^{2}-t,\label{eq:Xi}
\end{equation}
\begin{align}
\boldsymbol{\Delta}_{\perp}^{2}= & -t\frac{1-\xi}{1+\xi}-\frac{4\xi^{2}m_{N}^{2}}{\left(1+\xi\right)^{2}},\qquad\boldsymbol{p}_{\perp}\cdot\boldsymbol{\Delta}_{\perp}=\frac{2\xi m_{N}^{2}}{\left(1-\xi^{2}\right)}\,\frac{t'}{s}\label{eq:tDep-1-1}\\
\boldsymbol{p}_{\perp}^{2}= & -t'\left(\frac{1}{2}-\alpha_{\chi_{c}}\right)+\frac{1}{2}\left(\alpha_{\chi_{c}}M_{\gamma\chi_{c}}^{2}-M_{\chi_{c}}^{2}\right)+\frac{t}{4}\left(\frac{1-\xi}{1+\xi}-2\alpha_{\chi_{c}}\right)+\frac{\xi^{2}m_{N}^{2}}{\left(1+\xi\right)^{2}}.\label{eq:133-1}
\end{align}
\begin{align}
\alpha_{\chi_{c}} & =-\frac{1}{2\xi s}\left(t'-M_{\chi_{c}}^{2}-\frac{2\xi m_{N}^{2}}{s\left(1-\xi^{2}\right)}\left(M_{\gamma\chi_{c}}^{2}+t'-t\right)\right)-\frac{2\xi m_{N}^{2}}{s\left(1-\xi^{2}\right)},\label{eq:alpha}
\end{align}
which fix the kinematics of the process in terms of the variables
$W,\,t,t',M_{\gamma\chi_{c}}$ up to a global rotation in the transverse
plane. For high-energy processes we can simplify these relations using
the hierarchy of scales $W,\sqrt{s}\gg M_{\chi_{c}},M_{\gamma\chi_{c}}\gtrsim p_{\perp},\Delta_{\perp}$.
The parameter $\xi$ defined in (\ref{eq:Xi}) is also very small,
$\xi\sim M_{\gamma\chi_{c}}^{2}/W^{2}\ll1$, and can be disregarded
if accompanied by other terms which don't vanish in this limit. Similarly
the smallness of the ratio $t'/s\sim M_{\gamma\chi_{c}}^{2}/W^{2}\sim\mathcal{O}\left(\xi\right)\ll1$
in the right-hand side of~(\ref{eq:tDep-1-1}) implies that the vectors
$\boldsymbol{p}_{\perp},\,\boldsymbol{\Delta}_{\perp}$ are almost
orthogonal to each other. As was discussed long ago in~\cite{Lepage:1980fj}
and confirmed by various experimental studies, the cross-sections
of exclusive processes decrease rapidly as a function of the invariant
momentum transfer $t\sim\Delta_{\perp}^{2}$ to the target. Technically,
this happens because a large momentum transfer to the target won't
destruct it only if the momentum is shared equally between all partons,
and each momentum exchange between partons is suppressed by (perturbative)
propagators. As we will see below, such suppression also takes place
in our process, and for this reason in the kinematics of interest
the variables $\boldsymbol{\Delta}_{\perp}^{2},\,t$ are negligible
compared to typical scales $\sim M_{\chi_{c}}^{2}$. In this kinematic,
it is possible to rewrite~(\ref{eq:133-1},~\ref{eq:alpha}) in
a simpler form (see details in~\cite{Siddikov:2024bre,Siddikov:2024blb})
\begin{align}
-t' & \approx\alpha_{\chi_{c}}M_{\gamma\chi_{c}}^{2}-M_{\chi_{c}}^{2},\qquad\boldsymbol{p}_{\perp}^{2}=\bar{\alpha}_{\chi_{c}}\left[\alpha_{\chi_{c}}M_{\gamma\chi_{c}}^{2}-M_{\chi_{c}}^{2}\right]\approx-\bar{\alpha}_{\chi_{c}}t',\qquad M_{\gamma\chi_{c}}^{2}\approx2s\xi,\label{eq:KinApprox}
\end{align}
The physical constraint~$\boldsymbol{p}_{\perp}^{2}\ge0$ implies
that the variable $\alpha_{\chi_{c}}$ is bound by $\alpha_{\chi_{c}}\ge M_{\chi_{c}}^{2}/M_{\gamma\chi_{c}}^{2}$.
The cross-section of the photoproduction subprocess may be represented
as
\begin{equation}
\frac{d\sigma_{\gamma p\to\gamma\chi_{c}p}^{(T)}}{dt\,dt'\,dM_{\gamma\chi_{c}}}\approx\frac{\left|\mathcal{A}_{\gamma p\to\chi_{c}\gamma p}^{(\lambda;\,\sigma,H)}\right|^{2}}{128\pi^{3}M_{\gamma\chi_{c}}}\label{eq:Photo}
\end{equation}
where $\mathcal{A}_{\gamma p\to\chi_{c}\gamma p}^{(\lambda;\,\sigma,H)}$
is the process amplitude, and the superscript indices $\lambda,\,\sigma,\,H$
are the helicities of the incoming photon, outgoing photon and the
produced charmonium, respectively.The cross-section of the electroproduction
is given by~\cite{Mantysaari:2020lhf,Navas:2024X}
\begin{align}
\frac{d\sigma_{ep\to e\gamma\chi_{c}p}}{d\Omega} & =\epsilon\frac{d\sigma^{(L)}}{d\Omega}+\frac{d\sigma^{(T)}}{d\Omega}+\sqrt{\epsilon(1+\epsilon)}\cos\varphi\frac{d\sigma^{(LT)}}{d\Omega}+\sqrt{\epsilon(1+\epsilon)}\sin\varphi\frac{d\sigma^{(L'T)}}{d\Omega}+\label{eq:sigma_def}\\
 & +\epsilon\cos2\varphi\frac{d\sigma^{(TT)}}{d\Omega}+\epsilon\sin2\varphi\frac{d\sigma^{(T'T)}}{d\Omega},\nonumber 
\end{align}
where $\varphi$ is the angle between the lepton and proton scattering
planes (see Figure~\ref{fig:Planes} for definition), the notation
$d\Omega\equiv d\ln W^{2}dQ^{2}\,dt\,dt'\,dM_{\gamma\chi_{c}}d\varphi$
is used for the phase volume, and the variable $Q^{2}=-q^{2}$ is
the virtuality of the photon propagating from leptonic to hadronic
parts. The superscript letters $L,T,T'$ in the right-hand side of~(\ref{eq:sigma_def})
distinguish the contributions of the longitudinal and transverse photons
(as well as their possible interference). The parameter $\epsilon$
is the ratio of the longitudinal and transverse photon fluxes. The
latter may be expressed in terms of inelasticity parameter $y$ and
the invariant energy $\sqrt{s_{ep}}$ of the electron-proton collision
as 
\begin{equation}
\epsilon\approx\frac{1-y}{1-y+y^{2}/2},\qquad y=\frac{W^{2}+Q^{2}-m_{N}^{2}}{s_{ep}-m_{N}^{2}}.
\end{equation}

In what follows we'll focus on the kinematics of small $Q^{2}$, which
gives the dominant contribution to the total cross-section. In this
kinematics, the contribution of the longitudinally polarized photons
is negligible, and~(\ref{eq:sigma_def}) reduces to 
\begin{align}
\frac{d\sigma_{ep\to eM_{1}M_{2}p}}{d\Omega} & \approx\frac{d\sigma^{(T)}}{d\Omega}+\epsilon\cos2\varphi\frac{d\sigma^{(TT)}}{d\Omega}+\epsilon\sin2\varphi\frac{d\sigma^{(T'T)}}{d\Omega}\approx\frac{d\sigma^{(T)}}{d\Omega}\left(1+c_{2}\cos2\varphi+s_{2}\sin2\varphi\right),\label{eq:sigma_def-1}
\end{align}
where $d\sigma^{(T)}/d\Omega$ is related to the photoproduction cross-section~(\ref{eq:Photo})
as~\cite{Weizsacker:1934,Williams:1935,Budnev:1975poe}
\begin{equation}
\frac{d\sigma^{(T)}}{d\Omega}\approx\frac{\alpha_{{\rm em}}}{\pi\,Q^{2}}\,\left(1-y+\frac{y^{2}}{2}-(1-y)\frac{Q_{{\rm min}}^{2}}{Q^{2}}\right)\frac{d\sigma_{\gamma p\to\gamma\chi_{c}p}^{(T)}}{dt\,dt'\,dM_{\gamma\chi_{c}}},\label{eq:LTSep-1}
\end{equation}
the superscript index $(T)$ is used to remind that the dominant contribution
comes from the transversely polarized photon, $Q_{{\rm min}}^{2}=m_{e}^{2}y^{2}/\left(1-y\right)$,
$m_{e}$ is the mass of the electron, and the angular harmonics $c_{2},\,s_{2}$
are related to different components of the amplitude $\mathcal{A}_{\gamma p\to\chi_{c}\gamma p}^{(\lambda,\sigma,H)}$
as 
\begin{equation}
c_{2}=\frac{2\epsilon\,\,{\rm Re}\left(\mathcal{A}_{\gamma p\to\chi_{c}\gamma p}^{(-,+,H)*}\mathcal{A}_{\gamma p\to\chi_{c}\gamma p}^{(+,+,H)}\right)}{\left|\mathcal{A}_{\gamma p\to\chi_{c}\gamma p}^{(+,+)}\right|^{2}+\left|\mathcal{A}_{\gamma p\to\chi_{c}\gamma p}^{(-,+)}\right|^{2}},\quad s_{2}=\frac{2\epsilon\,\,{\rm Im}\left(\mathcal{A}_{\gamma p\to\chi_{c}\gamma p}^{(-,+,H)*}\mathcal{A}_{\gamma p\to\chi_{c}\gamma p}^{(+,+,H)}\right)}{\left|\mathcal{A}_{\gamma p\to\chi_{c}\gamma p}^{(+,+)}\right|^{2}+\left|\mathcal{A}_{\gamma p\to\chi_{c}\gamma p}^{(-,+)}\right|^{2}}.\label{eq:c2s2}
\end{equation}
As we will see below, these coefficients are small, so the cross-section~(\ref{eq:sigma_def-1})
is dominated by the angular-independent term (the so-called equivalent
photon approximation).

\subsection{The polarization vectors and wave functions}

\label{subsec:WFs}For the polarization vectors of the photons, we
will use the light-cone parametrization 
\begin{equation}
\varepsilon_{T}^{(\lambda=\pm1)}(\ell)=\left(\frac{\boldsymbol{\varepsilon}_{\lambda}\cdot\boldsymbol{\ell}_{\perp}}{\ell^{-}},\,0,\,\boldsymbol{\boldsymbol{\varepsilon}}_{\lambda}\right),\qquad\boldsymbol{\boldsymbol{\varepsilon}}_{\lambda}=\frac{\hat{\boldsymbol{x}}+i\lambda\,\hat{\boldsymbol{y}}}{\sqrt{2}}.\label{eq:PolVector}
\end{equation}
where $\ell$ is the 4-momentum of the corresponding photon ($\ell=q$
or $k$) in this process. For the polarization vector of the axial
vector quarkonium $\chi_{c1}$, we'll use a similar parametrization
suggested in~\cite{Benic:2024}
\begin{equation}
E_{T,\lambda}^{(\chi_{c1})}\left(p_{\chi_{c}}\right)=\left(\frac{\boldsymbol{\varepsilon}_{T}\cdot\boldsymbol{p}_{\perp}^{\chi_{c}}}{p_{\chi_{c}}^{-}},\,0,\,\boldsymbol{\boldsymbol{\varepsilon}}_{T}\right),\qquad E_{L}^{(\chi_{c1})}\left(p_{\chi_{c}}\right)=\left(\frac{\left(\boldsymbol{p}_{\perp}^{\chi_{c}}\right)^{2}-M_{\chi_{c}}^{2}}{2M_{\chi_{c}}p_{\chi_{c}}^{-}},\,\frac{p_{\chi_{c}}^{-}}{M_{\chi_{c}}},\,\frac{\boldsymbol{p}_{\perp}^{\chi_{c}}}{M_{\chi_{c}}}\right),\label{eq:PolVector-1}
\end{equation}
where the subindices $L$ and $T=\pm1$ distinguish the transverse
and longitudinal polarizations. This parametrization automatically
satisfies the transversality condition $p_{\chi_{c}}\cdot E^{(\chi_{c})}=0$.
The polarization tensor $E_{\mu\nu}$ of the charmonium $\chi_{c2}$
can be constructed from polarization vectors defined in~(\ref{eq:PolVector-1})
using conventional Clebsch-Gordan coefficients as
\begin{equation}
E_{\mu\nu}^{(\chi_{c2})}=\left\{ \begin{array}{cc}
\left(E_{T=\pm1}^{(\chi_{c1})}\right)_{\mu}\,\left(E_{T=\pm1}^{(\chi_{c1})}\right)_{\nu}, & H_{\chi_{c2}}=\pm2,\\
\frac{\left(E_{T=\pm1}^{(\chi_{c1})}\right)_{\mu}\,\left(E_{L}^{(\chi_{c1})}\right)_{\nu}+\left(E_{L}^{(\chi_{c1})}\right)_{\mu}\,\left(E_{T=\pm1}^{(\chi_{c1})}\right)_{\nu}}{\sqrt{2}},\qquad & H_{\chi_{c2}}=\pm1,\\
\frac{\left(E_{T=+1}^{(\chi_{c1})}\right)_{\mu}\,\left(E_{T=-1}^{(\chi_{c1})}\right)_{\nu}+\left(E_{T=-1}^{(\chi_{c1})}\right)_{\mu}\,\left(E_{T=+1}^{(\chi_{c1})}\right)_{\nu}+2\left(E_{L}^{(\chi_{c1})}\right)_{\mu}\,\left(E_{L}^{(\chi_{c1})}\right)_{\nu}}{\sqrt{6}},\qquad & H_{\chi_{c2}}=0,
\end{array}\right.\label{eq:PolVector-2}
\end{equation}
where $H_{\chi_{c2}}$ is the helicity projection of the corresponding
charmonium. The wave function of the $\chi_{c}$ mesons in helicity
basis in momentum space is given by 
\begin{equation}
\Psi_{H,h\bar{h}}=\bar{u}_{h}\left(k_{Q}\right)\Gamma_{\chi_{c}}\left(k_{Q},\,k_{\bar{Q}}\right)v_{\bar{h}}\left(k_{\bar{Q}}\right)\phi_{\chi_{cJ}}\left(z,\,\boldsymbol{k}_{\perp}^{({\rm rel})}\right),\quad\Gamma_{\chi_{c}}\left(k_{Q},\,k_{\bar{Q}}\right)=\left\{ \begin{array}{cc}
1, & J=0,\\
i\gamma_{5}\gamma^{\mu}E_{\mu}^{(\chi_{c1})},\qquad & J=1,\\
\frac{\gamma^{\mu}K^{\nu}+\gamma^{\nu}K^{\mu}}{2}E_{\mu\nu}^{(\chi_{c2})} & J=2.
\end{array}\right.\label{eq:PhiDef}
\end{equation}
where $h,\bar{h}$ are the helicities of the quark, $z$ and $\boldsymbol{k}_{\perp}^{({\rm rel})}$
are the light-cone fraction and transverse momentum of the quark (w.r.t
direction of $\chi_{c}$ meson), and the momentum of relative motion
$K^{\mu}\equiv k_{Q}^{\mu}-k_{\bar{Q}}^{\mu}$ is expressed in terms
of these variables as 
\begin{align}
K^{\mu} & =\left(\left(2z-1\right)\alpha_{\chi_{c}}\sqrt{\frac{s}{2}},\,\frac{2\boldsymbol{p}_{\perp}^{\chi_{c}}\cdot\boldsymbol{K}_{\perp}+\left(1-2z\right)\left[\left(\boldsymbol{p}_{\perp}^{\chi_{c}}\right)^{2}+M_{\chi_{c}}^{2}\right]}{\alpha_{\chi_{c}}\sqrt{2s}},\,\boldsymbol{K}_{\perp}\right).\label{eq:KVec}
\end{align}
The $K^{-}$-component of the vector~$K^{\mu}$ was fixed from the
orthogonality condition $K\cdot p_{\chi_{c}}=0$, which follows from
the fact that the quarks are (nearly) onshell in the heavy quark mass
limit. The transverse component of $K^{\mu}$ is related to $\boldsymbol{k}_{\perp}^{({\rm rel})}$
as 
\begin{equation}
\boldsymbol{k}_{\perp}^{({\rm rel})}=\bar{z}\boldsymbol{k}_{Q}-z\boldsymbol{k}_{\bar{Q}}=\boldsymbol{K}_{\perp}-(2z-1)\boldsymbol{p}_{\perp}^{\chi_{c}}.
\end{equation}
For $\chi_{c0}$ mesons, the Fourier transformation of~(\ref{eq:PhiDef})
over the transverse coordinates allows to obtain the light-cone wave
function 
\begin{equation}
\Psi_{h\bar{h}}^{\left(\chi_{c0}\right)}\left(z,\,r\right)=-\frac{1}{z\bar{z}}\left[ihe^{-ih\theta}\delta_{h,\bar{h}}\partial_{r}+m\left(z-\bar{z}\right)\delta_{h,-\bar{h}}\right]\Phi_{\chi_{c0}}\left(z,\,\boldsymbol{r}_{\perp}\right),\label{eq:Chic0}
\end{equation}
where $\boldsymbol{r}=\{r\cos\theta,\,r\sin\theta\}$ is the vector
of relative separation of the quark and antiquark in transverse plane,
$z$ is the fraction of the momentum carried by the quark, and the
configuration space wave function $\Phi_{\chi_{c}}$ is related to the function $\phi_{\chi_{c}}\left(z,\,\boldsymbol{k}_{\perp}^{({\rm rel})}\right)$
by relation
\begin{equation}
\Phi_{\chi_{cJ}}\left(z,\,\boldsymbol{r}_{\perp}\right)=\int\frac{d^{2}\boldsymbol{k}_{\perp}^{({\rm rel})}}{\left(2\,\pi\right)^{2}}\frac{\sqrt{z\bar{z}}\,\phi_{\chi_{cJ}}\left(z,\,\boldsymbol{k}_{\perp}^{({\rm rel})}\right)e^{i\boldsymbol{k}_{\perp}^{({\rm rel})}\cdot\boldsymbol{r}_{\perp}}}{\left(\boldsymbol{k}_{\perp}^{({\rm rel})}\right)^{2}-M_{\chi_{cJ}}^{2}z\left(1-z\right)+m^{2}}.\qquad J=0,1,2\label{eq:Ph}
\end{equation}
The integrand of (\ref{eq:Ph}) includes additional light-cone denominator
whose origin will become obvious in the next section (see the denominators
in the last lines of (\ref{eq:a-1}) and (\ref{eq:a-1-1}) below;
also see Appendix of~\cite{Siddikov:2024bre} for a more detailed
discussion). For the helicity components of the  $\chi_{c1}$
and $\chi_{c2}$ we may obtain in a similar fashion
\begin{align}
\Psi_{H=\pm1,\,h\bar{h}}^{\left(\chi_{c1}\right)}\left(z,\,r\right) & =\frac{\sqrt{2}}{z\bar{z}}\left[-e^{iH\theta}\left(\bar{z}\delta_{h,H}\delta_{\bar{h},-H}+z\delta_{h,-H}\delta_{\bar{h},H}\right)\partial_{r}-iHm\left(z-\bar{z}\right)\delta_{h,H}\delta_{\bar{h},H}\right]\Phi_{\chi_{c1}}\left(z,\,r\right),\label{eq:Chic1a}\\
\Psi_{H=0,\,h\bar{h}}^{\left(\chi_{c1}\right)}\left(z,\,r\right) & =\frac{2i}{z\bar{z}M_{\chi_{c1}}}\left[h\delta_{h,-\bar{h}}\,\Delta_{r}^{(+)}+ime^{-ih\theta}\delta_{h,\bar{h}}\partial_{r}\right]\Phi_{\chi_{c1}}\left(z,\,r\right),\label{eq:Chic1b}
\end{align}

where $\Delta_{r}^{(\pm)}=\partial_{r}^{2}\pm\left(1/r\right)\partial_{r}$, $\bar z= 1-z$, and
\begin{align}
\Psi_{H=+2,\,h\bar{h}}^{\left(\chi_{c2}\right)}\left(z,\,r\right) & =\frac{2}{z\bar{z}}\left[e^{2i\theta}\left(z\delta_{h,+}\delta_{\bar{h},-}-\bar{z}\delta_{h,-}\delta_{\bar{h},+}\right)\Delta_{r}^{(-)}+ime^{i\theta}\delta_{h,+}\delta_{\bar{h},+}\partial_{r}\right]\Phi_{\chi_{c2}}\left(z,\,r\right),\label{eq:Chic2a}\\
\Psi_{H=-2,\,h\bar{h}}^{\left(\chi_{c2}\right)}\left(z,\,r\right) & =\frac{2}{z\bar{z}}\left[-e^{-2i\theta}\left(\bar{z}\delta_{h,+}\delta_{\bar{h},-}-z\delta_{h,-}\delta_{\bar{h},+}\right)\Delta_{r}^{(-)}-ime^{-i\theta}\delta_{h,-}\delta_{\bar{h},-}\partial_{r}\right]\Phi_{\chi_{c2}}\left(z,\,r\right),\\
\Psi_{H=+1,\,h\bar{h}}^{\left(\chi_{c2}\right)}\left(z,\,r\right) & =\frac{M_{\chi_{c2}}}{z\bar{z}}\left[ie^{i\theta}\left(\left(3z-4z^{2}\right)\delta_{h,+}\delta_{\bar{h},-}+\left(3\bar{z}-4\bar{z}^{2}\right)\delta_{h,-}\delta_{\bar{h},+}\right)\partial_{r}+m\left(z-\bar{z}\right)\delta_{h,+}\delta_{\bar{h},+}\right]\Phi_{\chi_{c2}}\left(z,\,r\right),\\
\Psi_{H=-1,\,h\bar{h}}^{\left(\chi_{c2}\right)}\left(z,\,r\right) & =\frac{M_{\chi_{c2}}}{z\bar{z}}\left[-ie^{-i\theta}\left(\left(3\bar{z}-4\bar{z}^{2}\right)\delta_{h,+}\delta_{\bar{h},-}+\left(3z-4z^{2}\right)\delta_{h,-}\delta_{\bar{h},+}\right)\partial_{r}+m\left(z-\bar{z}\right)\delta_{h,-}\delta_{\bar{h},-}\right]\Phi_{\chi_{c2}}\left(z,\,r\right),\\
\Psi_{H=0,\,h\bar{h}}^{\left(\chi_{c2}\right)}\left(z,\,r\right) & =\frac{\sqrt{2/3}}{z\bar{z}}\left[\delta_{h,-\bar{h}}\,\left(-3\Delta_{r}^{(+)}+2m^{2}\left(z-\bar{z}\right)\right)-ime^{-ih\theta}\delta_{h,\bar{h}}\partial_{r}\right]\Phi_{\chi_{c2}}\left(z,\,r\right).\label{eq:Chic2c}
\end{align}

\subsubsection{Relation to NRQCD LDMEs}

In order to understand the relations of the light-cone distribution
amplitudes to long distance matrix elements (LDMEs) of NRQCD, we need
to analyze the behaviour of the wave functions in the limit when $K^{\mu}$
becomes small. Since we are interested in the $P$-wave quarkonia,
in the limit $K^{\mu}\to0$ the corresponding vertices should vanish.
This is obvious for $\chi_{c2}$ due to explicit $K^{\mu}$ in~(\ref{eq:PhiDef}).
To demonstrate that a similar property is valid for the states $J=0$
and $J=1$, we need to rewrite~(\ref{eq:PhiDef}) in equivalent form,
using the Dirac equation for the $\bar{u},\,v$ spinors and the orthogonality
of the polarization vector $E_{H}\cdot p_{\chi_{c}}=0$, namely 
\begin{align}
\bar{u}\left(k_{Q}\right) & \Gamma_{\chi_{cJ}}v\left(k_{\bar{Q}}\right)=\frac{1}{2m}\bar{u}\left(k_{Q}\right)\left(m\Gamma_{\chi_{cJ}}+\Gamma_{\chi_{cJ}}m\right)v\left(k_{\bar{Q}}\right)=\frac{1}{2m}\bar{u}\left(k_{Q}\right)\left(\hat{k}_{Q}\Gamma_{\chi_{cJ}}-\Gamma_{\chi_{cJ}}\hat{k}_{1}\right)v\left(k_{\bar{Q}}\right)\label{eq:Ga}\\
 & =\frac{1}{4m}\bar{u}\left(k_{Q}\right)\left(\hat{p}_{\chi_{c}}\Gamma_{\chi_{cJ}}-\Gamma_{\chi_{cJ}}\hat{p}_{\chi_{c}}\right)v\left(k_{\bar{Q}}\right)+\frac{1}{4m}\bar{u}\left(k_{Q}\right)\left(\hat{K}\Gamma_{\chi_{cJ}}+\Gamma_{\chi_{cJ}}\hat{K}\right)v\left(k_{\bar{Q}}\right)=\bar{u}\left(k_{Q}\right)\Gamma_{\chi_{cJ}}^{({\rm eff})}v\left(k_{\bar{Q}}\right),\nonumber 
\end{align}
where $\Gamma_{\chi_{cJ}}^{({\rm eff})}=K^{\mu}\left(\gamma^{\mu}\Gamma_{\chi_{cJ}}+\Gamma_{\chi_{cJ}}\gamma^{\mu}\right)/4m=\gamma_{\chi_{cJ}}^{\mu}K_{\mu}$,
and the momentum-independent matrices $\gamma_{\chi_{cJ}}^{\mu}$
for different quarkonia are given by $\gamma_{\chi_{c0}}^{\mu}\equiv\gamma^{\mu}/2m$,
$\gamma_{\chi_{c1}}^{\mu}\equiv\left[\gamma^{\mu},\,\hat{E}_{H}\right]\gamma^{5}/4m$,
and $\gamma_{\chi_{c2}}^{\mu}=E^{\mu\nu}\gamma_{\nu}$. In the amplitudes
of physical processes, the wave function~(\ref{eq:PhiDef}) contributes
in convolution with amplitude of hard partonic process. If the latter
possess a hard scale which exceeds significantly the relative momentum
$K^{\mu}$ ($\sim$inverse charmonium radius), we can disregard completely
the dependence on $z,\,\boldsymbol{k}_{\perp}^{({\rm rel})}$ in the
hard partonic amplitude, and the convolution over $z,\,\boldsymbol{K}_{\perp}$
will affect only the arguments of the wave function, yielding
\begin{align}
\int dz & \int\frac{\,d^{2}\boldsymbol{k}_{\perp}^{({\rm rel})}}{\left(2\pi\right)^{2}}\Phi_{\chi_{c}}\left(z,\,\boldsymbol{k}_{\perp}^{({\rm rel})}\right)=\label{eq:DAdefEta-1}\\
 & =\lim_{\lambda,\,\boldsymbol{r}_{\perp}\to0}\left\langle 0\left|\bar{\Psi}\left(-\frac{\lambda}{2}n_{+}-\frac{\boldsymbol{r}_{\perp}}{2}\right)\gamma_{\chi_{cJ}}^{\mu}i\overleftrightarrow{\partial}_{\mu}\,\mathcal{L}\left(-\frac{\lambda}{2}n_{+}-\frac{\boldsymbol{r}_{\perp}}{2},\,\frac{\lambda}{2}n_{+}+\frac{\boldsymbol{r}_{\perp}}{2}\right)\Psi\left(\frac{\lambda}{2}n_{+}+\frac{\boldsymbol{r}_{\perp}}{2}\right)\right|\chi_{c}(p)\right\rangle \nonumber \\
 & =\left\langle 0\left|\bar{\Psi}\left(0\right)\gamma_{\chi_{cJ}}^{\mu}\frac{i}{2}\overleftrightarrow{D}_{\mu}\,\Psi\left(0\right)\right|\chi_{c}(p)\right\rangle \nonumber 
\end{align}
where $\overleftrightarrow{D}_{\mu}=\overrightarrow{\partial_{\mu}}-\overleftarrow{\partial_{\mu}}+2igA_{\mu}^{a}t_{a}+2ieA_{\mu}$
is the covariant derivative, and we replaced $K_{\mu}$ with $i\left(\overrightarrow{\partial_{\mu}}-\overleftarrow{\partial_{\mu}}\right)$
in the configuration space. The square of the matrix element in the
last line of~(\ref{eq:DAdefEta-1}) in the nonrelativistic limit
reduces to NRQCD LDMEs. Indeed, using explicit form of $\gamma_{\chi_{cJ}}^{\mu}$
for $J=0,1,2$, we may recover the familiar NRQCD operators from~\cite{Ma:2006hc,Wang:2013ywc,Wang:2017bgv}~\footnote{In the rest frame of the proton, we may further simplify and rewrite
these operators as ~\cite{Bodwin:1994jh}
\begin{equation}
\left|\left\langle 0\left|\bar{\psi}\left(0\right)\gamma_{\chi_{cJ}}^{\mu}i\overleftrightarrow{D}_{\mu}\,\psi\left(0\right)\right|\chi_{c}(p)\right\rangle \right|^{2}=\left\{ \begin{array}{cc}
\left|\left\langle 0\left|\psi^{\dagger}\left(-\frac{i}{2}\overleftrightarrow{\boldsymbol{D}}\cdot\boldsymbol{\sigma}\right)\,\chi\right|\chi_{c}(p)\right\rangle \right|^{2}=\left\langle \mathcal{\mathcal{O}}_{\chi_{c}}^{[1]}\left(^{3}P_{0}^{[1]}\right)\right\rangle , & \qquad J=0\\
\left|\left\langle 0\left|\psi^{\dagger}\left(-\frac{i}{2}\overleftrightarrow{\boldsymbol{D}}\times\boldsymbol{\sigma}\right)\,\chi\right|\chi_{c}(p)\right\rangle \right|^{2}=\left\langle \mathcal{\mathcal{O}}_{\chi_{c}}^{[1]}\left(^{3}P_{1}^{[1]}\right)\right\rangle , & \qquad J=1\\
\left|\left\langle 0\left|\psi^{\dagger}\left(-\frac{i}{2}\overleftrightarrow{D^{(i}}\sigma^{j)}\right)\,\chi\right|\chi_{c}(p)\right\rangle \right|^{2}=\left\langle \mathcal{\mathcal{O}}_{\chi_{c}}^{[1]}\left(^{2}P_{2}^{[1]}\right)\right\rangle , & \qquad J=2
\end{array}\right.
\end{equation}
}
\begin{equation}
\left|\left\langle 0\left|\bar{\psi}\left(0\right)\gamma_{\chi_{cJ}}^{\mu}i\overleftrightarrow{D}_{\mu}\,\psi\left(0\right)\right|\chi_{c}(p)\right\rangle \right|^{2}=\frac{1}{4m^{2}}\times\left\{ \begin{array}{cc}
\left|\left\langle 0\left|\bar{\psi}\left(\frac{i}{2}\overleftrightarrow{D}_{\top}^{\mu}\gamma_{\mu}^{\top}\right)\,\chi\right|\chi_{c}(p)\right\rangle \right|^{2}=\left\langle \mathcal{\mathcal{O}}_{\chi_{c}}^{[1]}\left(^{3}P_{0}^{[1]}\right)\right\rangle , & \qquad J=0\\
\left|\left\langle 0\left|\psi^{\dagger}\left(-\frac{i}{4}\overleftrightarrow{D}_{\top}^{\nu}\left[\gamma_{\nu}^{\top},\,\gamma_{\mu}^{\top}\right]\right)\gamma_{5}\,\chi\right|\chi_{c}(p)\right\rangle \right|^{2}=\left\langle \mathcal{\mathcal{O}}_{\chi_{c}}^{[1]}\left(^{3}P_{1}^{[1]}\right)\right\rangle , & \qquad J=1\\
\left|\left\langle 0\left|\psi^{\dagger}\left(-\frac{i}{2}\overleftrightarrow{D_{\top}^{(\mu}}\gamma_{\top}^{\nu)}\right)\,\chi\right|\chi_{c}(p)\right\rangle \right|^{2}=\left\langle \mathcal{\mathcal{O}}_{\chi_{c}}^{[1]}\left(^{3}P_{2}^{[1]}\right)\right\rangle , & \qquad J=2
\end{array}\right.
\end{equation}
where $\top$ implies a part of the vector which is transverse to
$p_{\chi_{c}}^{\mu}$, namely $v_{\top}^{\mu}=\left(g^{\mu\nu}-p_{\chi_{c}}^{\mu}p_{\chi_{c}}^{\nu}/M_{\chi_{c}}^{2}\right)v^{\nu}$;
$a_{\top}^{(\mu}b_{\top}^{\nu)}=(a_{\top}^{\mu}b_{\top}^{\nu}+a_{\top}^{\nu}b_{\top}^{\mu})/2-a_{\top}\cdot b_{\top}\left(g^{\mu\nu}-p_{\chi_{c}}^{\mu}p_{\chi_{c}}^{\nu}/M_{\chi_{c}}^{2}\right)/3$,
the operators $\psi^{\dagger},\,\chi$ are the Pauli spinor operators
which create a quark and an antiquark, respectively. The presence
of the gauge fields in covariant derivative $\overleftrightarrow{D}_{\mu}$
implies that we should consider the $\chi_{c}$ meson as a mixture
of $|\bar{Q}Q\rangle$, $|\bar{Q}Qg\rangle$ and $|\bar{Q}Q\gamma\rangle$
Fock components unless we work in the light-cone gauge $A^{-}=0$,
$A_{a}^{-}=0$. In what follows we will use the latter gauge condition,
both for photons and gluons.

In the heavy quark mass, it is expected that the color singlet LDMEs
$\left\langle \mathcal{\mathcal{O}}_{\chi_{c}}^{[1]}\left(^{3}P_{J}^{[1]}\right)\right\rangle $
should coincide, and in the potential models can be represented in
terms of the slope of the radial wave functions $R'_{\chi_{c}}(0)$
as, 
\begin{equation}
\left\langle \mathcal{\mathcal{O}}_{\chi_{c}}^{[1]}\left(^{3}P_{J}^{[1]}\right)\right\rangle =6N_{c}\left(2J+1\right)\left|R'_{\chi_{c}}(0)\right|^{2}/4\pi.\label{eq:LDME}
\end{equation}
The latter parameter may be related to the light-cone function $\phi_{\chi_{cJ}}\left(z,\,\boldsymbol{k}_{\perp}^{({\rm rel})}\right)$
introduced in~(\ref{eq:PhiDef}): after straightforward but tedious
substitution of the variables, we may obtain~\cite{Babiarz:2020jkh}
\begin{equation}
3\sqrt{\frac{\pi}{2}}R'_{\chi_{c}}(0)=\sqrt{2M_{\chi_{cJ}}}\int\frac{dz\,d^{2}\boldsymbol{k}_{\perp}^{({\rm rel})}}{z(1-z)8\pi^{2}}\phi_{\chi_{cJ}}\left(z,\,\boldsymbol{k}_{\perp}^{({\rm rel})}\right)\left[M_{\chi_{cJ}}^{2}\left(z-\frac{1}{2}\right)^{2}+\boldsymbol{K}_{\perp}^{2}\right].\label{eq:Rel}
\end{equation}
In what follows we will use the explicit parametrization from~\cite{Benic:2024}
which in the light-cone representation is given by
\begin{equation}
\Phi_{\chi_{cJ}}\left(z,\,r\right)=\mathcal{N}_{\chi_{cJ}}z(1-z)\exp\left(-\frac{\mathcal{R}_{\chi_{cJ}}^{2}}{8}\left(\frac{m^{2}}{z(1-z)}-4m^{2}\right)-\frac{2z(1-z)r^2}{\mathcal{R}_{\chi_{cJ}}^{2}}\right),\label{eq:ExplicitX}
\end{equation}
and $\mathcal{N}_{\chi_{cJ}},\mathcal{R}_{\chi_{cJ}}$ are numerical
constants given explicitly by $\mathcal{R}_{\chi_{c0}}\approx1.54\,{\rm GeV}^{-1}$,
$\mathcal{R}_{\chi_{c1}}=\mathcal{R}_{\chi_{c2}}\approx1.48\,{\rm GeV^{-1}}$,
$\mathcal{N}_{\chi_{c0}}\approx1.15$ GeV$^{-1}$, $\mathcal{N}_{\chi_{c1}}\approx1.39$
GeV$^{-1}$, $\mathcal{N}_{\chi_{c2}}\approx0.60$ GeV$^{-1}$~\footnote{We disregard tiny $\sim1\%$ differences of these constants for different helicity states (polarizations) of the same meson, since these differences are smaller than the precision of experimental data for $\chi_{c}\to\gamma\gamma$
used in~~\cite{Benic:2024} (around 3-4\%); besides, a
comparable uncertainty may come from the theoretical NNLO corrections
$\sim\alpha_{s}^{2}\left(m_{c}\right)\approx10\%$.}. The values of $\left|R'(0)\right|$ obtained with this parametrization
are in reasonable agreement with the value $\left|R'(0)\right|^{2}=0.075\,{\rm GeV}^{5}$
found in potential models and widely used in the literature for numerical
estimates. For the momentum space wave function, we may obtain
\begin{equation}
\tilde\Phi_{\chi_{cJ}}\left(z,\,\boldsymbol{k}\right)\equiv\int d^2 \boldsymbol{r}e^{-i\boldsymbol{k}\cdot \boldsymbol{r}}\Phi_{\chi_{cJ}}\left(z,\,\boldsymbol{r}\right)=\frac{\sqrt{z\bar z}\phi_{\chi_{cJ}}\left(z,\,\boldsymbol{k}\right)}{ \boldsymbol{k}^{2}-M_{\chi_{cJ}}^{2}z\bar z+m^{2}}=\frac{\pi \mathcal{N}_{\chi_{cJ}}\mathcal{R}_{\chi_{cJ}}^{2}}{2}\exp\left(-\frac{\mathcal{R}_{\chi_{cJ}}^{2}}{8}\left(\frac{\boldsymbol{k}^2+m^{2}(1-4z\bar z)}{z\bar z}\right)\right).
 \label{eq:ExplicitX}
\end{equation}

\subsection{Scattering processes in the CGC framework}

\label{subsec:Derivation} In the CGC framework the interaction of
the high energy partons with the gluonic field of the target is described
in eikonal approximation by the Wilson line $U(\boldsymbol{x}_{\perp})$~
defined as
\begin{equation}
U\left(\boldsymbol{x}_{\perp}\right)=P\exp\left(ig\int dx^{-}A_{a}^{+}\left(x^{-},\,\boldsymbol{x}_{\perp}\right)t^{a}\right),\label{eq:Wilson}
\end{equation}
where we use notation $\boldsymbol{x}_{\perp}$ for the impact parameter
of the parton, $A_{a}^{\mu}$ for the gluonic field of the target,
and $t_{a}$ are the usual color group generators in the irreducible
representation which corresponds to the parton ( $\boldsymbol{3}$,
$\boldsymbol{\bar{3}}$ or $\boldsymbol{8}$ for quark, antiquark
or gluon, respectively). The interaction encoded in Wilson link~(\ref{eq:Wilson})
formally may be reformulated in terms of the CGC Feynman rules~~\cite{Blaizot:2004,Ayala:2017rmh,Caucal:2021ent,Caucal:2022ulg}
provided in Table~\ref{tab:FR}.

\begin{table}
\begin{tabular}{|c|c|}
\hline 
\begin{minipage}[t]{0.4\columnwidth}%
\textcolor{white}{.}\\
 \includegraphics[scale=0.7]{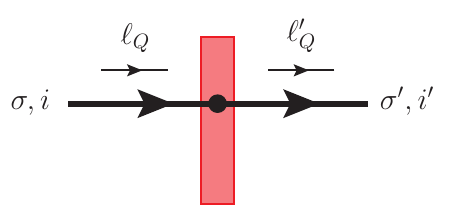}%
\end{minipage} & %
\begin{minipage}[t]{0.55\columnwidth}%
The interaction vertex of the quark with the shock wave: 
\[
T_{\sigma,\sigma';\,i,i'}^{Q}=2\pi\delta\left(\ell_{Q}^{+}-\ell'{}_{Q}^{+}\right)\gamma_{\sigma',\,\sigma}^{+}\int d^{2}\boldsymbol{z}e^{i\left(\boldsymbol{\ell}_{Q}-\boldsymbol{\ell}_{Q}'\right)\cdot\boldsymbol{z}}\left[U\left(\boldsymbol{z}\right)-1\right]_{i',i}
\]
\end{minipage}\tabularnewline
\hline 
\begin{minipage}[t]{0.4\columnwidth}%
\textcolor{white}{.}\\
 \includegraphics[scale=0.7]{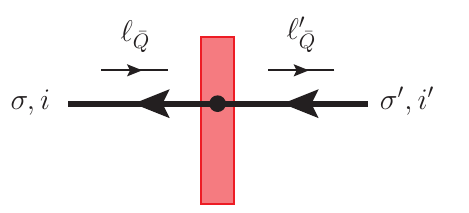}%
\end{minipage} & %
\begin{minipage}[t]{0.55\columnwidth}%
The interaction vertex of the antiquark with the shock wave: 
\[
T_{\sigma,\sigma';\,i,i'}^{\bar{Q}}=-2\pi\delta\left(\ell_{\bar{Q}}^{+}-\ell'{}_{\bar{Q}}^{+}\right)\gamma_{\sigma,\,\sigma'}^{+}\int d^{2}\boldsymbol{z}e^{i\left(\boldsymbol{\ell}_{\bar{Q}}-\boldsymbol{\ell}'_{\bar{Q}}\right)\cdot\boldsymbol{z}}\left[U^{\dagger}\left(\boldsymbol{z}\right)-1\right]_{i,i'}
\]
\end{minipage}\tabularnewline
\hline 
\begin{minipage}[t]{0.4\columnwidth}%
\textcolor{white}{.}\\
 \includegraphics[scale=0.7]{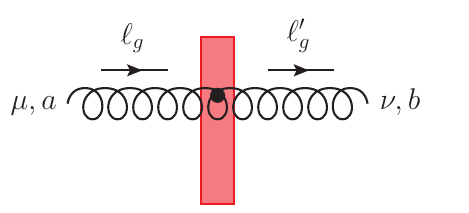}%
\end{minipage} & %
\begin{minipage}[t]{0.55\columnwidth}%
The interaction vertex of the gluon with the shock wave: 
\[
T_{\mu,\nu;a,b}^{g}=-2\pi\delta\left(\ell_{g}^{+}-\ell'{}_{g}^{+}\right)g_{\mu\nu}{\rm sgn}\left(\ell_{g}^{+}\right)\times
\]
\[
\quad\quad\times\int d^{2}\boldsymbol{z}e^{-i\left(\boldsymbol{\ell}'_{g}-\boldsymbol{\ell}_{g}\right)\cdot\boldsymbol{z}}\left[\mathcal{U}^{{\rm sgn}\left(\ell_{g}^{+}\right)}\left(\boldsymbol{z}\right)-1\right]_{b,a}
\]
\end{minipage}\tabularnewline
\hline 
\end{tabular}\caption{The CGC Feynman rules~~\cite{Blaizot:2004,Ayala:2017rmh,Caucal:2021ent,Caucal:2022ulg}
for the interaction of high energy partons with the target encoded
in Wilson link~(\ref{eq:Wilson}). The red-colored block stands for
the shock wave (Wilson line). We use notations $\ell_{i},\ell'_{i}$
($i=Q,\bar{Q},g$) for the momenta of the corresponding parton before
and after the interaction, $\sigma,\sigma'$ for Dirac indices of
the quarks, $i,\,i'$ for the color indices in the fundamental representations
$\boldsymbol{3}$ or $\bar{\boldsymbol{3}}$ of the color group, and
$a,\,b$ for the gluon color indices in the adjoint representation
$\boldsymbol{8}$. The matrices $U$ and $\mathcal{U}$ are the Wilson
lines in the fundamental and the adjoint representations. The prefactor
$2\pi\delta\left(\ell_{i}^{+}-\ell'{}_{i}^{+}\right)$ in the interaction
vertices reflects conservation of the longitudinal momentum of the
parton in eikonal approximation. We may also observe that interaction
with a shock wave does not change transverse coordinates of the partons
in the mixed (light-cone) representation, and using identities $\bar{u}_{h_{1}}(p)\gamma^{+}u_{h_{2}}(q)=\bar{v}_{h_{1}}(p)\gamma^{+}v_{h_{2}}(q)=2\delta_{h_{1},h_{2}}\sqrt{p^{+}q^{+}}$
from~\cite{Brodsky:1997de}, we may see that the interaction remains
diagonal in helicity basis.}
\label{tab:FR}
\end{table}

In CGC picture it is assumed that the gluonic field $A_{a}^{\mu}$
is created by randomly distributed color sources, and probability
to find a given density of color sources $\rho_{a}(x^{-},\boldsymbol{x}_{\perp})$
in the target is controlled by the weight functional $W[\rho]$~\cite{McLerran:1993ni,McLerran:1993ka,McLerran:1994vd}.
The component $A_{a}^{+}$ which appears in~(\ref{eq:Wilson}) may
be related to density of these source $\rho_{a}$ as $A_{a}^{+}(x)=-\frac{1}{\nabla_{\perp}^{2}}\rho_{a}(x^{-},\,\boldsymbol{x}_{\perp})$.
For any physical observable, the expected values may be obtained averaging
the observable $\mathcal{A[\rho]}$ found at fixed distribution of
charges $\rho(x^{-},\boldsymbol{x}_{\perp})$ over all possible configurations
$\rho(x^{-},\boldsymbol{x}_{\perp})$, namely 
\begin{equation}
\left\langle \mathcal{A}\right\rangle =\int\mathcal{D}\rho\,W[\rho]\,\mathcal{A}[\rho],\label{eq:Ave}
\end{equation}
where the angular brackets $\langle...\rangle$ stand for the above-mentioned
averaging. The analytic evaluation of the integral $\mathcal{D}\rho$
is possible only for a few simple choices of $W[\rho]$, as, for example,
gaussian parametrization~\cite{McLerran:1993ni,McLerran:1993ka},
and development of phenomenological models for the latter is very
challenging. Fortunately, in presence of hard scales which limit the
number of partons in the initial state, all physical amplitudes may
be represented as convolutions of process-dependent perturbative impact
factors and universal (target-dependent) correlators of Wilson lines
(multipole scattering amplitudes). For many exclusive processes, the
dominant contribution is controlled by the lowest order dipole scattering
amplitude,

\begin{equation}
\mathcal{N}\left(x,\,\boldsymbol{r},\,\boldsymbol{b}\right)=1-S_{2}\left(Y=\ln\left(\frac{1}{x}\right),\,\boldsymbol{x}_{q},\,\boldsymbol{x}_{\bar{q}}\right),\quad S_{2}\left(Y,\,\boldsymbol{x}_{q},\,\boldsymbol{x}_{\bar{q}}\right)=\frac{1}{N_{c}}\left\langle {\rm tr}\left(U\left(\boldsymbol{\boldsymbol{x}}_{q}\right)U^{\dagger}\left(\boldsymbol{x}_{\bar{q}}\right)\right)\right\rangle _{Y}\label{eq:NS}
\end{equation}
where $Y$ is the rapidity of the dipole, $\boldsymbol{x}_{q},\boldsymbol{x}_{\bar{q}}$
are the impact parameters of the quark, and the variables $\boldsymbol{r}\equiv\boldsymbol{x}_{q}-\boldsymbol{x}_{\bar{q}}$
and $\boldsymbol{b}\equiv\alpha_{q}\,\boldsymbol{x}_{q}+\alpha_{\bar{q}}\boldsymbol{x}_{\bar{q}}$
determine the transverse separation and the impact parameter of the
dipole's center of mass. In phenomenological studies the dipole amplitude
can be constrained from both exclusive and inclusive channels. Due
to optical theorem, the inclusive channels are sensitive only to the
imaginary part of the amplitude, and in order to have a consistent
description, for exclusive channels, the real part of the amplitude
should be taken into account. As was demonstrated long ago in~\cite{Bronzan:1974jh,Gotsman:1992ui},
if the amplitude grows as a function of invariant energy as $\sim s^{\alpha}$,
then the real part of the amplitude becomes proportional to the imaginary
part, and thus the full amplitude of exclusive process can be restored
from the imaginary part using a set of identities
\begin{equation}
\mathcal{A}=\left(\beta+i\right)\text{{\rm Im}}\,\mathcal{A},\qquad\left|\mathcal{A}\right|^{2}=\left(1+\beta^{2}\right)\left|{\rm Im}\mathcal{A}\right|^{2},\quad\beta\equiv\frac{{\rm Re\,\mathcal{A}}}{{\rm Im}\,\mathcal{A}}=\tan\left(\frac{\pi\alpha}{2}\right),\quad\alpha=\frac{\partial\ln\mathcal{A}}{\partial\ln s}.\label{eq:reA}
\end{equation}
Following existing conventions, we will assume that $\mathcal{N}\left(x,\,\boldsymbol{r},\,\boldsymbol{b}\right)$
corresponds to the dipole amplitude extracted from inclusive processes,
tacitly assuming that a factor $\left(\beta+i\right)$ should be added
in exclusive channels. We also take into account the so-called skewedness
factor $R_{g}$ suggested in~\cite{ske}, which takes into account
that the effective value of the parameter $x$ in the dipole amplitude
may be modified due to nonzero longitudinal momentum transfer in $t$-channel
and its possible unequal sharing between the gluons in the shock wave.
Explicitly, this factor is given by 
\begin{equation}
R_{g}(\gamma)=\frac{2^{2\gamma+3}}{\sqrt{\pi}}\frac{\Gamma(\gamma+5/2)}{\Gamma(\gamma+4)}\approx2.4^{\gamma},\quad\text{where}\quad\gamma\equiv\frac{\partial\ln\left[xg(x,\mu^{2})\right]}{\partial\ln(1/x)}\approx{\rm const,}\label{eq:Rg}
\end{equation}
$g\left(x,\mu^{2}\right)$ is the gluon PDF, and in approximate expression
for $R_{g}$ we took into account that in the kinematics where CGC
is applicable, the parameter $\gamma\le0.5$. While the derivation
of~(\ref{eq:Rg}) given in~\cite{ske} relies on a model which
is not valid in the deeply saturated regime $x\ll1$, at moderate
values of $x\gtrsim10^{-3}$ the factor $R_{g}$leads to a mild increase
of the cross-section, and improves the phenomenological description
of various exclusive processes (see e.g.~\cite{watt:bcgc,Kowalski:2008sa,RESH}
for more details).

\subsection{Photoproduction of the $\chi_{c}\gamma$ pairs}

\label{subsec:Formalism} 
\begin{figure}
\includegraphics[width=6cm]{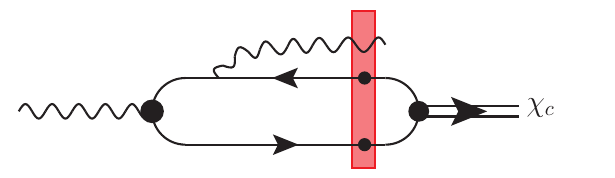}\includegraphics[width=6cm]{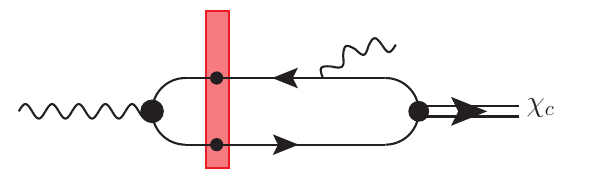}

\includegraphics[width=6cm]{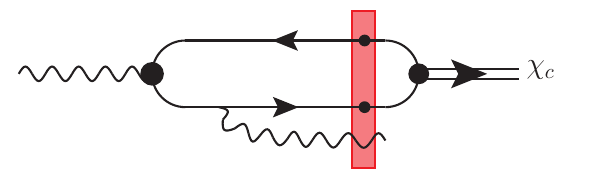}\includegraphics[width=6cm]{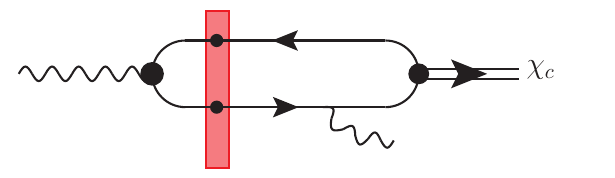}

\caption{The dominant mechanism of $\chi_{c}\gamma$ pair photoproduction in
the CGC framework in the leading order in $\alpha_{s}$. The red block
stands for a shock wave. The diagrams in the first and the second
row are related by charge conjugation (inversion of the quark line).
In the left column we disregard interaction of the emitted photon
with a shock wave as tiny $\mathcal{O}\left(\alpha_{{\rm em}}\right)$-correction,
and to emphasize this we don't put \textquotedblleft dot\textquotedblright{}
on a photon line which crosses the shock wave.}
\label{fig:CGCBasic-1}
\end{figure}

The evaluation of the exclusive photoproduction
of $\chi_{c}\gamma$ pairs resembles a similar evaluation of the $\eta_{c}\gamma$
photoproduction presented in~\cite{Siddikov:2024bre} and differs
only in the final state wave function, although yields a significantly
different results for the impact factors. Technically, it requires
evaluation of four diagrams shown in the Figure~\ref{fig:CGCBasic-1}.
In what follows we will separate the contributions with photon emission
before and after the interaction with a shockwave, so the amplitude
can be represented as
\begin{equation}
\mathcal{A}_{\gamma p\to\chi_{c}\gamma p}^{(\lambda;\,\sigma,H)}=\mathcal{A}_{1}^{(\lambda;\,\sigma,H)}+\mathcal{A}_{2}^{(\lambda;\,\sigma,H)}\label{eq:ASum}
\end{equation}
where $\mathcal{A}_{1}^{(\lambda,\sigma)}$ and $\mathcal{A}_{2}^{(\lambda,\sigma)}$
correspond to the left and right columns of the Figure~\ref{fig:CGCBasic-1}.
Since the interaction of the emitted photon with the shock wave may
be disregarded as $\mathcal{O}\left(\alpha_{{\rm em}}\right)$-correction,
both contributions are controlled by the dipole amplitude $\mathcal{N}\left(x,\,\boldsymbol{r},\,\boldsymbol{b}\right)$,
convoluted with appropriate impact factors. 

The detailed evaluation of these impact factors is straightforward
and is explained in detail in subsequent subsections. Following earlier
studies~\cite{Beuf:2021qqa,Beuf:2022ndu}, in all evaluations below
we will introduce the subindices 0,1,2 to distinguish the kinematic
variables associated with quark, antiquark and photon, respectively.
For example, the variables 
\begin{equation}
z_{0}\equiv\frac{k_{Q}^{+}}{q^{+}},\quad z_{1}\equiv\frac{k_{\bar{Q}}^{+}}{q^{+}},\quad z_{2}\equiv\frac{k_{\gamma}^{+}}{q^{+}}=1-\alpha_{\chi_{c}},\label{eq:lcFractions}
\end{equation}
are the fractions of the incoming photon's light-cone momentum $q^{+}$
carried by the quark, antiquark and photon when crossing the shock
wave, respectively, and $\boldsymbol{r}_{0},\,\boldsymbol{r}_{1},\,\boldsymbol{r}_{2}$
are their transverse coordinates (in configuration space). The light-cone
momentum conservation implies that for the diagrams in the left column
of the Figure~\ref{fig:CGCBasic-1}, the variables $z_{0},z_{1},z_{2}$
should satisfy 
\begin{equation}
z_{0}+z_{1}+z_{2}=1,\qquad z_{0}+z_{1}=\alpha_{\chi_{c}},
\end{equation}
whereas for the diagrams in the right column there is a similar constraint
\begin{equation}
z_{0}+z_{1}=1.
\end{equation}

\subsubsection{Evaluation of the amplitude $\mathcal{A}_{1}^{(\lambda;\,\sigma,H)}$}

\label{subsec:A1} 
\begin{figure}
\includegraphics[width=9cm]{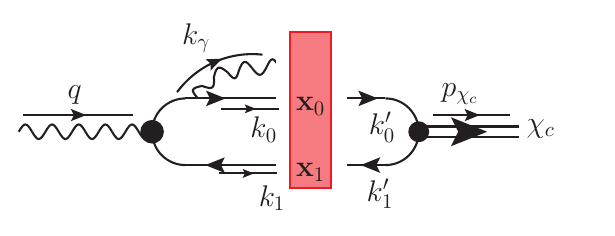}

\caption{The definitions of the parton momenta $k_{0},k_{1},k_{0}',k_{1}'$
before and after interaction with the shock wave. The variables $\boldsymbol{x}_{0},\boldsymbol{x}_{1}$
stand for the transverse coordinates of the quark and antiquark when
they pass through the shock wave.}
\label{fig:CGCBasic-2-1}
\end{figure}

The amplitude $\mathcal{A}_{1}^{(\lambda;\,\sigma,H)}$
obtains contributions of two charge conjugate diagrams shown in the
left column of the Figure~\ref{fig:CGCBasic-1}. We will start our
evaluation from the diagram shown in the upper left corner of the
Figure~\ref{fig:CGCBasic-1}, defining the partons' momenta as explained
in the Figure~\ref{fig:CGCBasic-2-1} and using the CGC Feynman rules
from Table~\ref{tab:FR}. \\
 
A straightforward evaluation yields
\begin{align}
\mathcal{A}_{1,1} & =4\pi\alpha_{{\rm em}}e_{c}^{2}\int d^{2}\boldsymbol{x}_{0}d^{2}\boldsymbol{x}_{1}\,\int\frac{d^{4}k_{0}}{\left(2\pi\right)^{4}}\frac{d^{4}k_{1}}{\left(2\pi\right)^{4}}\frac{d^{4}k_{0}'}{\left(2\pi\right)^{4}}\frac{d^{4}k_{1}'}{\left(2\pi\right)^{4}}\left(2\pi\right)^{4}\delta\left(k_{0}+k_{1}+k_{\gamma}-q\right)\left(2\pi\right)^{4}\delta\left(k_{0}'+k_{1}'-p_{\chi_{c}}\right)\times\label{eq:a}\\
 & \times\left(2\pi\right)\delta\left(k_{0}^{+}-k_{0}'{}^{+}\right)\left(2\pi\right)\delta\left(k_{1}^{+}-k_{1}'{}^{+}\right)e^{i\left(\boldsymbol{k}_{0}-\boldsymbol{k}_{0}'\right)\cdot\boldsymbol{x}_{0}}e^{i\left(\boldsymbol{k}_{1}-\boldsymbol{k}_{1}'\right)\cdot\boldsymbol{x}_{1}}\phi_{\chi_{cJ}}\left(z,\,\boldsymbol{k}_{\perp}^{({\rm rel})}\right)\times\nonumber \\
 & \times\left\langle {\rm Tr}\left[\hat{\varepsilon}_{\gamma}(q)S\left(-k_{1}\right)\gamma_{+}U^{\dagger}\left(\boldsymbol{x}_{1}\right)S\left(-k_{1}'\right)\Gamma_{\chi_{c}}\left(k_{1}',\,k_{0}'\right)S\left(k_{0}'\right)\gamma_{+}U\left(\boldsymbol{x}_{0}\right)S\left(k_{0}\right)\hat{\varepsilon}_{\gamma}^{*}\left(k_{\gamma}\right)S\left(k_{0}+k_{\gamma}\right)\right]\right\rangle \nonumber 
\end{align}
where $\varepsilon_{\gamma}(q),\,\varepsilon_{\gamma}^{*}\left(k_{\gamma}\right)$
are the polarization vectors of the photons, and a symbol $\hat{...}$
was defined in~(\ref{eq:gammaSlash}). We also use notation $e_{c}=2/3$
for the electric charge of the charm quark. The matrix $\Gamma_{\chi_{c}}$
was defined earlier in~(\ref{eq:PhiDef}) for different spins and
helicities of $\chi_{c}$. The wave function $\Phi$ depends on the
light-cone fraction $z_{\chi_{c}}=k_{0}^{'+}/p_{\chi_{c}}^{+}$ and
relative transverse momentum $\boldsymbol{k}_{\perp}^{({\rm rel})}=\left(z_{0}\boldsymbol{k}_{1}'-z_{1}\boldsymbol{k}_{0}'\right)/\left(z_{0}+z_{1}\right)$.
The propagator of free fermions is defined in a standard way as 
\begin{equation}
S(p)=i\frac{\hat{p}+m}{p^{2}-m^{2}}.\label{eq:freeProp}
\end{equation}
The numerators of the propagators may be rewritten as
\begin{align}
\hat{p}+m & =\left(\gamma^{+}\frac{\boldsymbol{p}_{\perp}^{2}+m^{2}}{2p^{+}}+\gamma^{-}p^{+}-\boldsymbol{\gamma}_{\perp}\cdot\boldsymbol{p}_{\perp}+m\right)+\gamma^{+}\left(p^{-}-\frac{\boldsymbol{p}_{\perp}^{2}+m^{2}}{2p^{+}}\right)=\label{eq:Nums}\\
 & =\theta\left(p^{+}\right)\sum_{h}u_{h}(\tilde{p})\bar{u}_{h}(\tilde{p})+\theta\left(-p^{+}\right)\sum_{\bar{h}}v_{\bar{h}}(-\tilde{p})\bar{v}_{\bar{h}}(\tilde{p})+\gamma^{+}\left(p^{-}-\frac{\boldsymbol{p}_{\perp}^{2}+m^{2}}{2p^{+}}\right),\qquad\tilde{p}=\left(p^{+},\frac{\boldsymbol{p}_{\perp}^{2}+m^{2}}{2p^{+}},\boldsymbol{p}_{\perp}\right).\nonumber 
\end{align}
The last term $\sim\gamma^{+}$ in~(\ref{eq:Nums}) drops out when
the quark is onshell, or when the propagator is multiplied by $\gamma^{+}$
(due to identity $\gamma^{+}\gamma^{+}=0$), so effectively (\ref{eq:Nums})
reduces to a helicity sum of quark or antiquark spinors. Since the
matrices $U,U^{\dagger}$ in~(\ref{eq:a}) don't have any spinorial
indices, whereas all the other objects don't depend on color, the
evaluation of traces over color and spinorial indices can be done
independently, thus reducing~~(\ref{eq:a}) to a convolution of
the dipole scattering amplitude~(\ref{eq:NS}) with the so-called
(target-independent) impact factor 
\begin{align}
\mathcal{A}_{1,1} & =\int d^{2}\boldsymbol{x}_{0}d^{2}\boldsymbol{x}_{1}\,\mathcal{N}\left(Y,\,\boldsymbol{x}_{0},\,\boldsymbol{x}_{1}\right)R_{1,1}\left(Y,\boldsymbol{x}_{0},\,\boldsymbol{x}_{1},q,\,p_{\chi_{c}},\,k_{\gamma}\right),\label{eq:a-2-1}
\end{align}
where $R_{1,1}$ is given by
\begin{align}
 & R_{1,1}\left(Y,\boldsymbol{x}_{0},\,\boldsymbol{x}_{1},q,\,p_{\chi_{c}},\,k_{\gamma}\right)=4\pi\alpha_{{\rm em}}e_{c}^{2}N_{c}\times\label{eq:R11}\\
 & \times\int\frac{d^{4}k_{0}}{\left(2\pi\right)^{4}}\frac{d^{4}k_{1}}{\left(2\pi\right)^{4}}\frac{d^{4}k_{0}'}{\left(2\pi\right)^{4}}\frac{d^{4}k_{1}'}{\left(2\pi\right)^{4}}\left(2\pi\right)^{4}\delta\left(k_{0}+k_{1}+k_{\gamma}-q\right)\left(2\pi\right)^{4}\delta\left(k_{0}'+k_{1}'-p_{\chi_{c}}\right)\times\nonumber \\
 & \times\left(2\pi\right)\delta\left(k_{0}^{+}-k_{0}'{}^{+}\right)\left(2\pi\right)\delta\left(k_{1}^{+}-k_{1}'{}^{+}\right)e^{i\left(\boldsymbol{k}_{0\perp}-\boldsymbol{k}_{0\perp}'\right)\cdot\boldsymbol{x}_{0}}e^{i\left(\boldsymbol{k}_{1\perp}-\boldsymbol{k}_{1\perp}'\right)\cdot\boldsymbol{x}_{1}}\phi_{\chi_{cJ}}\left(z,\,\boldsymbol{k}_{\perp}^{({\rm rel})}\right)\times\nonumber \\
 & \times\left\langle {\rm Tr}\left[\hat{\varepsilon}_{\gamma}(q)S\left(-k_{1}\right)\gamma_{+}S\left(-k_{1}'\right)\Gamma_{\chi_{c}}\left(k_{1}',\,k_{0}'\right)S\left(k_{0}'\right)\gamma_{+}S\left(k_{0}\right)\hat{\varepsilon}_{\gamma}^{*}\left(k_{\gamma}\right)S\left(k_{0}+k_{\gamma}\right)\right]\right\rangle \nonumber 
\end{align}
The integrals over the momenta $d^{4}k_{1}$ and $d^{4}k_{1}'$ in~(\ref{eq:R11})
can be calculated using $\delta$-functions, which effectively fix
$k_{1}$ and $k_{1}'$ to values 
\begin{equation}
k_{1}=q-k_{\gamma}-k_{0},\qquad k_{1}'=p_{\chi_{c}}-k_{0}'.\label{eq:k1}
\end{equation}
Due to identity $\gamma^{+}\gamma^{+}=0$, we may replace the numerators
of propagators connected to the shockwave with onshell spinors, as
explained in the text under Eq.~(\ref{eq:Nums}). Furthermore, using
the well-known spinor identities~\cite{Lepage:1980fj,Brodsky:1997de}
\begin{equation}
\bar{u}_{h}\left(\ell\right)\gamma_{+}u_{h'}\left(\ell'\right)=\bar{v}_{h}\left(\ell\right)\gamma_{+}v_{h'}\left(\ell'\right)=2\sqrt{\ell^{+}\ell'{}^{+}}\delta_{h,h'}
\end{equation}
where $h,h'$ are helicities of partons, and conservation of the plus-component
light-cone, we may rewrite~(\ref{eq:R11}) as
\begin{align}
 & R_{1,1}\left(Y,\boldsymbol{x}_{0},\,\boldsymbol{x}_{1},q,\,p_{\chi_{c}},\,k_{\gamma}\right)=16\,\pi\alpha_{{\rm em}}e_{c}^{2}N_{c}\left(2\pi\right)\delta\left(q^{+}-k_{\gamma}^{+}-p_{\chi_{c}}^{+}\right)\times\label{eq:R11-2}\\
 & \times\int\frac{dk_{0}^{+}dk_{0}^{-}d^{2}\boldsymbol{k}_{0}}{\left(2\pi\right)^{4}}\frac{dk'{}_{0}^{-}d^{2}\boldsymbol{k}'_{0}}{\left(2\pi\right)^{3}}k_{0}^{+}\left(q^{+}-k_{\gamma}^{+}-k_{0}^{+}\right)e^{i\left(\boldsymbol{k}_{0\perp}-\boldsymbol{k}_{0\perp}'\right)\cdot\boldsymbol{x}_{0}}e^{i\left(\boldsymbol{k}_{1\perp}-\boldsymbol{k}_{1\perp}'\right)\cdot\boldsymbol{x}_{1}}\times\nonumber \\
 & \times\sum_{h,\bar{h}}\frac{\left[\bar{u}_{h}\left(\tilde{k}_{0}\right)\hat{\varepsilon}_{\gamma}^{*}\left(k_{\gamma}\right)S\left(k_{0}+k_{\gamma}\right)\hat{\varepsilon}_{\gamma}(q)v_{\bar{h}}\left(\tilde{k}_{1}\right)\right]}{\left(k_{0}^{2}-m^{2}\right)\left(k_{1}^{2}-m^{2}\right)}\,\frac{\left[\bar{v}_{\bar{h}}\left(\tilde{k}_{1}'\right)\Gamma_{\chi_{c}}\left(k_{1}',\,k_{0}'\right)u_{h}\left(\tilde{k}_{0}'\right)\right]\phi_{\chi_{cJ}}\left(z,\,\boldsymbol{k}_{\perp}^{({\rm rel})}\right)}{\left(\left(k_{0}'\right)^{2}-m^{2}\right)\left(\left(k_{1}'\right)^{2}-m^{2}\right)}.\nonumber 
\end{align}
For the sake of brevity, in~(\ref{eq:R11-2}) we keep notations $k_{1},\,k_{1}'$,
tacitly assuming that they are linear combinations of $k_{0},k_{0}'$
defined in~(\ref{eq:k1}). The dependence on $k_{0}^{-}$ and $k_{0}^{'-}$
in the integrand of~(\ref{eq:R11-2}) is isolated in the first and
the second terms of the last line, for this reason integration over
these variables becomes straightforward and can be realized using
the Cauchy's theorem and reduced to a sum of residues of the integrands
at the poles, namely~ 
\begin{align}
 & \int_{-\infty}^{+\infty}\frac{dk_{0}^{-}}{2\pi}\frac{1}{\left(2k_{0}^{+}k{}_{0}^{-}-\boldsymbol{k}_{0\perp}^{2}-m^{2}+i0\right)\left(2\left(k_{\gamma}^{+}+k_{0}^{+}\right)\left(k_{\gamma}^{-}+k_{0}^{-}\right)-\left(\boldsymbol{k}_{\perp}^{\gamma}+\boldsymbol{k}_{0\perp}\right)^{2}-m^{2}+i0\right)}\times\label{eq:Cauchy_1}\\
 & \times\frac{1}{\left(2\left(q^{+}-k_{\gamma}^{+}-k_{0}^{+}\right)\left(q^{-}-k_{\gamma}^{-}-k_{0}^{-}\right)-\left(\boldsymbol{k}_{\gamma\perp}+\boldsymbol{k}_{0\perp}\right)^{2}-m^{2}+i0\right)}=\nonumber \\
 & =-\frac{i\Theta\left(q^{+}-k_{\gamma}^{+}-k_{0}^{+}\right)}{8k_{0}^{+}\left(k_{0}^{+}+k_{\gamma}^{+}\right)\left(q^{+}-k_{\gamma}^{+}-k_{0}^{+}\right)\left[-\frac{q^{+}\left(\left(\boldsymbol{k}_{\gamma\perp}+\boldsymbol{k}_{0\perp}\right)^{2}+m^{2}\right)}{2\left(q^{+}-k_{\gamma}^{+}-k_{0}^{+}\right)\left(k_{\gamma}^{+}+k_{0}^{+}\right)}\right]\left[-\frac{\boldsymbol{k}_{\gamma\perp}^{2}}{2k_{\gamma}^{+}}-\frac{\boldsymbol{k}_{0\perp}^{2}+m^{2}}{2k_{0}^{+}}-\frac{\left(\boldsymbol{k}_{\gamma\perp}+\boldsymbol{k}_{0\perp}\right)^{2}+m^{2}}{2\left(q^{+}-k_{\gamma}^{+}-k_{0}^{+}\right)}\right]},\nonumber 
\end{align}
\begin{align}
 & \int_{-\infty}^{+\infty}\frac{dk_{0}^{-}}{2\pi}\frac{\left(k_{\gamma}^{-}+k_{0}^{-}\right)}{\left(2k_{0}^{+}k{}_{0}^{-}-\boldsymbol{k}_{0\perp}^{2}-m^{2}+i0\right)\left(2\left(k_{\gamma}^{+}+k_{0}^{+}\right)\left(k_{\gamma}^{-}+k_{0}^{-}\right)-\left(\boldsymbol{k}_{\perp}^{\gamma}+\boldsymbol{k}_{0\perp}\right)^{2}-m^{2}+i0\right)}\times\label{eq:Cauchy_1-1}\\
 & \times\frac{1}{\left(2\left(q^{+}-k_{\gamma}^{+}-k_{0}^{+}\right)\left(q^{-}-k_{\gamma}^{-}-k_{0}^{-}\right)-\left(\boldsymbol{k}_{\gamma\perp}+\boldsymbol{k}_{0\perp}\right)^{2}-m^{2}+i0\right)}=\nonumber \\
 & =-\frac{i\Theta\left(q^{+}-k_{\gamma}^{+}-k_{0}^{+}\right)\frac{\left(\boldsymbol{k}_{\gamma\perp}+\boldsymbol{k}_{0\perp}\right)^{2}+m^{2}}{2\left(k_{\gamma}^{+}+k_{0}^{+}\right)}}{8k_{0}^{+}\left(k_{0}^{+}+k_{\gamma}^{+}\right)\left(q^{+}-k_{\gamma}^{+}-k_{0}^{+}\right)\left[-\frac{q^{+}\left(\left(\boldsymbol{k}_{\gamma\perp}+\boldsymbol{k}_{0\perp}\right)^{2}+m^{2}\right)}{2\left(q^{+}-k_{\gamma}^{+}-k_{0}^{+}\right)\left(k_{\gamma}^{+}+k_{0}^{+}\right)}\right]\left[-\frac{\boldsymbol{k}_{\gamma\perp}^{2}}{2k_{\gamma}^{+}}-\frac{\boldsymbol{k}_{0\perp}^{2}+m^{2}}{2k_{0}^{+}}-\frac{\left(\boldsymbol{k}_{\gamma\perp}+\boldsymbol{k}_{0\perp}\right)^{2}+m^{2}}{2\left(q^{+}-k_{\gamma}^{+}-k_{0}^{+}\right)}\right]}\nonumber \\
 & -\frac{i\Theta\left(q^{+}-k_{\gamma}^{+}-k_{0}^{+}\right)}{8k_{0}^{+}\left(k_{0}^{+}+k_{\gamma}^{+}\right)\left(q^{+}-k_{\gamma}^{+}-k_{0}^{+}\right)\left[-\frac{\boldsymbol{k}_{\gamma\perp}^{2}}{2k_{\gamma}^{+}}-\frac{\boldsymbol{k}_{0\perp}^{2}+m^{2}}{2k_{0}^{+}}-\frac{\left(\boldsymbol{k}_{\gamma\perp}+\boldsymbol{k}_{0\perp}\right)^{2}+m^{2}}{2\left(q^{+}-k_{\gamma}^{+}-k_{0}^{+}\right)}\right]},\nonumber 
\end{align}
\begin{align}
 & \int_{-\infty}^{+\infty}\frac{dk'{}_{0}^{-}}{2\pi}\frac{1}{\left(2k'{}_{0}^{+}k'{}_{0}^{-}-\boldsymbol{k}'{}_{0\perp}^{2}-m^{2}+i0\right)\left(2\left(p_{\chi_{c}}^{-}-k'{}_{0}^{+}\right)\left(p_{\chi_{c}}^{-}-k'{}_{0}^{-}\right)-\left(\boldsymbol{p}_{\perp}^{\chi_{c}}-\boldsymbol{k}'_{0\perp}\right)^{2}-m^{2}+i0\right)}=\label{eq:Cauchy_2}\\
 & =-\frac{\Theta\left(p_{\chi_{c}}^{+}-k'{}_{0}^{+}\right)}{4k'{}_{0}^{+}\left(p_{\chi_{c}}^{+}-k'{}_{0}^{+}\right)}\,\frac{i}{p_{\chi_{c}}^{-}-\frac{\boldsymbol{k}'{}_{0\perp}^{2}+m^{2}}{2k'{}_{0}^{+}}-\frac{\left(\boldsymbol{p}_{\perp}^{\chi_{c}}-\boldsymbol{k}'_{0\perp}\right)^{2}+m^{2}}{2\left(p_{\chi_{c}}^{+}-k'{}_{0}^{+}\right)}}.\nonumber 
\end{align}
The Heaviside step functions $\Theta(...)$ in ~(\ref{eq:Cauchy_1}-\ref{eq:Cauchy_2})
introduce the upper cutoff for the integral over the plus-component
$k_{0}^{+}$ and stem from the requirement that the integrands should
have poles on both sides of the real axis (${\rm Im}\,k^{-}=0$) for
nonzero result. The identities~(\ref{eq:Cauchy_1},\ref{eq:Cauchy_2})
allow us to rewrite~(\ref{eq:R11}) as

\begin{align}
 & R_{1,1}\left(Y,\boldsymbol{x}_{0},\,\boldsymbol{x}_{1},q,\,p_{\chi_{c}},\,k_{\gamma}\right)=\delta\left(q^{+}-k_{\gamma}^{+}-p_{\chi_{c}}^{+}\right)e^{i\left(\boldsymbol{q}_{\perp}-\boldsymbol{k}_{\perp}^{\gamma}-\boldsymbol{p}_{\perp}^{\chi_{c}}\right)\cdot\boldsymbol{x}_{1}}\int_{0}^{q^{+}-k_{\gamma}^{+}}dk_{0}^{+}\times\label{eq:a-1}\\
 & \times\left[\frac{2\pi\alpha_{{\rm em}}e_{c}^{2}}{k_{0}^{+}+k_{\gamma}^{+}}\int\frac{d^{2}\boldsymbol{k}_{0\perp}}{\left(2\pi\right)^{2}}\frac{\left.\bar{u}_{h}\left(k_{0}\right)\hat{\varepsilon}_{\gamma}^{*}\left(k_{\gamma}\right)\left(\hat{k}_{0}+\hat{k}_{\gamma}+m\right)\hat{\varepsilon}_{\gamma}(q)v_{\bar{h}}\left(q-k_{\gamma}-k_{0}\right)e^{i\boldsymbol{k}_{0}^{\perp}\cdot\left(\boldsymbol{x}_{0}-\boldsymbol{x}_{1}\right)}\right|_{k_{0}^{-}=\frac{\left(\boldsymbol{k}_{\gamma\perp}+\boldsymbol{k}_{0\perp}\right)^{2}+m^{2}}{2\left(k_{\gamma}^{+}+k_{0}^{+}\right)}-k_{\gamma}^{-}}}{\sqrt{k_{0}^{+}\left(q^{+}-k_{\gamma}^{+}-k_{0}^{+}\right)}\left[-\frac{q^{+}\left(\boldsymbol{k}_{0\perp}^{2}+m^{2}\right)}{2\left(q^{+}-k_{\gamma}^{+}-k_{0}^{+}\right)\left(k_{\gamma}^{+}+k_{0}^{+}\right)}\right]\left[-\frac{\boldsymbol{k}_{\gamma\perp}^{2}}{2k_{\gamma}^{+}}-\frac{\boldsymbol{k}_{0\perp}^{2}+m^{2}}{2k_{0}^{+}}-\frac{\left(\boldsymbol{k}_{\gamma\perp}+\boldsymbol{k}_{0\perp}\right)^{2}+m^{2}}{2\left(q^{+}-k_{\gamma}^{+}-k_{0}^{+}\right)}\right]}\right.\nonumber \\
 & +\left.\frac{2\pi\alpha_{{\rm em}}e_{c}^{2}}{k_{0}^{+}+k_{\gamma}^{+}}\int\frac{d^{2}\boldsymbol{k}_{0\perp}}{\left(2\pi\right)^{2}}\frac{\bar{u}_{h}\left(k_{0}\right)\hat{\varepsilon}_{\gamma}^{*}\left(k_{\gamma}\right)\gamma^{+}\hat{\varepsilon}_{\gamma}(q)v_{\bar{h}}\left(q-k_{\gamma}-k_{0}\right)e^{i\boldsymbol{k}_{0}^{\perp}\cdot\left(\boldsymbol{x}_{0}-\boldsymbol{x}_{1}\right)}}{\sqrt{k_{0}^{+}\left(q^{+}-k_{\gamma}^{+}-k_{0}^{+}\right)}\left[-\frac{\boldsymbol{k}_{\gamma\perp}^{2}}{2k_{\gamma}^{+}}-\frac{\boldsymbol{k}_{0\perp}^{2}+m^{2}}{2k_{0}^{+}}-\frac{\left(\boldsymbol{k}_{\gamma\perp}+\boldsymbol{k}_{0\perp}\right)^{2}+m^{2}}{2\left(q^{+}-k_{\gamma}^{+}-k_{0}^{+}\right)}\right]}\right]\nonumber \\
 & \times\int\frac{d^{2}\boldsymbol{k}'_{0\perp}}{\left(2\pi\right)^{2}}\frac{\bar{v}_{\bar{h}}\left(p_{\chi_{c}}^{+}-k_{0}^{+},\,\boldsymbol{p}_{\perp}^{\chi_{c}}-\boldsymbol{k}'_{0\perp}\right)\Gamma_{\chi_{c}}\left(p_{\chi_{c}}-k_{0}',\,k_{0}'\right)u_{h}\left(k_{0}^{+},\,\boldsymbol{k}'_{0\perp}\right)\phi_{\chi_{cJ}}\left(z,\,\boldsymbol{K}_{\perp}\right)e^{-i\boldsymbol{k}'_{0\perp}\cdot\left(\boldsymbol{x}_{0}-\boldsymbol{x}_{1}\right)}}{2\sqrt{k_{0}^{+}\left(p_{\chi_{c}}^{+}-k_{0}^{+}\right)}\left[p_{\chi_{c}}^{-}-\frac{\boldsymbol{k}'{}_{0\perp}^{2}+m^{2}}{2k_{0}^{+}}-\frac{\left(\boldsymbol{p}_{\perp}^{\chi_{c}}-\boldsymbol{k}'_{0\perp}\right)^{2}+m^{2}}{2\left(p_{\chi_{c}}^{+}-k_{0}^{+}\right)}\right]}.\nonumber 
\end{align}
where the $\delta$-function $\delta\left(q^{+}-k_{\gamma}^{+}-p_{\chi_{c}}^{+}\right)$
in~(\ref{eq:a-1}) reflects smallness of the light-cone momentum
transfer $\Delta^{+}$ compared to transverse momentum transfer to
the target $\boldsymbol{\Delta}_{\perp}=\boldsymbol{q}_{\perp}-\boldsymbol{k}_{\perp}^{\gamma}-\boldsymbol{p}_{\perp}^{\chi_{c}}$
in the high energy kinematics. The expressions in square brackets
in denominators of~(\ref{eq:a-1}) coincide with the energy denominators
of the light-cone perturbation theory~\cite{Brodsky:1997de}. The
analysis of the other diagrams is done similarly. The evaluation of
the lower left diagram in Figure~\ref{fig:CGCBasic-1} yields for
the impact factor 
\begin{align}
 & R_{1,2}\left(Y,\boldsymbol{x}_{0},\,\boldsymbol{x}_{1},q,\,p_{\chi_{c}},\,k_{\gamma}\right)=-\delta\left(q^{+}-k_{\gamma}^{+}-p_{\chi_{c}}^{+}\right)e^{i\left(\boldsymbol{q}_{\perp}-\boldsymbol{k}_{\perp}^{\gamma}-\boldsymbol{p}_{\perp}^{\chi_{c}}\right)\cdot\boldsymbol{x}_{1}}\int_{0}^{q^{+}-k_{\gamma}^{+}}dk_{0}^{+}\times\label{eq:a-1-1}\\
 & \times\left[\frac{2\pi\alpha_{{\rm em}}e_{c}^{2}}{q^{+}-k_{0}^{+}}\int\frac{d^{2}\boldsymbol{k}_{0\perp}}{\left(2\pi\right)^{2}}\frac{\left.\bar{u}_{h}\left(k_{0}\right)\hat{\varepsilon}_{\gamma}(q)\left(\hat{q}-\hat{k}_{0}-m\right)\hat{\varepsilon}_{\gamma}^{*}\left(k_{\gamma}\right)v_{\bar{h}}\left(q-k_{\gamma}-k_{0}\right)e^{i\boldsymbol{k}_{0}^{\perp}\cdot\left(\boldsymbol{x}_{0}-\boldsymbol{x}_{1}\right)}\right|_{k_{0}^{-}=q^{-}-\frac{\boldsymbol{k}_{0\perp}^{2}+m^{2}}{2\left(q^{+}-k_{0}^{+}\right)}}}{\sqrt{k_{0}^{+}\left(q^{+}-k_{\gamma}^{+}-k_{0}^{+}\right)}\left[q^{-}-\frac{q^{+}\left(\boldsymbol{k}_{0\perp}^{2}+m^{2}\right)}{2\left(q^{+}-k_{\gamma}^{+}\right)k_{0}^{+}}\right]\left[q^{-}-\frac{\boldsymbol{k}_{\gamma\perp}^{2}}{2k_{\gamma}^{+}}-\frac{\boldsymbol{k}_{0\perp}^{2}+m^{2}}{2k_{0}^{+}}-\frac{\left(\boldsymbol{k}_{\gamma\perp}+\boldsymbol{k}_{0\perp}\right)^{2}+m^{2}}{2\left(q^{+}-k_{\gamma}^{+}-k_{0}^{+}\right)}\right]}\right.\nonumber \\
 & +\left.\frac{2\pi\alpha_{{\rm em}}e_{c}^{2}}{q^{+}-k_{0}^{+}}\int\frac{d^{2}\boldsymbol{k}_{0\perp}}{\left(2\pi\right)^{2}}\frac{\bar{u}_{h}\left(k_{0}\right)\hat{\varepsilon}_{\gamma}(q)\gamma^{+}\hat{\varepsilon}_{\gamma}^{*}\left(k_{\gamma}\right)v_{\bar{h}}\left(q-k_{\gamma}-k_{0}\right)e^{i\boldsymbol{k}_{0}^{\perp}\cdot\left(\boldsymbol{x}_{0}-\boldsymbol{x}_{1}\right)}}{\sqrt{k_{0}^{+}\left(q^{+}-k_{\gamma}^{+}-k_{0}^{+}\right)}\left[q^{-}-\frac{\boldsymbol{k}_{\gamma\perp}^{2}}{2k_{\gamma}^{+}}-\frac{\boldsymbol{k}_{0\perp}^{2}+m^{2}}{2k_{0}^{+}}-\frac{\left(\boldsymbol{k}_{\gamma\perp}+\boldsymbol{k}_{0\perp}\right)^{2}+m^{2}}{2\left(q^{+}-k_{\gamma}^{+}-k_{0}^{+}\right)}\right]}\right]\nonumber \\
 & \times\int\frac{d^{2}\boldsymbol{k}'_{0\perp}}{\left(2\pi\right)^{2}}\frac{\bar{v}_{\bar{h}}\left(p_{\chi_{c}}^{+}-k_{0}^{+},\,\boldsymbol{p}_{\perp}^{\chi_{c}}-\boldsymbol{k}'_{0\perp}\right)\Gamma_{\chi_{c}}\left(p_{\chi_{c}}-k_{0}',\,k_{0}'\right)u_{h}\left(k_{0}^{+},\,\boldsymbol{k}'_{0\perp}\right)e^{-i\boldsymbol{k}'_{0\perp}\cdot\left(\boldsymbol{x}_{0}-\boldsymbol{x}_{1}\right)}\phi_{\chi_{cJ}}\left(z,\,\boldsymbol{K}_{\perp}\right)}{2\sqrt{k_{0}^{+}\left(p_{\chi_{c}}^{+}-k_{0}^{+}\right)}\left[p_{\chi_{c}}^{-}-\frac{\boldsymbol{k}'{}_{0\perp}^{2}+m^{2}}{2k_{0}^{+}}-\frac{\left(\boldsymbol{p}_{\perp}^{\chi_{c}}-\boldsymbol{k}'_{0\perp}\right)^{2}+m^{2}}{2\left(p_{\chi_{c}}^{+}-k_{0}^{+}\right)}\right]}.\nonumber 
\end{align}

\begin{figure}
\includegraphics[width=5cm]{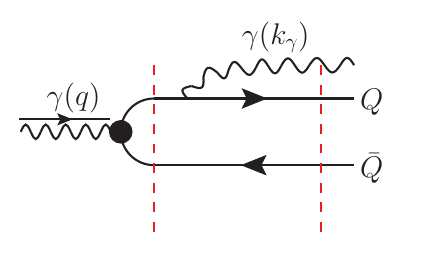}\includegraphics[width=5cm]{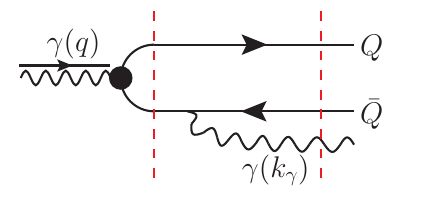}

\includegraphics[width=5cm]{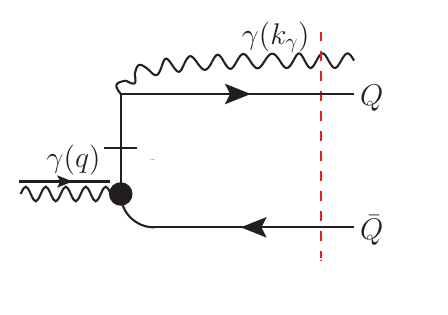}\includegraphics[width=5cm]{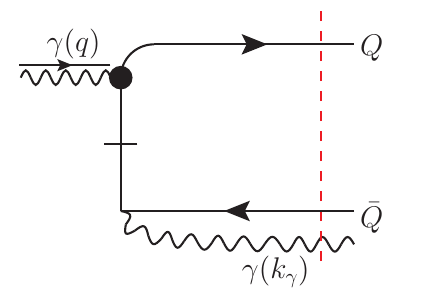}

\caption{The diagrams which correspond to the $\gamma\to\bar{Q}Q\gamma$ amplitude
in the leading order in strong coupling $\alpha_{s}$. The contributions
of the two diagrams in the first row correspond to expressions in
the second lines of~(\ref{eq:a-1}) and (\ref{eq:a-1-1}). Similarly,
the contributions of the diagrams in the last row are given by the
expressions in the third lines of the same Equations (\ref{eq:a-1})
and (\ref{eq:a-1-1}). The vertical fermion line with a short horizontal
dash in the diagrams of the last row denotes instantaneous part of
the quark propagator. The colored vertical dashed lines stand for
the light-cone energy denominators of the canonical light-cone perturbation
theory~\cite{Brodsky:1997de,Lepage:1980fj}.}
\label{fig:Diags-2}
\end{figure}

The result for (\ref{eq:a-1-1}) differs from (\ref{eq:a-1}) only
by the structure of the second and the third lines. After integration
over $\boldsymbol{k}_{0\perp}'$ in the last lines of~(\ref{eq:a-1})
and (\ref{eq:a-1-1}), we may obtain the wave function of $\chi_{c}$
meson in the light-cone representation introduced earlier in~(\ref{eq:Ph}).
Likewise, the result of integration over $\boldsymbol{k}_{0\perp}$
in the second and the third lines of~(\ref{eq:a-1},\ref{eq:a-1-1})
can be interpreted as a light-cone wave function of the $\bar{Q}Q\gamma$
Fock component in the photon, which in the leading order is given
by a set of diagrams shown in the Figure~\ref{fig:Diags-2}. These
diagrams nearly coincide with the diagrams $j,k,\ell,m$ in~~\cite{Beuf:2022ndu}
which contribute to the $\gamma\to\bar{Q}Qg$ wave function, and for
this reason the result of integration over $\boldsymbol{k}_{0\perp}$
differs from expressions found in~~\cite{Beuf:2022ndu} only by
color factor. The structure of~(\ref{eq:a-1}) and (\ref{eq:a-1-1})
implies that the amplitude $\mathcal{A}_{1}$ can be rewritten as
a convolution of the dipole amplitude $\mathcal{N}\left(x,\,\boldsymbol{r},\,\boldsymbol{b}\right)$,
the wave function $\Psi_{\chi_{c}}^{\dagger}$of the produced $\chi_{c}$
meson, and the wave function $\Psi_{\gamma\to\gamma\bar{Q}Q}$ of
the $\bar{Q}Q\gamma$ Fock component of the incoming photon (the amplitude
of the subprocess $\gamma\to\bar{Q}Q\gamma$), namely 

\begin{align}
\mathcal{A}_{1}^{(\lambda;\,\sigma,\,H)} & =\int_{0}^{\alpha_{\chi_{c}}}dz_{0}\prod_{k=0}^{2}\left(d^{2}\boldsymbol{x}_{k}\right)\,\,\Psi_{\chi_{c}}^{(h,\bar{h})\dagger}\left(\frac{z_{0}}{z_{0}+z_{1}},\,\boldsymbol{r}_{10}\right)\,\Psi_{\gamma\to\gamma\bar{Q}Q}^{(\lambda,\sigma,h,\bar{h})}\left(z_{0},\,z_{1}=\alpha_{\chi_{c}}-z_{0},\,z_{2}\equiv\bar{\alpha}_{\chi_{c}},\boldsymbol{x}_{0},\,\boldsymbol{x}_{1},\,\boldsymbol{x}_{2}\right)\times\label{eq:A}\\
 & \times\mathcal{N}\left(x,\,\boldsymbol{r}_{10},\,\,\boldsymbol{b}_{10}\right)\exp\left[-i\boldsymbol{p}_{\perp}^{\chi_{c}}\cdot\left(\boldsymbol{b}_{10}-\frac{\bar{\alpha}_{\chi_{c}}}{\alpha_{\chi_{c}}}\boldsymbol{r}_{\gamma}\right)-i\boldsymbol{k}_{\perp}^{\gamma}\cdot\boldsymbol{x}_{2}\right],\nonumber 
\end{align}
where the dummy integration variable $z_{0}$ is related to $k_{0}^{+}$
using identities~(\ref{eq:lcFractions}), the parameters $\lambda,\sigma$
are the helicities of the photon before and after interaction; $h,\bar{h}$
are the helicities of the heavy quarks when they cross the shock wave,
$\boldsymbol{r}_{10}\equiv\boldsymbol{x}_{1}-\boldsymbol{x}_{0}$
is the relative distance between the quarks, $\boldsymbol{b}_{10}=\left(z_{0}\boldsymbol{x}_{0}+z_{1}\boldsymbol{x}_{1}\right)/\left(z_{0}+z_{1}\right)$
is the position (impact parameter) of the center of mass of the quark-antiquark
pair, $\boldsymbol{r}_{\gamma}=\boldsymbol{x}_{2}-\boldsymbol{b}_{10}$
is the distance between the emitted photon and the center of mass
of the dipole. The argument $x$ of the dipole cross-section $\mathcal{N}$
corresponds to the fraction of the target's momentum in minus-direction
which is taken by heavy quarks, $x\approx\Delta^{-}/P^{-}=2\xi/(1+\xi)\approx M_{\gamma\chi_{c}}^{2}/W^{2}$
. As we mentioned earlier, the impact factors $R_{1,1}$ and $R_{1,2}$
are related to each other by charge conjugation ($C$-parity). In
order to fully exploit this symmetry , it is convenient to introduce
the variable $\zeta$ defined as a fraction of the photon's momentum
carried by virtual parton (quark or antiquark) which emits secondary
(final-state) photon afterwards, namely
\begin{equation}
z_{0}=\zeta-\bar{\alpha}_{\chi_{c}},\qquad z_{1}=1-\zeta,\qquad z_{2}=1-\alpha_{\chi_{c}}
\end{equation}
for the contribution $R_{1,1}$, and
\begin{equation}
z_{0}=1-\zeta,\qquad z_{1}=\zeta-\bar{\alpha}_{\chi_{c}},\qquad z_{2}=1-\alpha_{\chi_{c}}
\end{equation}
for the contribution $R_{1,2}$. While the expression~(\ref{eq:A})
presents a lot of interest for conceptual understanding, unfortunately
it is not feasible to use it directly for numerical phenomenological
studies due to multidimensional integration of the oscillating function;
besides the wave function $\Psi_{\gamma\to\gamma\bar{Q}Q}^{(...)}$in
light-cone (configuration) representation include new special functions
$\mathcal{I},\,\mathcal{J}$ defined in~\cite{Beuf:2021qqa,Beuf:2022ndu}
as nontrivial 2-dimensional integrals. This suggests that evaluation
of the amplitude should be done in the momentum space. However, evaluation
of the Fourier image of $\mathcal{N}\left(x,\,\boldsymbol{r}_{10},\,\,\boldsymbol{b}_{10}\right)$
is not well-defined due to saturation of the latter at large distances
$r_{10}$ and non-commutativity of the limits $r_{10}\to\infty$ and
$b_{10}\to\infty$. At the same time, we may observe that for the
product $\Psi_{\chi_{cJ}}^{\dagger}\left(\boldsymbol{r}_{10}\right)\mathcal{N}\left(x,\,\boldsymbol{r}_{10},\,\,\boldsymbol{b}_{10}\right)$
this problem does not exist, because $\Psi_{\chi_{cJ}}$ is proportional
to $\Phi_{\chi_{c}}$ and its derivatives, which are strongly suppressed
at large distances. For this reason, in what follows we will introduce
new functions~\footnote{For phenomenological parametrizations of $\mathcal{N}\left(x,\,\boldsymbol{r},\,\,\boldsymbol{b}\right)$
that do not depend explicitly on orientations of $\boldsymbol{r},\boldsymbol{b}$,
the integration over orientations of these vector can be done analytically,
yielding $\left(2\pi\right)^{2}J_{0}(\ell r)\,J_{0}(sb)$ for $n_{\Phi_{\chi_{c}}}^{(2,\,0)}$,
$\left(2\pi\right)^{2}iJ_{1}(\ell r)\,J_{0}(sb)$ for $n_{\Phi_{\chi_{c}}}^{(\pm)}$,
and $-\left(2\pi\right)^{2}J_{2}(\ell r)\,J_{0}(sb)$ for $n_{\Phi_{\chi_{c}}}^{(2,\,\pm2)}$.}
\begin{equation}
n_{\Phi_{\chi_{c}}}\left(x,\boldsymbol{\ell},\,\boldsymbol{s},\,z\right)=\int d^{2}\boldsymbol{b}\,d^{2}\boldsymbol{r}e^{-i\boldsymbol{\ell}\cdot\boldsymbol{r}}e^{-i\boldsymbol{s}\cdot\boldsymbol{b}}\mathcal{N}\left(x,\boldsymbol{r},\boldsymbol{b}\right)\Phi_{\chi_{c}}\left(\boldsymbol{r},\,z\right),\label{eq:nPhi}
\end{equation}
\begin{align}
n_{\Phi_{\chi_{c}}}^{(\pm)}\left(x,\,\boldsymbol{\ell},\,\boldsymbol{s},\,z\right) & =\int d^{2}\boldsymbol{b}\,d^{2}\boldsymbol{r}e^{-i\boldsymbol{\ell}\cdot\boldsymbol{r}}e^{-i\boldsymbol{s}\cdot\boldsymbol{b}}\mathcal{N}\left(x,\,\boldsymbol{r},\,\,\boldsymbol{b}\right)e^{\pm i\theta}\partial_{r}\Phi_{\chi_{c}}\left(\boldsymbol{r},\,z\right),\label{eq:nPhi1}
\end{align}
\begin{align}
n_{\Phi_{\chi_{c}}}^{(2,\,\pm2)}\left(x,\,\boldsymbol{\ell},\,\boldsymbol{s},\,z\right) & =\int d^{2}\boldsymbol{b}\,d^{2}\boldsymbol{r}e^{-i\boldsymbol{\ell}\cdot\boldsymbol{r}}e^{-i\boldsymbol{s}\cdot\boldsymbol{b}}\mathcal{N}\left(x,\,\boldsymbol{r},\,\,\boldsymbol{b}\right)e^{\pm2i\theta}\Delta_{r}^{(-)}\Phi_{\chi_{c}}\left(\boldsymbol{r},\,z\right),\label{eq:nPhi2a}\\
n_{\Phi_{\chi_{c}}}^{(2,\,0)}\left(x,\,\boldsymbol{\ell},\,\boldsymbol{s},\,z\right) & =\int d^{2}\boldsymbol{b}\,d^{2}\boldsymbol{r}e^{-i\boldsymbol{\ell}\cdot\boldsymbol{r}}e^{-i\boldsymbol{s}\cdot\boldsymbol{b}}\mathcal{N}\left(x,\,\boldsymbol{r},\,\,\boldsymbol{b}\right)\Delta_{r}^{(+)}\Phi_{\chi_{c}}\left(\boldsymbol{r},\,z\right),\label{eq:nPhi2c}
\end{align}
and, using explicit expressions for the wave functions in helicity
basis~(\ref{eq:Chic0}-\ref{eq:Chic2c}), will replace the corresponding
products $\Psi_{\chi_{cJ}}^{\dagger}\left(\boldsymbol{r}_{10}\right)\mathcal{N}\left(x,\,\boldsymbol{r}_{10},\,\,\boldsymbol{b}_{10}\right)$
with (inverse) Fourier images of the functions $n_{\Phi_{\chi_{c}}},\,n_{\Phi_{\chi_{c}}}^{(\pm)},\,n_{\Phi_{\chi_{c}}}^{(2,\,0)},\,n_{\Phi_{\chi_{c}}}^{(2,\,\pm2)}$.
After integration over the variables $\boldsymbol{r}_{10}$ and $\boldsymbol{b}_{10}$,
we may obtain two additional $\delta$-functions
\[
\left(2\pi\right)^{4}\delta\left(\boldsymbol{s}+\boldsymbol{\Delta}_{\perp}\right)\delta\left(\boldsymbol{\ell}-\boldsymbol{k}_{0}^{\perp}-\frac{\zeta-\bar{\alpha}_{\chi_{c}}}{\alpha}\boldsymbol{\Delta}_{\perp}\right)
\]
 which allow to get rid of the integrals over $d^{2}\boldsymbol{k}_{0}$
and $d^{2}\boldsymbol{s}$. The final result of this procedure depends
on the spin and helicity of $\chi_{cJ}$ meson. We performed these
evaluations in helicity basis, using explicit expressions for spinors
from~\cite{Lepage:1980fj,Brodsky:1997de} and performing the algebraic
simplifications using \emph{Mathematica}. In view of the space constraints,
we provide below the results only for  components with ``+'' helicity
of the incoming photon, tacitly assuming that the remaining components
may be recovered using $\left[\mathcal{A}_{1}^{(\lambda;\,\sigma,\,H)}\right]^{*}=\mathcal{A}_{1}^{(-\lambda;\,-\sigma,\,-H)}$.
For $\chi_{c0}$ meson, the helicity amplitudes read as
\begin{align}
\mathcal{A}_{1,\,\chi_{c0}}^{(+;\,+)} & =4\pi\bar{\kappa}m\int_{\bar{\alpha}_{\chi_{c}}}^{1}d\zeta\,\int\frac{d^{2}\ell}{(2\pi)^{2}}\,\sqrt{\frac{\bar{\zeta}}{\zeta-\bar{\alpha}_{\chi_{c}}}}\times\label{eq:Amplitude_PP-1}\\
 & \times\left\{ \frac{\,\alpha_{\chi_{c}}\bar{\alpha}_{\chi_{c}}\left(\alpha_{\chi_{c}}-2\bar{\zeta}\right)\left[\left(\zeta-\bar{\alpha}_{\chi_{c}}\right)\left(\boldsymbol{P}_{(1)}^{2}+m^{2}\right)-\bar{\alpha}_{\chi_{c}}m^{2}\right]n_{\Phi_{\chi_{c}}}\left(x,\,\boldsymbol{\ell},\,\,-\boldsymbol{\Delta}_{\perp},\,z_{\chi_{c}}\right)}{\left(\boldsymbol{P}_{(1)}^{2}+m^{2}\right)\left(\bar{\zeta}\,\zeta^{2}\boldsymbol{K}_{(1)}^{2}+\left(\zeta-\bar{\alpha}_{\chi_{c}}\right)\,\bar{\alpha}_{\chi_{c}}\boldsymbol{P}_{(1)}^{2}+\alpha_{\chi_{c}}\bar{\alpha}_{\chi_{c}}\zeta m^{2}\right)}\right.\nonumber \\
 & \quad\left.-\frac{i\,\alpha_{\chi_{c}}\bar{\alpha}_{\chi_{c}}^{2}\zeta\,n_{\Phi_{\chi_{c}}}^{(-)}\left(x,\,\boldsymbol{\ell},\,\,-\boldsymbol{\Delta}_{\perp},\,z_{\chi_{c}}\right)\left(P_{(1)}+iP_{(1)y}\right)}{\left(\boldsymbol{P}_{(1)}^{2}+m^{2}\right)\left(\bar{\zeta}\,\zeta^{2}\boldsymbol{K}_{(1)}^{2}+\left(\zeta-\bar{\alpha}_{\chi_{c}}\right)\,\bar{\alpha}_{\chi_{c}}\boldsymbol{P}_{(1)}^{2}+\alpha_{\chi_{c}}\bar{\alpha}_{\chi_{c}}\zeta m^{2}\right)}\right\} ,\nonumber 
\end{align}
\begin{align}
\mathcal{A}_{1,\,\chi_{c0}}^{(+;\,-)} & =-4i\pi\bar{\kappa}m\alpha_{\chi_{c}}^{2}\bar{\alpha}_{\chi_{c}}^{2}\int_{\bar{\alpha}_{\chi_{c}}}^{1}d\zeta\,\int\frac{d^{2}\ell}{(2\pi)^{2}}\,\sqrt{\frac{\bar{\zeta}^{3}}{\zeta-\bar{\alpha}_{\chi_{c}}}}n_{\Phi_{\chi_{c}}}^{(+)}\left(x,\,\boldsymbol{\ell},\,-\boldsymbol{\Delta}_{\perp},\,z_{\chi_{c}}\right)\times\label{eq:Amplitude_PM-1}\\
 & \frac{\left(P_{(1)x}+iP_{(1)y}\right)}{\left(\boldsymbol{P}_{(1)}^{2}+m^{2}\right)\left(\bar{\zeta}\,\zeta^{2}\boldsymbol{K}_{(1)}^{2}+\left(\zeta-\bar{\alpha}_{\chi_{c}}\right)\,\bar{\alpha}_{\chi_{c}}\boldsymbol{P}_{(1)}^{2}+\alpha_{\chi_{c}}\bar{\alpha}_{\chi_{c}}\zeta m^{2}\right)},\nonumber 
\end{align}

where we introduced shorthand notations

\begin{align}
\boldsymbol{K}_{(1)} & =\frac{\boldsymbol{\ell}\bar{\alpha}_{\chi_{c}}+\zeta\left(\alpha_{\chi_{c}}\boldsymbol{k}_{\perp}^{\gamma}-\bar{\alpha}_{\chi_{c}}\boldsymbol{p}_{\perp}^{\chi_{c}}\right)}{\zeta-\bar{\alpha}_{\chi_{c}}}=\frac{\boldsymbol{P}_{(1)}\bar{\alpha}_{\chi_{c}}}{\zeta-\bar{\alpha}_{\chi_{c}}}+\frac{\left(\alpha_{\chi_{c}}\boldsymbol{k}_{\perp}^{\gamma}-\bar{\alpha}_{\chi_{c}}\boldsymbol{p}_{\perp}^{\chi_{c}}\right)}{\alpha_{\chi_{c}}},\label{K1}\\
\boldsymbol{P}_{(1)} & =\boldsymbol{\ell}+\frac{\bar{\zeta}}{\alpha_{\chi_{c}}}\left(\alpha_{\chi_{c}}\boldsymbol{k}_{\perp}^{\gamma}-\bar{\alpha}_{\chi_{c}}\boldsymbol{p}_{\perp}^{\chi_{c}}\right),\qquad z_{\chi_{c}}=\frac{\zeta-\bar{\alpha}_{\chi_{c}}}{\alpha_{\chi_{c}}}.\label{eq:P1}
\end{align}
For $\chi_{c1}$ meson, the corresponding amplitudes are given by

\begin{align}
\mathcal{A}_{1,\,\chi_{c1}}^{(+;\,+,+1)} & =4\pi\bar{\kappa}\alpha_{\chi_{c}}\int_{\bar{\alpha}_{\chi_{c}}}^{1}d\zeta\,\int\frac{d^{2}\ell}{(2\pi)^{2}}\sqrt{\frac{2\bar{\zeta}}{\zeta-\bar{\alpha}_{\chi_{c}}}}\zeta\left(K_{(1)x}-iK_{(1)y}\right)\times\\
 & \frac{\left(\bar{\zeta}\bar{\alpha}_{\chi_{c}}\left(P_{(1)x}+iP_{(1)y}\right)(2\zeta-\bar{\alpha}_{\chi_{c}})n_{\Phi_{\chi_{c}}}^{(-)}\left(x,\,\boldsymbol{\ell},\,\,-\boldsymbol{\Delta}_{\perp},\,z_{\chi_{c}}\right)-i\zeta m^{2}n_{\Phi_{\chi_{c}}}\left(x,\,\boldsymbol{\ell},\,\,-\boldsymbol{\Delta}_{\perp},\,z_{\chi_{c}}\right)\left(\alpha_{\chi_{c}}-2\bar{\zeta}\right)\right)}{\left(\boldsymbol{P}_{(1)}^{2}+m^{2}\right)\left(\bar{\zeta}\,\zeta^{2}\boldsymbol{K}_{(1)}^{2}+\left(\zeta-\bar{\alpha}_{\chi_{c}}\right)\,\bar{\alpha}_{\chi_{c}}\boldsymbol{P}_{(1)}^{2}+\alpha_{\chi_{c}}\bar{\alpha}_{\chi_{c}}\zeta m^{2}\right)},\nonumber 
\end{align}
\begin{align}
\mathcal{A}_{1,\,\chi_{c1}}^{(+;\,-,+1)} & =-4i\pi\bar{\kappa}m^{2}\alpha_{\chi_{c}}\int_{\bar{\alpha}_{\chi_{c}}}^{1}d\zeta\,\int\frac{d^{2}\ell}{(2\pi)^{2}}n_{\Phi_{\chi_{c}}}\left(x,\,\boldsymbol{\ell},\,\,-\boldsymbol{\Delta}_{\perp},\,z_{\chi_{c}}\right)\times\\
 & \frac{\zeta\left(K_{(1)x}+iK_{(1)y}\right)\sqrt{2\bar{\zeta}\left(\zeta-\bar{\alpha}_{\chi_{c}}\right)}\left(\alpha_{\chi_{c}}-2\bar{\zeta}\right)}{\left(\boldsymbol{P}_{(1)}^{2}+m^{2}\right)\left(\bar{\zeta}\,\zeta^{2}\boldsymbol{K}_{(1)}^{2}+\left(\zeta-\bar{\alpha}_{\chi_{c}}\right)\,\bar{\alpha}_{\chi_{c}}\boldsymbol{P}_{(1)}^{2}+\alpha_{\chi_{c}}\bar{\alpha}_{\chi_{c}}\zeta m^{2}\right)},\nonumber 
\end{align}

\begin{align}
\mathcal{A}_{1,\,\chi_{c1}}^{(+;\,+,0)} & =-\frac{8\bar{\alpha}_{\chi_{c}}\alpha_{\chi_{c}}^{2}}{M_{\chi_{c1}}}\pi\bar{\kappa}\int_{\bar{\alpha}_{\chi_{c}}}^{1}d\zeta\,\int\frac{d^{2}\ell}{(2\pi)^{2}}\sqrt{\frac{\bar{\zeta}}{\zeta-\bar{\alpha}_{\chi_{c}}}}\times\\
 & \left\{ \frac{m^{2}\bar{\alpha}_{\chi_{c}}\zeta\left(P_{(1)x}+iP_{(1)y}\right)n_{\Phi_{\chi_{c}}}^{(-)}\left(x,\,\boldsymbol{\ell},\,\,-\boldsymbol{\Delta}_{\perp},\,z_{\chi_{c}}\right)}{\left(\boldsymbol{P}_{(1)}^{2}+m^{2}\right)\left(\bar{\zeta}\,\zeta^{2}\boldsymbol{K}_{(1)}^{2}+\left(\zeta-\bar{\alpha}_{\chi_{c}}\right)\,\bar{\alpha}_{\chi_{c}}\boldsymbol{P}_{(1)}^{2}+\alpha_{\chi_{c}}\bar{\alpha}_{\chi_{c}}\zeta m^{2}\right)}\right.\nonumber \\
 & \left.-i\frac{n_{\Phi_{\chi_{c}}}^{(2,\,0)}\left(x,\,\boldsymbol{\ell},\,\,-\boldsymbol{\Delta}_{\perp},\,z_{\chi_{c}}\right)\left[\left(\zeta-\bar{\alpha}_{\chi_{c}}\right)\left(\boldsymbol{P}_{(1)}^{2}+m^{2}\right)+m^{2}\bar{\alpha}_{\chi_{c}}\right]}{\left(\boldsymbol{P}_{(1)}^{2}+m^{2}\right)\left(\bar{\zeta}\,\zeta^{2}\boldsymbol{K}_{(1)}^{2}+\left(\zeta-\bar{\alpha}_{\chi_{c}}\right)\,\bar{\alpha}_{\chi_{c}}\boldsymbol{P}_{(1)}^{2}+\alpha_{\chi_{c}}\bar{\alpha}_{\chi_{c}}\zeta m^{2}\right)}\right\} \nonumber 
\end{align}
\begin{align}
\mathcal{A}_{1,\,\chi_{c1}}^{(+;\,-,0)} & =-\frac{8m^{2}}{M_{\chi_{c1}}}\pi\bar{\kappa}\bar{\alpha}_{\chi_{c}}^{2}\alpha_{\chi_{c}}^{2}\int_{\bar{\alpha}_{\chi_{c}}}^{1}d\zeta\,\int\frac{d^{2}\ell}{(2\pi)^{2}}\sqrt{\frac{\bar{\zeta}^{3}}{\zeta-\bar{\alpha}_{\chi_{c}}}}n_{\Phi_{\chi_{c}}}^{(+)}\left(x,\,\boldsymbol{\ell},\,\,-\boldsymbol{\Delta}_{\perp},\,z_{\chi_{c}}\right)\\
 & \times\frac{\left(P_{(1)x}+iP_{(1)y}\right)}{\left(\boldsymbol{P}_{(1)}^{2}+m^{2}\right)\left(\bar{\zeta}\,\zeta^{2}\boldsymbol{K}_{(1)}^{2}+\left(\zeta-\bar{\alpha}_{\chi_{c}}\right)\,\bar{\alpha}_{\chi_{c}}\boldsymbol{P}_{(1)}^{2}+\alpha_{\chi_{c}}\bar{\alpha}_{\chi_{c}}\zeta m^{2}\right)},\nonumber 
\end{align}

\begin{align}
\mathcal{A}_{1,\,\chi_{c1}}^{(+;\,+,-1)} & =4\pi\bar{\kappa}\alpha_{\chi_{c}}\int_{\bar{\alpha}_{\chi_{c}}}^{1}d\zeta\,\int\frac{d^{2}\ell}{(2\pi)^{2}}\,n_{\Phi_{\chi_{c}}}^{(+)}\left(x,\,\boldsymbol{\ell},\,\,-\boldsymbol{\Delta}_{\perp},\,z_{\chi_{c}}\right)\\
 & \times\frac{\left(K_{(1)x}-iK_{(1)y}\right)\left(P_{(1)x}+iP_{(1)y}\right)\zeta\left(2\zeta-1\right)\sqrt{2\bar{\zeta}\left(\zeta-\bar{\alpha}_{\chi_{c}}\right)}}{\left(\boldsymbol{P}_{(1)}^{2}+m^{2}\right)\left(\bar{\zeta}\,\zeta^{2}\boldsymbol{K}_{(1)}^{2}+\left(\zeta-\bar{\alpha}_{\chi_{c}}\right)\,\bar{\alpha}_{\chi_{c}}\boldsymbol{P}_{(1)}^{2}+\alpha_{\chi_{c}}\bar{\alpha}_{\chi_{c}}\zeta m^{2}\right)},\nonumber 
\end{align}
\begin{align}
\mathcal{A}_{1,\,\chi_{c1}}^{(+;\,-,-1)} & =4\pi\bar{\kappa}\alpha_{\chi_{c}}^{2}\int_{\bar{\alpha}_{\chi_{c}}}^{1}d\zeta\,\int\frac{d^{2}\ell}{(2\pi)^{2}}\,\sqrt{\frac{2\bar{\zeta}}{\zeta-\bar{\alpha}_{\chi_{c}}}}\\
 & \times\frac{\zeta^{2}\,\left(K_{(1)x}+iK_{(1)y}\right)\left(P_{(1)x}+iP_{(1)y}\right)\left(\alpha_{\chi_{c}}-2\bar{\zeta}\right)n_{\Phi_{\chi_{c}}}^{(+)}\left(x,\,\boldsymbol{\ell},\,\,-\boldsymbol{\Delta}_{\perp},\,z_{\chi_{c}}\right)}{\left(\boldsymbol{P}_{(1)}^{2}+m^{2}\right)\left(\bar{\zeta}\,\zeta^{2}\boldsymbol{K}_{(1)}^{2}+\left(\zeta-\bar{\alpha}_{\chi_{c}}\right)\,\bar{\alpha}_{\chi_{c}}\boldsymbol{P}_{(1)}^{2}+\alpha_{\chi_{c}}\bar{\alpha}_{\chi_{c}}\zeta m^{2}\right)},\nonumber 
\end{align}
\begin{equation}
\mathcal{A}_{1,\,\chi_{c1}}^{(+;\,-,-1)}=0.
\end{equation}
Finally, for $\chi_{c2}$ meson we obtained

\begin{align}
\mathcal{A}_{1,\,\chi_{c2}}^{(+;\,+,+2)} & =8i\pi\bar{\kappa}\,\alpha_{\chi_{c}}\bar{\alpha}_{\chi_{c}}\int_{\bar{\alpha}_{\chi_{c}}}^{1}d\zeta\,\int\frac{d^{2}\ell}{(2\pi)^{2}}\sqrt{\frac{\bar{\zeta}^{3}}{\zeta-\bar{\alpha}_{\chi_{c}}}}\\
 & \times\frac{\left[m^{2}\bar{\alpha}_{\chi_{c}}+\left(\boldsymbol{P}_{(1)}^{2}+m^{2}\right)\left(\zeta-\bar{\alpha}_{\chi_{c}}\right)\right]n_{\Phi_{\chi_{c}}}^{(2,\,-2)}\left(x,\,\boldsymbol{\ell},\,\,-\boldsymbol{\Delta}_{\perp},\,z_{\chi_{c}}\right)}{\left(\boldsymbol{P}_{(1)}^{2}+m^{2}\right)\left(\bar{\zeta}\,\zeta^{2}\boldsymbol{K}_{(1)}^{2}+\left(\zeta-\bar{\alpha}_{\chi_{c}}\right)\,\bar{\alpha}_{\chi_{c}}\boldsymbol{P}_{(1)}^{2}+\alpha_{\chi_{c}}\bar{\alpha}_{\chi_{c}}\zeta m^{2}\right)},\nonumber 
\end{align}
\begin{align}
\mathcal{A}_{1,\,\chi_{c2}}^{(+;\,-,+2)} & =8\pi\bar{\kappa}m^{2}\bar{\alpha}_{\chi_{c}}^{2}\alpha_{\chi_{c}}^{2}\int_{\bar{\alpha}_{\chi_{c}}}^{1}d\zeta\,\int\frac{d^{2}\ell}{(2\pi)^{2}}\sqrt{\frac{\bar{\zeta}^{3}}{\zeta-\bar{\alpha}_{\chi_{c}}}}n_{\Phi_{\chi_{c}}}^{(-)}\left(x,\,\boldsymbol{\ell},\,\,-\boldsymbol{\Delta}_{\perp},\,z_{\chi_{c}}\right)\\
 & \times\frac{\left(P_{(1)x}+iP_{(1)y}\right)}{\left(\bar{\zeta}+\bar{\alpha}_{\chi_{c}}\right)^{2}\left(\boldsymbol{P}_{(1)}^{2}+m^{2}\right)\left(\bar{\zeta}\,\zeta^{2}\boldsymbol{K}_{(1)}^{2}+\left(\zeta-\bar{\alpha}_{\chi_{c}}\right)\,\bar{\alpha}_{\chi_{c}}\boldsymbol{P}_{(1)}^{2}+\alpha_{\chi_{c}}\bar{\alpha}_{\chi_{c}}\zeta m^{2}\right)}\nonumber 
\end{align}
\begin{align}
\mathcal{A}_{1,\,\chi_{c2}}^{(+;\,+,+1)} & =4iM_{\chi_{c2}}\pi\bar{\kappa}\int_{\bar{\alpha}_{\chi_{c}}}^{1}d\zeta\,\int\frac{d^{2}\ell}{(2\pi)^{2}}\left(K_{(1)x}-iK_{(1)y}\right)\sqrt{\frac{\bar{\zeta}}{\zeta-\bar{\alpha}_{\chi_{c}}}}\zeta\\
 & \times\left\{ \frac{\left(P_{(1)x}+iP_{(1)y}\right)\left(\zeta-\bar{\alpha}_{\chi_{c}}\right)\left(8\zeta^{3}+4\zeta^{2}(\alpha_{\chi_{c}}-4)-6\zeta(\alpha_{\chi_{c}}-2)+3\alpha_{\chi_{c}}-4\right)n_{\Phi_{\chi_{c}}}^{(-)}\left(x,\,\boldsymbol{\ell},\,\,-\boldsymbol{\Delta}_{\perp},\,z_{\chi_{c}}\right)}{\left(\boldsymbol{P}_{(1)}^{2}+m^{2}\right)\left(\bar{\zeta}\,\zeta^{2}\boldsymbol{K}_{(1)}^{2}+\left(\zeta-\bar{\alpha}_{\chi_{c}}\right)\,\bar{\alpha}_{\chi_{c}}\boldsymbol{P}_{(1)}^{2}+\alpha_{\chi_{c}}\bar{\alpha}_{\chi_{c}}\zeta m^{2}\right)}\right.\nonumber \\
 & -i\left.\frac{\zeta m^{2}\alpha_{\chi_{c}}n_{\Phi_{\chi_{c}}}\left(x,\,\boldsymbol{\ell},\,\,-\boldsymbol{\Delta}_{\perp},\,z_{\chi_{c}}\right)\left(\alpha_{\chi_{c}}-2\bar{\zeta}\right)}{\left(\boldsymbol{P}_{(1)}^{2}+m^{2}\right)\left(\bar{\zeta}\,\zeta^{2}\boldsymbol{K}_{(1)}^{2}+\left(\zeta-\bar{\alpha}_{\chi_{c}}\right)\,\bar{\alpha}_{\chi_{c}}\boldsymbol{P}_{(1)}^{2}+\alpha_{\chi_{c}}\bar{\alpha}_{\chi_{c}}\zeta m^{2}\right)}\right\} ,\nonumber 
\end{align}
\begin{align}
\mathcal{A}_{1,\,\chi_{c2}}^{(+;\,-,+1)} & =4\pi\bar{\kappa}M_{\chi_{c2}}\int_{\bar{\alpha}_{\chi_{c}}}^{1}d\zeta\,\int\frac{d^{2}\ell}{(2\pi)^{2}}\zeta\sqrt{\frac{\bar{\zeta}}{\zeta-\bar{\alpha}_{\chi_{c}}}}\left(\alpha_{\chi_{c}}-2\bar{\zeta}\right)\\
 & \times\left\{ \frac{\alpha_{\chi_{c}}\left(\alpha_{\chi_{c}}-\bar{\zeta}\right)m^{2}n_{\Phi_{\chi_{c}}}\left(x,\,\boldsymbol{\ell},\,\,-\boldsymbol{\Delta}_{\perp},\,z_{\chi_{c}}\right)}{\left(\boldsymbol{P}_{(1)}^{2}+m^{2}\right)\left(\bar{\zeta}\,\zeta^{2}\boldsymbol{K}_{(1)}^{2}+\left(\zeta-\bar{\alpha}_{\chi_{c}}\right)\,\bar{\alpha}_{\chi_{c}}\boldsymbol{P}_{(1)}^{2}+\alpha_{\chi_{c}}\bar{\alpha}_{\chi_{c}}\zeta m^{2}\right)}\right.\nonumber \\
 & \left.+\frac{i\zeta\left(P_{(1)x}+iP_{(1)y}\right)\left(\alpha_{\chi_{c}}-2\bar{\zeta}\right)^{2}n_{\Phi_{\chi_{c}}}^{(-)}\left(x,\,\boldsymbol{\ell},\,\,-\boldsymbol{\Delta}_{\perp},\,z_{\chi_{c}}\right)}{\left(\boldsymbol{P}_{(1)}^{2}+m^{2}\right)\left(\bar{\zeta}\,\zeta^{2}\boldsymbol{K}_{(1)}^{2}+\left(\zeta-\bar{\alpha}_{\chi_{c}}\right)\,\bar{\alpha}_{\chi_{c}}\boldsymbol{P}_{(1)}^{2}+\alpha_{\chi_{c}}\bar{\alpha}_{\chi_{c}}\zeta m^{2}\right)}\right\} ,\nonumber 
\end{align}

\begin{align}
\mathcal{A}_{1,\,\chi_{c2}}^{(+;\,+,0)} & =-4\sqrt{\frac{2}{3}}\pi\bar{\kappa}\alpha_{\chi_{c}}\int_{\bar{\alpha}_{\chi_{c}}}^{1}d\zeta\,\int\frac{d^{2}\ell}{(2\pi)^{2}}\sqrt{\frac{\bar{\zeta}}{\zeta-\bar{\alpha}_{\chi_{c}}}}\\
 & \times\left\{ \frac{2m^{2}\left(\alpha_{\chi_{c}}-2\bar{\zeta}\right)n_{\Phi_{\chi_{c}}}\left(x,\,\boldsymbol{\ell},\,\,-\boldsymbol{\Delta}_{\perp},\,z_{\chi_{c}}\right)\left[\bar{\alpha}_{\chi_{c}}\left(\zeta-\bar{\alpha}_{\chi_{c}}\right)\left(\boldsymbol{P}_{(1)}^{2}+m^{2}\right)+m\left(m\bar{\alpha}_{\chi_{c}}^{2}-\zeta^{2}\left(K_{(1)x}-iK_{(1)y}\right)\right)\right]}{\left(\boldsymbol{P}_{(1)}^{2}+m^{2}\right)\left(\bar{\zeta}\,\zeta^{2}\boldsymbol{K}_{(1)}^{2}+\left(\zeta-\bar{\alpha}_{\chi_{c}}\right)\,\bar{\alpha}_{\chi_{c}}\boldsymbol{P}_{(1)}^{2}+\alpha_{\chi_{c}}\bar{\alpha}_{\chi_{c}}\zeta m^{2}\right)}\right.+\nonumber \\
 & -\frac{2im^{2}\zeta\alpha_{\chi_{c}}\left(P_{(1)x}+iP_{(1)y}\right)\bar{\alpha}_{\chi_{c}}^{2}n_{\Phi_{\chi_{c}}}^{(-)}\left(x,\,\boldsymbol{\ell},\,\,-\boldsymbol{\Delta}_{\perp},\,z_{\chi_{c}}\right)}{\left(\boldsymbol{P}_{(1)}^{2}+m^{2}\right)\left(\bar{\zeta}\,\zeta^{2}\boldsymbol{K}_{(1)}^{2}+\left(\zeta-\bar{\alpha}_{\chi_{c}}\right)\,\bar{\alpha}_{\chi_{c}}\boldsymbol{P}_{(1)}^{2}+\alpha_{\chi_{c}}\bar{\alpha}_{\chi_{c}}\zeta m^{2}\right)}\nonumber \\
 & -\left.\frac{3\zeta\alpha_{\chi_{c}}\left(P_{(1)x}+iP_{(1)y}\right)\left(K_{(1)x}-iK_{(1)y}\right)\left[\bar{\zeta}\left(2\zeta+\alpha_{\chi_{c}}\right)-1\right]n_{\Phi_{\chi_{c}}}^{(2,\,0)}\left(x,\,\boldsymbol{\ell},\,\,-\boldsymbol{\Delta}_{\perp},\,z_{\chi_{c}}\right)}{\left(\boldsymbol{P}_{(1)}^{2}+m^{2}\right)\left(\bar{\zeta}\,\zeta^{2}\boldsymbol{K}_{(1)}^{2}+\left(\zeta-\bar{\alpha}_{\chi_{c}}\right)\,\bar{\alpha}_{\chi_{c}}\boldsymbol{P}_{(1)}^{2}+\alpha_{\chi_{c}}\bar{\alpha}_{\chi_{c}}\zeta m^{2}\right)}\right\} ,\nonumber 
\end{align}
\begin{align}
\mathcal{A}_{1,\,\chi_{c2}}^{(+;\,-,0)} & =4\sqrt{\frac{2}{3}}\pi\bar{\kappa}m^{2}\alpha_{\chi_{c}}\int_{\bar{\alpha}_{\chi_{c}}}^{1}d\zeta\,\int\frac{d^{2}\ell}{(2\pi)^{2}}\sqrt{\frac{\bar{\zeta}}{\zeta-\bar{\alpha}_{\chi_{c}}}}\\
 & \times\left\{ \frac{2\zeta m^{3}\left(K_{(1)x}+iK_{(1)y}\right)\left(\zeta-\bar{\alpha}_{\chi_{c}}\right)\left(\alpha_{\chi_{c}}-2\bar{\zeta}\right)n_{\Phi_{\chi_{c}}}\left(x,\,\boldsymbol{\ell},\,\,-\boldsymbol{\Delta}_{\perp},\,z_{\chi_{c}}\right)}{\left(\boldsymbol{P}_{(1)}^{2}+m^{2}\right)\left(\bar{\zeta}\,\zeta^{2}\boldsymbol{K}_{(1)}^{2}+\left(\zeta-\bar{\alpha}_{\chi_{c}}\right)\,\bar{\alpha}_{\chi_{c}}\boldsymbol{P}_{(1)}^{2}+\alpha_{\chi_{c}}\bar{\alpha}_{\chi_{c}}\zeta m^{2}\right)}\right.+\nonumber \\
 & \quad-\frac{2i\alpha_{\chi_{c}}\bar{\alpha}_{\chi_{c}}^{2}\bar{\zeta}m^{2}\left(P_{(1)x}+iP_{(1)y}\right)n_{\Phi_{\chi_{c}}}^{(+)}\left(x,\,\boldsymbol{\ell},\,\,-\boldsymbol{\Delta}_{\perp},\,z_{\chi_{c}}\right)}{\left(\boldsymbol{P}_{(1)}^{2}+m^{2}\right)\left(\bar{\zeta}\,\zeta^{2}\boldsymbol{K}_{(1)}^{2}+\left(\zeta-\bar{\alpha}_{\chi_{c}}\right)\,\bar{\alpha}_{\chi_{c}}\boldsymbol{P}_{(1)}^{2}+\alpha_{\chi_{c}}\bar{\alpha}_{\chi_{c}}\zeta m^{2}\right)}\nonumber \\
 & \left.-\frac{3\alpha_{\chi_{c}}\zeta^{2}\left(\alpha_{\chi_{c}}-2\bar{\zeta}\right)\left(K_{(1)x}+iK_{(1)y}\right)\left(P_{(1)x}+iP_{(1)y}\right)n_{\Phi_{\chi_{c}}}^{(2,\,0)}\left(x,\,\boldsymbol{\ell},\,\,-\boldsymbol{\Delta}_{\perp},\,z_{\chi_{c}}\right)}{\left(\boldsymbol{P}_{(1)}^{2}+m^{2}\right)\left(\bar{\zeta}\,\zeta^{2}\boldsymbol{K}_{(1)}^{2}+\left(\zeta-\bar{\alpha}_{\chi_{c}}\right)\,\bar{\alpha}_{\chi_{c}}\boldsymbol{P}_{(1)}^{2}+\alpha_{\chi_{c}}\bar{\alpha}_{\chi_{c}}\zeta m^{2}\right)}\right\} ,\nonumber 
\end{align}

\begin{align}
\mathcal{A}_{1,\,\chi_{c2}}^{(+;\,+,-1)} & =4i\pi\bar{\kappa}M_{\chi_{c2}}\int_{\bar{\alpha}_{\chi_{c}}}^{1}d\zeta\,\int\frac{d^{2}\ell}{(2\pi)^{2}}\zeta\bar{\zeta}\sqrt{\frac{\bar{\zeta}}{\zeta-\bar{\alpha}_{\chi_{c}}}}n_{\Phi_{\chi_{c}}}^{(+)}\left(x,\,\boldsymbol{\ell},\,\,-\boldsymbol{\Delta}_{\perp},\,z_{\chi_{c}}\right)\\
 & \times\frac{\left(K_{(1)x}-iK_{(1)y}\right)\left(P_{(1)x}+iP_{(1)y}\right)\left(8\zeta^{3}+4\zeta^{2}(3\alpha_{\chi_{c}}-4)-6\zeta(\alpha_{\chi_{c}}-2)\bar{\alpha}_{\chi_{c}}+(\alpha_{\chi_{c}}-4)\bar{\alpha}_{\chi_{c}}^{2}\right)}{\left(\boldsymbol{P}_{(1)}^{2}+m^{2}\right)\left(\bar{\zeta}\,\zeta^{2}\boldsymbol{K}_{(1)}^{2}+\left(\zeta-\bar{\alpha}_{\chi_{c}}\right)\,\bar{\alpha}_{\chi_{c}}\boldsymbol{P}_{(1)}^{2}+\alpha_{\chi_{c}}\bar{\alpha}_{\chi_{c}}\zeta m^{2}\right)},\nonumber 
\end{align}
\begin{align}
\mathcal{A}_{1,\,\chi_{c2}}^{(+;\,-,-1)} & =16i\pi\bar{\kappa}M_{\chi_{c2}}\int_{\bar{\alpha}_{\chi_{c}}}^{1}d\zeta\,\int\frac{d^{2}\ell}{(2\pi)^{2}}\zeta^{2}\bar{\zeta}\sqrt{\bar{\zeta}\left(\zeta-\bar{\alpha}_{\chi_{c}}\right)}n_{\Phi_{\chi_{c}}}^{(+)}\left(x,\,\boldsymbol{\ell},\,\,-\boldsymbol{\Delta}_{\perp},\,z_{\chi_{c}}\right)\\
 & \times\frac{\left(K_{(1)x}+iK_{(1)y}\right)\left(P_{(1)x}+iP_{(1)y}\right)\left(\alpha_{\chi_{c}}-2\bar{\zeta}\right)}{\left(\boldsymbol{P}_{(1)}^{2}+m^{2}\right)\left(\bar{\zeta}\,\zeta^{2}\boldsymbol{K}_{(1)}^{2}+\left(\zeta-\bar{\alpha}_{\chi_{c}}\right)\,\bar{\alpha}_{\chi_{c}}\boldsymbol{P}_{(1)}^{2}+\alpha_{\chi_{c}}\bar{\alpha}_{\chi_{c}}\zeta m^{2}\right)},\nonumber 
\end{align}

\begin{align}
\mathcal{A}_{1,\,\chi_{c2}}^{(+;\,+,-2)} & =-8i\alpha_{\chi_{c}}\bar{\alpha}_{\chi_{c}}\pi\bar{\kappa}\int_{\bar{\alpha}_{\chi_{c}}}^{1}d\zeta\,\int\frac{d^{2}\ell}{(2\pi)^{2}}\sqrt{\frac{\bar{\zeta}}{\zeta-\bar{\alpha}_{\chi_{c}}}}\\
 & \times\left\{ \frac{n_{\Phi_{\chi_{c}}}^{(2,\,+2)}\left(x,\,\boldsymbol{\ell},\,\,-\boldsymbol{\Delta}_{\perp},\,z_{\chi_{c}}\right)\left[\left(\zeta-\bar{\alpha}_{\chi_{c}}\right)\left(\boldsymbol{P}_{(1)}^{2}+m^{2}\right)+m^{2}\bar{\alpha}_{\chi_{c}}\right]\left(\zeta-\bar{\alpha}_{\chi_{c}}\right)}{\left(\boldsymbol{P}_{(1)}^{2}+m^{2}\right)\left(\bar{\zeta}\,\zeta^{2}\boldsymbol{K}_{(1)}^{2}+\left(\zeta-\bar{\alpha}_{\chi_{c}}\right)\,\bar{\alpha}_{\chi_{c}}\boldsymbol{P}_{(1)}^{2}+\alpha_{\chi_{c}}\bar{\alpha}_{\chi_{c}}\zeta m^{2}\right)}\right.\nonumber \\
 & \left.+\frac{im^{2}\zeta\alpha_{\chi_{c}}\bar{\alpha}_{\chi_{c}}\left(P_{(1)x}+iP_{(1)y}\right)n_{\Phi_{\chi_{c}}}^{(+)}\left(x,\,\boldsymbol{\ell},\,\,-\boldsymbol{\Delta}_{\perp},\,z_{\chi_{c}}\right)}{\left(\boldsymbol{P}_{(1)}^{2}+m^{2}\right)\left(\bar{\zeta}\,\zeta^{2}\boldsymbol{K}_{(1)}^{2}+\left(\zeta-\bar{\alpha}_{\chi_{c}}\right)\,\bar{\alpha}_{\chi_{c}}\boldsymbol{P}_{(1)}^{2}+\alpha_{\chi_{c}}\bar{\alpha}_{\chi_{c}}\zeta m^{2}\right)}\right\} ,\nonumber 
\end{align}
\begin{equation}
\mathcal{A}_{1,\,\chi_{c2}}^{(+;\,-,-2)}=0.\label{eq:Amplitude_PP--1}
\end{equation}

\subsubsection{Evaluation of the amplitude $\mathcal{A}_{2}^{(\lambda;\,\sigma,\,H)}$}

\label{subsec:A2}The amplitude $\mathcal{A}_{2}^{(\lambda;\,\sigma,\,H)}$
corresponds to contribution of the diagrams in the right column of
the Figure~\ref{fig:CGCBasic-1} and can be evaluated using the procedure
outlined in the previous subsection~\ref{subsec:A1}. We may represent
the amplitude as a convolution of the dipole amplitude and the impact
factors $R_{2,i}$, \\
 
\begin{align}
\mathcal{A}_{2,i} & =\int d^{2}\boldsymbol{x}_{0}d^{2}\boldsymbol{x}_{1}\,\mathcal{N}\left(Y,\,\boldsymbol{x}_{0},\,\boldsymbol{x}_{1}\right)R_{2,i}\left(Y,\boldsymbol{x}_{0},\,\boldsymbol{x}_{1},q,\,p_{\chi_{c}},\,k_{\gamma}\right),\qquad i=1,2.\label{eq:a-2-1-1}
\end{align}
The corresponding expressions for $R_{2,i}$ in~(\ref{eq:a-2-1-1})
differ from the impact factors $R_{1,i}$ only by permutation of the
$\chi_{c}$ and the incoming photon, namely
\begin{align}
 & R_{2,1}\left(Y,\boldsymbol{x}_{0},\,\boldsymbol{x}_{1},q,\,p_{\chi_{c}},\,k_{\gamma}\right)=4\pi\alpha_{{\rm em}}e_{c}^{2}N_{c}\label{eq:R11-1-1}\\
 & \times\int\frac{d^{4}k_{0}}{\left(2\pi\right)^{4}}\frac{d^{4}k_{1}}{\left(2\pi\right)^{4}}\frac{d^{4}k_{0}'}{\left(2\pi\right)^{4}}\frac{d^{4}k_{1}'}{\left(2\pi\right)^{4}}\left(2\pi\right)^{4}\delta\left(k_{0}+k_{1}-q\right)\left(2\pi\right)^{4}\delta\left(k_{0}'+k_{1}'-p_{\chi_{c}}-k_{\gamma}\right)\nonumber \\
 & \times\left(2\pi\right)\delta\left(k_{0}^{+}-k_{0}'{}^{+}\right)\left(2\pi\right)\delta\left(k_{1}^{+}-k_{1}'{}^{+}\right)e^{i\left(\boldsymbol{k}_{0\perp}-\boldsymbol{k}_{0\perp}'\right)\cdot\boldsymbol{x}_{0}}e^{i\left(\boldsymbol{k}_{1\perp}-\boldsymbol{k}_{1\perp}'\right)\cdot\boldsymbol{x}_{1}}\phi_{\chi_{cJ}}\left(z,\,\boldsymbol{k}_{\perp}^{({\rm rel})}\right)\nonumber \\
 & \times\left\langle {\rm Tr}\left[\Gamma_{\chi_{c}}\left(k_{1}',\,k_{0}'-k_{\gamma}\right)S\left(k_{0}'-k_{\gamma}\right)\hat{\varepsilon}_{\gamma}^{*}\left(k_{\gamma}\right)S\left(k_{0}'\right)\gamma_{+}S\left(k_{0}\right)\hat{\varepsilon}_{\gamma}(q)S\left(-k_{1}\right)\gamma^{+}S\left(-k_{1}'\right)\right]\right\rangle \nonumber 
\end{align}
\begin{align}
 & R_{2,2}\left(Y,\boldsymbol{x}_{0},\,\boldsymbol{x}_{1},q,\,p_{\chi_{c}},\,k_{\gamma}\right)=4\pi\alpha_{{\rm em}}e_{c}^{2}N_{c}\label{eq:R11-1}\\
 & \times\int\frac{d^{4}k_{0}}{\left(2\pi\right)^{4}}\frac{d^{4}k_{1}}{\left(2\pi\right)^{4}}\frac{d^{4}k_{0}'}{\left(2\pi\right)^{4}}\frac{d^{4}k_{1}'}{\left(2\pi\right)^{4}}\left(2\pi\right)^{4}\delta\left(k_{0}+k_{1}-q\right)\left(2\pi\right)^{4}\delta\left(k_{0}'+k_{1}'-p_{\chi_{c}}-k_{\gamma}\right)\nonumber \\
 & \times\left(2\pi\right)\delta\left(k_{0}^{+}-k_{0}'{}^{+}\right)\left(2\pi\right)\delta\left(k_{1}^{+}-k_{1}'{}^{+}\right)e^{i\left(\boldsymbol{k}_{0\perp}-\boldsymbol{k}_{0\perp}'\right)\cdot\boldsymbol{x}_{0}}e^{i\left(\boldsymbol{k}_{1\perp}-\boldsymbol{k}_{1\perp}'\right)\cdot\boldsymbol{x}_{1}}\phi_{\chi_{cJ}}\left(z,\,\boldsymbol{k}_{\perp}^{({\rm rel})}\right)\nonumber \\
 & \left\langle {\rm Tr}\left[\Gamma_{\chi_{c}}\left(k_{1}'+k_{\gamma},\,k_{0}'\right)S\left(k_{0}'\right)\gamma_{+}S\left(k_{0}\right)\hat{\varepsilon}_{\gamma}(q)S\left(-k_{1}\right)\gamma_{+}S\left(-k_{1}'\right)\hat{\varepsilon}_{\gamma}^{*}\left(k_{\gamma}\right)S\left(-k_{1}'-k_{\gamma}\right)\right]\right\rangle \nonumber
\end{align}

\begin{figure}
\includegraphics[width=6cm]{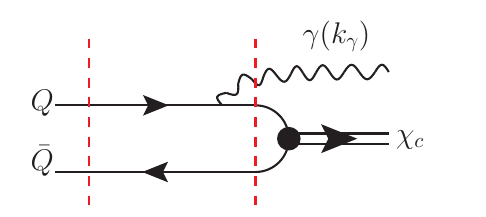}\includegraphics[width=6cm]{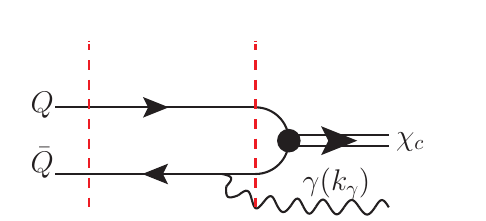}

\includegraphics[width=6cm]{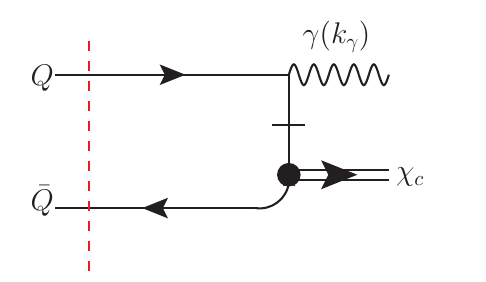}\includegraphics[width=6cm]{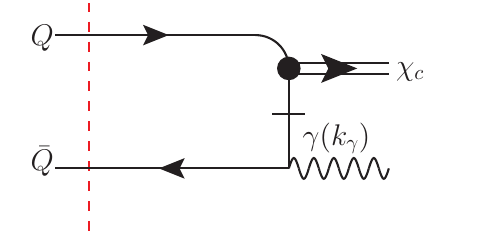}

\caption{The leading order diagrams which contribute to the amplitude of the
subprocess $Q\bar{Q}\to\gamma\chi_{c}$. The vertical dashed lines denote the energy denominators of the light-cone perturbation theory\cite{Lepage:1980fj,Brodsky:1997de}.}
\label{fig:Diags}
\end{figure}

Repeating the procedure outlined in the previous Section~\ref{subsec:A1}, we may rewrite
the impact factors $R_{2,i}$ in terms of the photon wave function
$\psi_{\gamma\to\bar{Q}Q}$ and the amplitude (``wave function'')
$\Psi_{\bar{Q}Q\to\gamma\,\chi_{c}}^{\dagger}$ of the $\bar{Q}Q\to\chi_{c}\gamma$
subprocess, contracted over helicities $h,\bar{h}$ of the quarks
and convoluted over the light-cone component $k_{Q}^{+}$
of the quark momentum. The final result for the amplitude $\mathcal{A}_{2}^{(\lambda,\sigma)}$
may be rewritten as 
\begin{align}
\mathcal{A}_{2}^{(\lambda,\sigma)} & =\int_{0}^{\alpha_{\chi_{c}}}dz_{0}\prod_{k=1}^{3}\left(d^{2}\boldsymbol{r}_{k}\right)\,\Psi_{\bar{Q}Q\to\gamma\,\chi_{c}}^{(\sigma,h,\bar{h})\dagger}\left(z_{0},\,z_{1}=\alpha_{\chi_{c}}-z_{0},\,z_{2}\equiv\bar{\alpha}_{\chi_{c}},\,\boldsymbol{r}_{0},\,\boldsymbol{r}_{1},\,\boldsymbol{r}_{2}\right)\psi_{\gamma\to\bar{Q}Q}^{(\lambda,\,h,\bar{h})}\left(\frac{z_{0}}{z_{0}+z_{1}},\,\boldsymbol{r}_{10}\right)\times\label{eq:A-2}\\
 & \times\mathcal{N}\left(x,\,\boldsymbol{r}_{10},\,\,\boldsymbol{b}_{10}\right)\exp\left[-i\boldsymbol{p}_{\perp}^{\chi_{c}}\cdot\left(\boldsymbol{b}_{10}-\frac{\bar{\alpha}_{\chi_{c}}}{\alpha_{\chi_{c}}}\boldsymbol{r}_{\gamma}\right)-i\boldsymbol{k}_{\perp}^{\gamma}\cdot\left(\boldsymbol{r}_{\gamma}+\boldsymbol{b}_{10}\right)\right].\nonumber 
\end{align}
where we use the same notations as in the previous section (see the
text under Eq.~(\ref{eq:A})). The photon wave function $\psi_{\gamma\to\bar{Q}Q}^{(\lambda,\,h,\bar{h})}$
in the leading order in $\alpha_{s}$ is given by~\cite{Bjorken:1970ah,Dosch:1996ss}
\begin{align}
\psi_{\gamma\to\bar{Q}Q}^{(\lambda,\,h,\bar{h})}\left(\zeta,\,r\right) & =-iee_{c}\int\frac{d^{2}\boldsymbol{k}_{Q\perp}}{\left(2\pi\right)^{2}}\left.\frac{\bar{u}_{h}\left(k_{Q}^{+},\,\boldsymbol{k}_{Q\perp}\right)\hat{\varepsilon}_{\lambda}(q)v_{\bar{h}}\left(q^{+}-k_{Q}^{+},\,-\boldsymbol{k}_{Q\perp}\right)}{q^{-}-\frac{q^{+}\left(\boldsymbol{k}_{Q\perp}^{2}+m^{2}\right)}{2k_{Q}^{+}\left(q^{+}-k_{Q}^{+}\right)}}e^{i\boldsymbol{k}_{Q\perp}\cdot\boldsymbol{r}_{10}}\right|_{q^{-}=0,\quad k_{Q}^{+}=\zeta q^{+}}=\label{eq:WF_gamma}\\
 & =\frac{\sqrt{2}}{2\pi}ee_{f}\,\left[-i\lambda e^{-i\lambda\phi_{r}}\left(\zeta\delta_{h,\lambda}\delta_{\bar{h},-\lambda}-(1-\zeta)\delta_{h,-\lambda}\delta_{\bar{h},\lambda}\right)mK_{1}\left(mr\right)+m\delta_{h,\lambda}\delta_{\bar{h},\lambda}K_{0}\left(mr\right)\right],\nonumber 
\end{align}
The evaluation of the amplitude $\Psi_{\bar{Q}Q\to\gamma\,\chi_{c}}^{(\sigma,h,\bar{h})}$
in the leading order in the coupling $\alpha_{s}$ resembles a similar
evaluation of the wave function $\Psi_{\gamma\to\gamma\bar{Q}Q}^{(\lambda,\sigma,h,\bar{h})}$
discussed earlier and requires evaluation of the diagrams shown in
the right column of the Figure~\ref{fig:Diags} using the light-cone
rules from~\cite{Lepage:1980fj}. The evaluation is straighforward, however, the final expression
is very lengthy, introduces a number of new special functions defined
via 2-dimensional integrals, and for this reason evaluation of~(\ref{eq:A-2})
in configuration space is not feasible. As we mentioned earlier, the
Fourier transformation of the dipole amplitude $\mathcal{N}\left(x,\,\boldsymbol{r},\,\,\boldsymbol{b}\right)$
presents certain challenges due to saturation at large $r$ and non-commutativity
of limits $r\to\infty$ and $b\to\infty$. However, the product of
the dipole amplitude and wave function $\mathcal{N}\left(x,\,r,\,\,b\right)\psi_{\gamma\to\bar{Q}Q}^{(\lambda,\,h,\bar{h})}\left(\zeta,\,r\right)$
has a well-defined Fourier image: the functions $K_{0},K_{1}$ suppress
the product at large $r$, whereas a color transparency (suppression
of $\mathcal{N}$ at small $r$) guarantees that the Fourier integrals
remain convergent at small $r$. Similar to~(\ref{eq:nPhi}-\ref{eq:nPhi2c}),
we may introduce new functions
\begin{equation}
n_{0}\left(x,\,\boldsymbol{\ell},\,\,\boldsymbol{s}\right)=\int d^{2}\boldsymbol{b}\,d^{2}\boldsymbol{r}e^{-i\boldsymbol{\ell}\cdot\boldsymbol{r}}e^{-i\boldsymbol{s}\cdot\boldsymbol{b}}\mathcal{N}\left(x,\,\boldsymbol{r},\,\,\boldsymbol{b}\right)K_{0}\left(mr\right),\label{eq:n0}
\end{equation}

\begin{equation}
n_{1}^{(\pm)}\left(x,\,\boldsymbol{\ell},\,\,\boldsymbol{s}\right)=\int d^{2}\boldsymbol{b}\,d^{2}\boldsymbol{r}e^{-i\boldsymbol{\ell}\cdot\boldsymbol{r}}e^{-i\boldsymbol{s}\cdot\boldsymbol{b}}\mathcal{N}\left(x,\,\boldsymbol{r},\,\,\boldsymbol{b}\right)K_{1}\left(mr\right)e^{\pm i\phi_{r}},\label{eq:n1}
\end{equation}
where the angle $\phi_{r}={\rm arg}(r_{x}+ir_{y})$ characterizes
the azimuthal orientation of the vector $\boldsymbol{r}$, and express
the product $\mathcal{N}\left(x,\,r,\,\,b\right)\psi_{\gamma\to\bar{Q}Q}^{(\lambda,\,h,\bar{h})}\left(\zeta,\,r\right)$
in terms of the (inverse) Fourier transforms of these functions. We
found that the functions $n_{0},n_{1}^{(\pm)}$ have a mild dependence
on all their arguments and are strongly suppressed at large $\ell,\,s$.
For this reason, it is possible to significantly speed up the evaluations
using caching (interpolation of pre-evaluated) on a modestly sized
grid.

In order to reduce the charge conjugate contributions in the amplitude
$\mathcal{A}_{2}^{(\lambda;\,\sigma,H)}$ to the common denominator,
we will replace the dummy integration variable $z_{0}$ with a new
variable $\zeta$ defined as a fraction of the incoming photon's momentum
carried by the active fermion before emission of the secondary photon,
namely
\begin{equation}
\zeta=\frac{k_{0}^{+}}{k_{0}^{+}+k_{1}^{+}}=\frac{z_{0}}{z_{0}+z_{1}},
\end{equation}
for the contribution with photon emission from the quark, and with
\begin{equation}
\zeta=\frac{k_{1}^{+}}{k_{0}^{+}+k_{1}^{+}}=1-\frac{z_{0}}{z_{0}+z_{1}},
\end{equation}
for the other contribution. In both cases, the variable $\zeta$ cannot
be less than $k_{\gamma}^{+}/q^{+}=\bar{\alpha}_{\chi_{c}}$. The
evaluation of the amplitudes is straightforward and largely repeats
the evaluation of~(\ref{eq:Amplitude_PP-1}-\ref{eq:Amplitude_PP--1}).
Due to space limitation, we will provide only the final results. For
the $\chi_{c0}$ mesons, the corresponding helicity amplitudes are
given by
\begin{align}
\mathcal{A}_{2,\,\chi_{c0}}^{(+;\,+)} & =-2m\alpha_{\chi_{c}}\bar{\alpha}_{\chi_{c}}\bar{\kappa}\int_{\bar{\alpha}_{\chi_{c}}}^{1}d\zeta\,\int\frac{d^{2}\ell}{(2\pi)^{2}}\,\frac{\sqrt{\zeta\,\bar{\zeta}^{3}}}{\left(\zeta-\bar{\alpha}_{\chi_{c}}\right)^{2}}\tilde\Phi_{\chi_{c}}\left(z_{\chi_{c}},\,\boldsymbol{P}_{(2)}\right)\times\label{eq:Amplitude_PP-1-1}\\
 & \times\left\{ \frac{\zeta n_{0}\left(x,\,\boldsymbol{\ell},\,\,-\boldsymbol{\Delta}_{\perp}\right)\left(\boldsymbol{P}_{(2)}^{2}+m^{2}\left(z_{c}-\frac{1}{2}\right)^{2}\right)}{\left(\bar{\zeta}\left(\zeta-\bar{\alpha}_{\chi_{c}}\right)^{2}\boldsymbol{K}_{(2)}^{2}-\zeta\,\alpha_{\chi_{c}}\bar{\alpha}_{\chi_{c}}\boldsymbol{P}_{(2)}^{2}-\frac{\bar{\alpha}_{\chi_{c}}}{\alpha_{\chi_{c}}}\left(\zeta-\bar{\alpha}_{\chi_{c}}\right)\left(\alpha_{\chi_{c}}-4\zeta\,\bar{\zeta}\right)m^{2}\right)}\right.\nonumber \\
 & \left.+\frac{m\bar{\alpha}_{\chi_{c}}\left(mn_{0}\left(x,\,\boldsymbol{\ell},\,\,-\boldsymbol{\Delta}_{\perp}\right)\left(z_{c}-\frac{1}{2}\right)+i\zeta\left(P_{(2)x}+iP_{(2)y}\right)n_{1}^{(-)}\left(x,\,\boldsymbol{\ell},\,\,-\boldsymbol{\Delta}_{\perp}\right)\right)}{\left(\bar{\zeta}\left(\zeta-\bar{\alpha}_{\chi_{c}}\right)^{2}\boldsymbol{K}_{(2)}^{2}-\zeta\,\alpha_{\chi_{c}}\bar{\alpha}_{\chi_{c}}\boldsymbol{P}_{(2)}^{2}-\frac{\bar{\alpha}_{\chi_{c}}}{\alpha_{\chi_{c}}}\left(\zeta-\bar{\alpha}_{\chi_{c}}\right)\left(\alpha_{\chi_{c}}-4\zeta\,\bar{\zeta}\right)m^{2}\right)}\right\} ,\nonumber 
\end{align}
\begin{align}
\mathcal{A}_{2,\,\chi_{c0}}^{(+;\,-)} & =2i\bar{\kappa}m^{2}\alpha_{\chi_{c}}\bar{\alpha}_{\chi_{c}}^{2}\int_{\bar{\alpha}_{\chi_{c}}}^{1}d\zeta\,\int\frac{d^{2}\ell}{(2\pi)^{2}}\,\sqrt{\zeta\,\bar{\zeta}^{5}}\label{eq:Amplitude_PM-1-1}\\
 & \times\frac{\left(P_{(2)x}-iP_{(2)y}\right)n_{1}^{(-)}\left(x,\,\boldsymbol{\ell},\,\,-\boldsymbol{\Delta}_{\perp}\right)\tilde\Phi_{\chi_{c}}\left(z_{\chi_{c}},\,\boldsymbol{P}_{(2)}\right)}{\left(\bar{\zeta}\left(\zeta-\bar{\alpha}_{\chi_{c}}\right)^{2}\boldsymbol{K}_{(2)}^{2}-\zeta\,\alpha_{\chi_{c}}\bar{\alpha}_{\chi_{c}}\boldsymbol{P}_{(2)}^{2}-\frac{\bar{\alpha}_{\chi_{c}}}{\alpha_{\chi_{c}}}\left(\zeta-\bar{\alpha}_{\chi_{c}}\right)\left(\alpha_{\chi_{c}}-4\zeta\,\bar{\zeta}\right)m^{2}\right)},\nonumber
\end{align}

where we introduced the shorthand notations

\begin{align}
 & \boldsymbol{K}_{(2)}=-\frac{\boldsymbol{\ell}\bar{\alpha}_{\chi_{c}}}{\zeta}+\frac{\left(\zeta-\bar{\alpha}_{\chi_{c}}\right)}{\alpha_{\chi_{c}}^{2}\zeta}\left(\alpha_{\chi_{c}}\boldsymbol{k}_{\perp}^{\gamma}-\bar{\alpha}_{\chi_{c}}\boldsymbol{p}_{\perp}^{\chi_{c}}\right)=-\frac{\boldsymbol{P}_{(2)}\bar{\alpha}_{\chi_{c}}}{\zeta}+\frac{\left(\alpha_{\chi_{c}}\boldsymbol{k}_{\perp}^{\gamma}-\bar{\alpha}_{\chi_{c}}\boldsymbol{p}_{\perp}^{\chi_{c}}\right)}{\alpha_{\chi_{c}}},\label{eq:K2}\\
 & \boldsymbol{P}_{(2)}=\boldsymbol{\ell}+\frac{\bar{\zeta}}{\alpha_{\chi_{c}}^{2}}\left(\alpha_{\chi_{c}}\boldsymbol{k}_{\perp}^{\gamma}-\bar{\alpha}_{\chi_{c}}\boldsymbol{p}_{\perp}^{\chi_{c}}\right),\qquad z_{\chi_{c}}=\frac{\zeta-\bar{\alpha}_{\chi_{c}}}{\alpha_{\chi_{c}}}.\label{eq:P2}
\end{align}

For $\chi_{c1}$ meson, the corresponding amplitudes are given by

\begin{align}
\mathcal{A}_{2,\,\chi_{c1}}^{(+;\,+,+1)} & =2m\bar{\kappa}\int_{\bar{\alpha}_{\chi_{c}}}^{1}d\zeta\,\int\frac{d^{2}\ell}{(2\pi)^{2}}\frac{\tilde\Phi_{\chi_{c}}\left(z_{\chi_{c}},\,\boldsymbol{P}_{(2)}\right)\sqrt{2\,\zeta\,\bar{\zeta}^{3}}\left(K_{x}+iK_{y}\right)}{\left(\bar{\zeta}\left(\zeta-\bar{\alpha}_{\chi_{c}}\right)^{2}\boldsymbol{K}_{(2)}^{2}-\zeta\,\alpha_{\chi_{c}}\bar{\alpha}_{\chi_{c}}\boldsymbol{P}_{(2)}^{2}-\frac{\bar{\alpha}_{\chi_{c}}}{\alpha_{\chi_{c}}}\left(\zeta-\bar{\alpha}_{\chi_{c}}\right)\left(\alpha_{\chi_{c}}-4\zeta\,\bar{\zeta}\right)m^{2}\right)}\\
 & \times\left[\left((2\zeta-1)\left(P_{(2)x}+iP_{(2)y}\right)\left(\zeta-\bar{\alpha}_{\chi_{c}}\right)n_{1}^{(-)}\left(x,\,\boldsymbol{\ell},\,\,-\boldsymbol{\Delta}_{\perp}\right)-i\zeta mn_{0}\left(x,\,\boldsymbol{\ell},\,\,-\boldsymbol{\Delta}_{\perp}\right)(2\zeta+\alpha_{\chi_{c}}-2)\right)\right],\nonumber 
\end{align}
\begin{align}
\mathcal{A}_{2,\,\chi_{c1}}^{(+;\,-,+1)} & =-2im\bar{\kappa}\int_{\bar{\alpha}_{\chi_{c}}}^{1}d\zeta\,\int\frac{d^{2}\ell}{(2\pi)^{2}}\frac{\sqrt{2\,\zeta\,\bar{\zeta}^{3}}\left(K_{(2)x}-iK_{(2)y}\right)\left(2\zeta+\alpha_{\chi_{c}}-2\right)\tilde\Phi_{\chi_{c}}\left(z_{\chi_{c}},\,\boldsymbol{P}_{(2)}\right)}{\left(\bar{\zeta}\left(\zeta-\bar{\alpha}_{\chi_{c}}\right)^{2}\boldsymbol{K}_{(2)}^{2}-\zeta\,\alpha_{\chi_{c}}\bar{\alpha}_{\chi_{c}}\boldsymbol{P}_{(2)}^{2}-\frac{\bar{\alpha}_{\chi_{c}}}{\alpha_{\chi_{c}}}\left(\zeta-\bar{\alpha}_{\chi_{c}}\right)\left(\alpha_{\chi_{c}}-4\zeta\,\bar{\zeta}\right)m^{2}\right)}\\
 & \times\left[mn_{0}\left(x,\,\boldsymbol{\ell},\,\,-\boldsymbol{\Delta}_{\perp}\right)\left(\zeta-\bar{\alpha}_{\chi_{c}}\right)+i\zeta\alpha_{\chi_{c}}\left(P_{(2)x}+iP_{(2)y}\right)n_{1}^{(-)}\left(x,\,\boldsymbol{\ell},\,\,-\boldsymbol{\Delta}_{\perp}\right)\right],\nonumber 
\end{align}

\begin{align}
\mathcal{A}_{2,\,\chi_{c1}}^{(+;\,+,0)} & =i\alpha_{\chi_{c}}\,\bar{\alpha}_{\chi_{c}}\bar{\kappa}\int_{\bar{\alpha}_{\chi_{c}}}^{1}d\zeta\,\int\frac{d^{2}\ell}{(2\pi)^{2}}\frac{\tilde\Phi_{\chi_{c}}\left(z_{\chi_{c}},\,\boldsymbol{P}_{(2)}\right)\sqrt{\zeta\,\bar{\zeta}}}{\left(\bar{\zeta}\left(\zeta-\bar{\alpha}_{\chi_{c}}\right)^{2}\boldsymbol{K}_{(2)}^{2}-\zeta\,\alpha_{\chi_{c}}\bar{\alpha}_{\chi_{c}}\boldsymbol{P}_{(2)}^{2}-\frac{\bar{\alpha}_{\chi_{c}}}{\alpha_{\chi_{c}}}\left(\zeta-\bar{\alpha}_{\chi_{c}}\right)\left(\alpha_{\chi_{c}}-4\zeta\,\bar{\zeta}\right)m^{2}\right)}\\
 & \times\left\{ mn_{0}\left(x,\,\boldsymbol{\ell},\,\,-\boldsymbol{\Delta}_{\perp}\right)\left[\bar{\alpha}_{\chi_{c}}\alpha_{\chi_{c}}^{2}\bar{\zeta}\left(m^{2}(z_{c}-1/2)^{2}-\boldsymbol{P}_{(2)}^{2}\right)-\alpha_{\chi_{c}}\zeta\left(\boldsymbol{P}_{(2)}^{2}+m^{2}\left(z_{c}-\frac{1}{2}\right)^{2}\right)\right]\right.+\nonumber \\
 & +\left.i\,n_{1}^{(-)}\left(x,\,\boldsymbol{\ell},\,\,-\boldsymbol{\Delta}_{\perp}\right)\left(P_{(2)x}+iP_{(2)y}\right)\,\,\,\,\left(2m^{2}\bar{\alpha}_{\chi_{c}}\alpha_{\chi_{c}}^{2}\zeta\,\bar{\zeta}-\alpha_{\chi_{c}}\zeta^{2}\left(\boldsymbol{P}_{(2)}^{2}+m^{2}\left(z_{c}-\frac{1}{2}\right)^{2}\right)\right)\right\} ,\nonumber 
\end{align}
\begin{align}
\mathcal{A}_{2,\,\chi_{c1}}^{(+;\,-,0)} & =\bar{\kappa}\alpha_{\chi_{c}}\bar{\alpha}_{\chi_{c}}\int_{\bar{\alpha}_{\chi_{c}}}^{1}d\zeta\,\int\frac{d^{2}\ell}{(2\pi)^{2}}\frac{\tilde\Phi_{\chi_{c}}\left(z_{\chi_{c}},\,\boldsymbol{P}_{(2)}\right)\sqrt{\zeta\,\bar{\zeta}^{3}}\left(P_{(2)x}-iP_{(2)y}\right)}{\left(\bar{\zeta}\left(\zeta-\bar{\alpha}_{\chi_{c}}\right)^{2}\boldsymbol{K}_{(2)}^{2}-\zeta\,\alpha_{\chi_{c}}\bar{\alpha}_{\chi_{c}}\boldsymbol{P}_{(2)}^{2}-\frac{\bar{\alpha}_{\chi_{c}}}{\alpha_{\chi_{c}}}\left(\zeta-\bar{\alpha}_{\chi_{c}}\right)\left(\alpha_{\chi_{c}}-4\zeta\,\bar{\zeta}\right)m^{2}\right)}\\
 & \times n_{1}^{(-)}\left(x,\,\boldsymbol{\ell},\,\,-\boldsymbol{\Delta}_{\perp}\right)\left(\zeta\alpha_{\chi_{c}}\left(\boldsymbol{P}_{(2)}^{2}+m^{2}\left(z_{c}-\frac{1}{2}\right)^{2}\right)-2\bar{\zeta}m^{2}\bar{\alpha}_{\chi_{c}}\right),\nonumber 
\end{align}

\begin{align}
\mathcal{A}_{2,\,\chi_{c1}}^{(+;\,+,-1)} & =-2\bar{\kappa}m\bar{\alpha}_{\chi_{c}}\int_{\bar{\alpha}_{\chi_{c}}}^{1}d\zeta\,\int\frac{d^{2}\ell}{(2\pi)^{2}}\tilde\Phi_{\chi_{c}}\left(z_{\chi_{c}},\,\boldsymbol{P}_{(2)}\right)\times\\
 & \frac{\left[\sqrt{2\zeta\,\bar{\zeta}^{5}}\left(\zeta-\bar{\alpha}_{\chi_{c}}\right)\left(K_{(2)x}+iK_{(2)y}\right)\left(P_{(2)x}-iP_{(2)y}\right)\left(2\zeta-\bar{\alpha}_{\chi_{c}}\right)n_{1}^{(-)}\left(x,\,\boldsymbol{\ell},\,\,-\boldsymbol{\Delta}_{\perp}\right)\right]}{\left(\bar{\zeta}\left(\zeta-\bar{\alpha}_{\chi_{c}}\right)^{2}\boldsymbol{K}_{(2)}^{2}-\zeta\,\alpha_{\chi_{c}}\bar{\alpha}_{\chi_{c}}\boldsymbol{P}_{(2)}^{2}-\frac{\bar{\alpha}_{\chi_{c}}}{\alpha_{\chi_{c}}}\left(\zeta-\bar{\alpha}_{\chi_{c}}\right)\left(\alpha_{\chi_{c}}-4\zeta\,\bar{\zeta}\right)m^{2}\right)},\nonumber 
\end{align}
\begin{equation}
\mathcal{A}_{2\,\chi_{c1}}^{(+;\,-,-1)}=0.
\end{equation}
Finally, for $\chi_{c2}$ meson we obtained

\begin{equation}
\mathcal{A}_{2,\,\chi_{c2}}^{(+;\,+,+2)}=4m^{2}\bar{\alpha}_{\chi_{c}}^{2}\bar{\kappa}\int_{\bar{\alpha}_{\chi_{c}}}^{1}d\zeta\,\int\frac{d^{2}\ell}{(2\pi)^{2}}\frac{\tilde\Phi_{\chi_{c}}\left(z_{\chi_{c}},\,\boldsymbol{P}_{(2)}\right)\sqrt{\zeta\,\bar{\zeta}^{5}}\left(P_{(2)x}+iP_{(2)y}\right)^{2}n_{0}\left(x,\,\boldsymbol{\ell},\,\,-\boldsymbol{\Delta}_{\perp}\right)}{\left(\bar{\zeta}\left(\zeta-\bar{\alpha}_{\chi_{c}}\right)^{2}\boldsymbol{K}_{(2)}^{2}-\zeta\,\alpha_{\chi_{c}}\bar{\alpha}_{\chi_{c}}\boldsymbol{P}_{(2)}^{2}-\frac{\bar{\alpha}_{\chi_{c}}}{\alpha_{\chi_{c}}}\left(\zeta-\bar{\alpha}_{\chi_{c}}\right)\left(\alpha_{\chi_{c}}-4\zeta\,\bar{\zeta}\right)m^{2}\right)},
\end{equation}
\begin{align}
\mathcal{A}_{2,\,\chi_{c2}}^{(+;\,-,+2)} & =4i\bar{\kappa}m\int_{\bar{\alpha}_{\chi_{c}}}^{1}d\zeta\,\int\frac{d^{2}\ell}{(2\pi)^{2}}\frac{\alpha_{\chi_{c}}\bar{\alpha}_{\chi_{c}}\tilde\Phi_{\chi_{c}}\left(z_{\chi_{c}},\,\boldsymbol{P}_{(2)}\right)\sqrt{\zeta\,\,\bar{\zeta}^{5}}\left(P_{(2)x}+iP_{(2)y}\right)n_{1}^{(-)}\left(x,\,\boldsymbol{\ell},\,\,-\boldsymbol{\Delta}_{\perp}\right)}{\left(\bar{\zeta}\left(\zeta-\bar{\alpha}_{\chi_{c}}\right)^{2}\boldsymbol{K}_{(2)}^{2}-\zeta\,\alpha_{\chi_{c}}\bar{\alpha}_{\chi_{c}}\boldsymbol{P}_{(2)}^{2}-\frac{\bar{\alpha}_{\chi_{c}}}{\alpha_{\chi_{c}}}\left(\zeta-\bar{\alpha}_{\chi_{c}}\right)\left(\alpha_{\chi_{c}}-4\zeta\,\bar{\zeta}\right)m^{2}\right)}\\
 & \times\left(\zeta\left(\boldsymbol{P}_{(2)}^{2}+m^{2}\left(z_{c}-\frac{1}{2}\right)^{2}\right)-m^{2}\bar{\alpha}_{\chi_{c}}\right),\nonumber 
\end{align}
\begin{align}
\mathcal{A}_{2,\,\chi_{c2}}^{(+;\,+,+1)} & =-\frac{\bar{\kappa}}{2\alpha_{\chi_{c}}}\int_{\bar{\alpha}_{\chi_{c}}}^{1}d\zeta\,\int\frac{d^{2}\ell}{(2\pi)^{2}}\frac{\tilde\Phi_{\chi_{c}}\left(z_{\chi_{c}},\,\boldsymbol{P}_{(2)}\right)\sqrt{\zeta\bar{\zeta}}\left(K_{(2)x}+iK_{(2)y}\right)}{\left(\bar{\zeta}\left(\zeta-\bar{\alpha}_{\chi_{c}}\right)^{2}\boldsymbol{K}_{(2)}^{2}-\zeta\,\alpha_{\chi_{c}}\bar{\alpha}_{\chi_{c}}\boldsymbol{P}_{(2)}^{2}-\frac{\bar{\alpha}_{\chi_{c}}}{\alpha_{\chi_{c}}}\left(\zeta-\bar{\alpha}_{\chi_{c}}\right)\left(\alpha_{\chi_{c}}-4\zeta\,\bar{\zeta}\right)m^{2}\right)}\times\\
 & \times\left(m^{2}\left(-4\bar{\zeta}^{2}+\alpha_{\chi_{c}}^{2}+4\bar{\zeta}\alpha_{\chi_{c}}\right)+\alpha_{\chi_{c}}^{2}\boldsymbol{P}_{(2)}^{2}\right)\left\{ \zeta m\alpha_{\chi_{c}}n_{0}\left(x,\,\boldsymbol{\ell},\,\,-\boldsymbol{\Delta}_{\perp}\right)(2\zeta+\alpha_{\chi_{c}}-2)\right.\nonumber \\
 & \left.+i\left(\zeta-\bar{\alpha}_{\chi_{c}}\right)\left(8\zeta^{3}+4\zeta^{2}(\alpha_{\chi_{c}}-4)-6\zeta(\alpha_{\chi_{c}}-2)+3\alpha_{\chi_{c}}-4\right)\left(P_{(2)x}+iP_{(2)y}\right)n_{1}^{(-)}\left(x,\,\boldsymbol{\ell},\,\,-\boldsymbol{\Delta}_{\perp}\right)\right\} ,\nonumber 
\end{align}
\begin{align}
\mathcal{A}_{2,\,\chi_{c2}}^{(+;\,-,+1)} & =-\frac{\bar{\kappa}}{2\alpha_{\chi_{c}}}\int_{\bar{\alpha}_{\chi_{c}}}^{1}d\zeta\,\int\frac{d^{2}\ell}{(2\pi)^{2}}\frac{\tilde\Phi_{\chi_{c}}\left(z_{\chi_{c}},\,\boldsymbol{P}_{(2)}\right)\sqrt{\zeta\,\bar{\zeta}}\left(K_{(2)x}+iK_{(2)y}\right)(2\zeta+\alpha_{\chi_{c}}-2)}{\left(\bar{\zeta}\left(\zeta-\bar{\alpha}_{\chi_{c}}\right)^{2}\boldsymbol{K}_{(2)}^{2}-\zeta\,\alpha_{\chi_{c}}\bar{\alpha}_{\chi_{c}}\boldsymbol{P}_{(2)}^{2}-\frac{\bar{\alpha}_{\chi_{c}}}{\alpha_{\chi_{c}}}\left(\zeta-\bar{\alpha}_{\chi_{c}}\right)\left(\alpha_{\chi_{c}}-4\zeta\,\bar{\zeta}\right)m^{2}\right)}\\
 & \times\left(m^{2}\left(-4\bar{\zeta}^{2}+\alpha_{\chi_{c}}^{2}+4\bar{\zeta}\alpha_{\chi_{c}}\right)+\alpha_{\chi_{c}}^{2}\boldsymbol{P}_{(2)}^{2}\right)\nonumber \\
 & \times\left(m\alpha_{\chi_{c}}n_{0}\left(x,\,\boldsymbol{\ell},\,\,-\boldsymbol{\Delta}_{\perp}\right)\left(\zeta-\bar{\alpha}_{\chi_{c}}\right)+i\zeta(2\zeta+\alpha_{\chi_{c}}-2)^{2}\left(P_{(2)x}+iP_{(2)y}\right)n_{1}^{(-)}\left(x,\,\boldsymbol{\ell},\,\,-\boldsymbol{\Delta}_{\perp}\right)\right),\nonumber 
\end{align}

\begin{align}
\mathcal{A}_{2,\,\chi_{c2}}^{(+;\,+,0)} & =\frac{\bar{\kappa}\bar{\alpha}_{\chi_{c}}}{2\sqrt{6}\alpha_{\chi_{c}}}\int_{\bar{\alpha}_{\chi_{c}}}^{1}d\zeta\,\int\frac{d^{2}\ell}{(2\pi)^{2}}\frac{\tilde\Phi_{\chi_{c}}\left(z_{\chi_{c}},\,\boldsymbol{P}_{(2)}\right)\sqrt{\zeta/\bar{\zeta}}}{\left(\bar{\zeta}\left(\zeta-\bar{\alpha}_{\chi_{c}}\right)^{2}\boldsymbol{K}_{(2)}^{2}-\zeta\,\alpha_{\chi_{c}}\bar{\alpha}_{\chi_{c}}\boldsymbol{P}_{(2)}^{2}-\frac{\bar{\alpha}_{\chi_{c}}}{\alpha_{\chi_{c}}}\left(\zeta-\bar{\alpha}_{\chi_{c}}\right)\left(\alpha_{\chi_{c}}-4\zeta\,\bar{\zeta}\right)m^{2}\right)}\\
 & \times\Bigg\{ n_{0}\left(x,\,\boldsymbol{\ell},\,\,-\boldsymbol{\Delta}_{\perp}\right)\left(1-2z_{c}\right)\nonumber \\
 & \qquad\times\left(m^{4}\left(-4\bar{\zeta}^{2}+\alpha_{\chi_{c}}^{2}+4\bar{\zeta}\alpha_{\chi_{c}}\right)\left(-4\bar{\zeta}^{2}+\alpha_{\chi_{c}}^{2}+4\bar{\zeta}(1-2\zeta)\alpha_{\chi_{c}}\right)+2m^{2}\alpha_{\chi_{c}}^{4}\boldsymbol{P}_{(2)}^{2}+\alpha_{\chi_{c}}^{4}\left(\boldsymbol{P}_{(2)}^{2}\right)^{2}\right)\nonumber \\
 & \qquad+in_{1}^{(-)}\left(x,\,\boldsymbol{\ell},\,\,-\boldsymbol{\Delta}_{\perp}\right)\alpha_{\chi_{c}}^{2}\left(\left(P_{(2)x}+iP_{(2)y}\right)/m\right)\Bigg[8\zeta\,\bar{\zeta}^{2}m^{4}\bar{\alpha}_{\chi_{c}}\left(\zeta-\bar{\alpha}_{\chi_{c}}\right)\nonumber \\
 & \qquad\qquad\left.\left.+\zeta^{2}\alpha_{\chi_{c}}\left(m^{2}\left(-4\bar{\zeta}^{2}+\alpha_{\chi_{c}}^{2}+2\bar{\zeta}\alpha_{\chi_{c}}\right)+\alpha_{\chi_{c}}\left(\alpha_{\chi_{c}}-2\bar{\zeta}\right)\boldsymbol{P}_{(2)}^{2}\right)\left(m^{2}\left(z_{c}-\frac{1}{2}\right)^{2}+\boldsymbol{P}_{(2)}^{2}\right)\right]\right\} ,\nonumber 
\end{align}
\begin{align}
\mathcal{A}_{2,\,\chi_{c2}}^{(+;\,-,0)} & =\frac{\bar{\kappa}\alpha_{\chi_{c}}\bar{\alpha}_{\chi_{c}}}{2\sqrt{6}m}\int_{\bar{\alpha}_{\chi_{c}}}^{1}d\zeta\,\int\frac{d^{2}\ell}{(2\pi)^{2}}\frac{i\tilde\Phi_{\chi_{c}}\left(z_{\chi_{c}},\,\boldsymbol{P}_{(2)}\right)\left(P_{(2)x}-iP_{(2)y}\right)n_{1}^{(-)}\left(x,\,\boldsymbol{\ell},\,\,-\boldsymbol{\Delta}_{\perp}\right)\sqrt{\zeta\,\bar{\zeta}}}{\left(\zeta-\bar{\alpha}_{\chi_{c}}\right)}\\
 & \times\left\{ \frac{\zeta\alpha_{\chi_{c}}\left(m^{2}\left(-4\bar{\zeta}^{2}+\alpha_{\chi_{c}}^{2}+2\bar{\zeta}\alpha_{\chi_{c}}\right)+\alpha_{\chi_{c}}\left(\alpha_{\chi_{c}}-2\bar{\zeta}\right)\boldsymbol{P}_{(2)}^{2}\right)\left(\boldsymbol{P}_{(2)}^{2}+m^{2}\left(z_{c}-\frac{1}{2}\right)^{2}\right)}{\left(\bar{\zeta}\left(\zeta-\bar{\alpha}_{\chi_{c}}\right)^{2}\boldsymbol{K}_{(2)}^{2}-\zeta\,\alpha_{\chi_{c}}\bar{\alpha}_{\chi_{c}}\boldsymbol{P}_{(2)}^{2}-\frac{\bar{\alpha}_{\chi_{c}}}{\alpha_{\chi_{c}}}\left(\zeta-\bar{\alpha}_{\chi_{c}}\right)\left(\alpha_{\chi_{c}}-4\zeta\,\bar{\zeta}\right)m^{2}\right)}\right.\nonumber \\
 & \left.-\frac{8\bar{\zeta}^{2}m^{4}\bar{\alpha}_{\chi_{c}}\left(\zeta-\bar{\alpha}_{\chi_{c}}\right)}{\left(\bar{\zeta}\left(\zeta-\bar{\alpha}_{\chi_{c}}\right)^{2}\boldsymbol{K}_{(2)}^{2}-\zeta\,\alpha_{\chi_{c}}\bar{\alpha}_{\chi_{c}}\boldsymbol{P}_{(2)}^{2}-\frac{\bar{\alpha}_{\chi_{c}}}{\alpha_{\chi_{c}}}\left(\zeta-\bar{\alpha}_{\chi_{c}}\right)\left(\alpha_{\chi_{c}}-4\zeta\,\bar{\zeta}\right)m^{2}\right)}\right\} ,\nonumber 
\end{align}

\begin{align}
\mathcal{A}_{2,\,\chi_{c2}}^{(+;\,+,-1)} & =\bar{\kappa}\int_{\bar{\alpha}_{\chi_{c}}}^{1}d\zeta\,\int\frac{d^{2}\ell}{(2\pi)^{2}}\frac{\tilde\Phi_{\chi_{c}}\left(z_{\chi_{c}},\,\boldsymbol{P}_{(2)}\right)\sqrt{\bar{\zeta}\zeta}\left(m^{2}\left(-4\bar{\zeta}^{2}+\alpha_{\chi_{c}}^{2}+4\bar{\zeta}\alpha_{\chi_{c}}\right)+\alpha_{\chi_{c}}^{2}\boldsymbol{P}_{(2)}^{2}\right)}{4\alpha_{\chi_{c}}\left(\bar{\zeta}\left(\zeta-\bar{\alpha}_{\chi_{c}}\right)^{2}\boldsymbol{K}_{(2)}^{2}-\zeta\,\alpha_{\chi_{c}}\bar{\alpha}_{\chi_{c}}\boldsymbol{P}_{(2)}^{2}-\frac{\bar{\alpha}_{\chi_{c}}}{\alpha_{\chi_{c}}}\left(\zeta-\bar{\alpha}_{\chi_{c}}\right)\left(\alpha_{\chi_{c}}-4\zeta\,\bar{\zeta}\right)m^{2}\right)}\times\\
 & \left\{ mn_{0}\left(x,\,\boldsymbol{\ell},\,\,-\boldsymbol{\Delta}_{\perp}\right)\left[\bar{\alpha}_{\chi_{c}}^{2}\left(P_{(2)x}-iP_{(2)y}\right)\left(\alpha_{\chi_{c}}-4\bar{\zeta}\right)-\alpha_{\chi_{c}}\left(K_{(2)x}+iK_{(2)y}\right)\left(\zeta-\bar{\alpha}_{\chi_{c}}\right)\left(\alpha_{\chi_{c}}-2\bar{\zeta}\right)\right]\right.\nonumber \\
 & -in_{1}^{(-)}\left(x,\,\boldsymbol{\ell},\,\,-\boldsymbol{\Delta}_{\perp}\right)\left[2m^{2}\bar{\alpha}_{\chi_{c}}^{2}\alpha_{\chi_{c}}\right.\nonumber \\
 & \qquad\left.\left.-(2\zeta-1)\left(K_{(2)x}+iK_{(2)y}\right)\left(P_{(2)x}-iP_{(2)y}\right)\left(4\bar{\zeta}^{2}(2\zeta-1)+\alpha_{\chi_{c}}^{3}-6\bar{\zeta}\alpha_{\chi_{c}}^{2}-3\bar{\zeta}^{2}(4\zeta-3)\alpha_{\chi_{c}}\right)\right]\right\} ,\nonumber 
\end{align}
\begin{align}
\mathcal{A}_{2,\,\chi_{c2}}^{(+;\,-,-1)} & =2i\bar{\kappa}\int_{\bar{\alpha}_{\chi_{c}}}^{1}d\zeta\,\int\frac{d^{2}\ell}{(2\pi)^{2}}\frac{\tilde\Phi_{\chi_{c}}\left(z_{\chi_{c}},\,\boldsymbol{P}_{(2)}\right)n_{1}^{(-)}\left(x,\,\boldsymbol{\ell},\,\,-\boldsymbol{\Delta}_{\perp}\right)\left(\zeta\bar{\zeta}\right)^{3/2}}{\left(\bar{\zeta}\left(\zeta-\bar{\alpha}_{\chi_{c}}\right)^{2}\boldsymbol{K}_{(2)}^{2}-\zeta\,\alpha_{\chi_{c}}\bar{\alpha}_{\chi_{c}}\boldsymbol{P}_{(2)}^{2}-\frac{\bar{\alpha}_{\chi_{c}}}{\alpha_{\chi_{c}}}\left(\zeta-\bar{\alpha}_{\chi_{c}}\right)\left(\alpha_{\chi_{c}}-4\zeta\,\bar{\zeta}\right)m^{2}\right)}\\
 & \times\left(K_{(2)x}-iK_{(2)y}\right)\left(P_{(2)x}-iP_{(2)y}\right)\left(\zeta-\bar{\alpha}_{\chi_{c}}\right)\left(z_{c}-1/2\right)\left(m^{2}\left(-4\bar{\zeta}^{2}+\alpha_{\chi_{c}}^{2}+4\bar{\zeta}\alpha_{\chi_{c}}\right)+\alpha_{\chi_{c}}^{2}\boldsymbol{P}_{(2)}^{2}\right),\nonumber 
\end{align}

\begin{align}
\mathcal{A}_{2,\,\chi_{c2}}^{(+;\,+,-2)} & =-4m\bar{\alpha}_{\chi_{c}}\bar{\kappa}\int_{\bar{\alpha}_{\chi_{c}}}^{1}d\zeta\,\int\frac{d^{2}\ell}{(2\pi)^{2}}\tilde\Phi_{\chi_{c}}\left(z_{\chi_{c}},\,\boldsymbol{P}_{(2)}\right)\sqrt{\zeta\,\bar{\zeta}^{3}}\left(P_{(2)x}-iP_{(2)y}\right)\\
 & \times\left\{ \frac{m\bar{\alpha}_{\chi_{c}}\left(\left(P_{(2)x}-iP_{(2)y}\right)n_{0}\left(x,\,\boldsymbol{\ell},\,\,-\boldsymbol{\Delta}_{\perp}\right)\left(\zeta-\bar{\alpha}_{\chi_{c}}\right)-i\zeta m\alpha_{\chi_{c}}n_{1}^{(-)}\left(x,\,\boldsymbol{\ell},\,\,-\boldsymbol{\Delta}_{\perp}\right)\right)}{\left(\bar{\zeta}\left(\zeta-\bar{\alpha}_{\chi_{c}}\right)^{2}\boldsymbol{K}_{(2)}^{2}-\zeta\,\alpha_{\chi_{c}}\bar{\alpha}_{\chi_{c}}\boldsymbol{P}_{(2)}^{2}-\frac{\bar{\alpha}_{\chi_{c}}}{\alpha_{\chi_{c}}}\left(\zeta-\bar{\alpha}_{\chi_{c}}\right)\left(\alpha_{\chi_{c}}-4\zeta\,\bar{\zeta}\right)m^{2}\right)}\right.\nonumber \\
 & \left.+\frac{i\alpha_{\chi_{c}}\zeta^{2}n_{1}^{(-)}\left(x,\,\boldsymbol{\ell},\,\,-\boldsymbol{\Delta}_{\perp}\right)\left(\boldsymbol{P}_{(2)}^{2}+m^{2}\left(z_{c}-\frac{1}{2}\right)^{2}\right)}{\left(\bar{\zeta}\left(\zeta-\bar{\alpha}_{\chi_{c}}\right)^{2}\boldsymbol{K}_{(2)}^{2}-\zeta\,\alpha_{\chi_{c}}\bar{\alpha}_{\chi_{c}}\boldsymbol{P}_{(2)}^{2}-\frac{\bar{\alpha}_{\chi_{c}}}{\alpha_{\chi_{c}}}\left(\zeta-\bar{\alpha}_{\chi_{c}}\right)\left(\alpha_{\chi_{c}}-4\zeta\,\bar{\zeta}\right)m^{2}\right)}\right\} ,\nonumber 
\end{align}
\begin{equation}
\mathcal{A}_{2,\,\chi_{c2}}^{(+;\,-,-2)}=0.\label{eq:Amplitude_PP-1--1}
\end{equation}
At the extremes of the integration domain $\zeta=1$ and $\zeta=\bar{\alpha}_{\chi_{c}}$,
the integrand is strongly suppressed by the wave function $\tilde\Phi_{\chi_{c}}\left(z_{\chi_{c}},\,\boldsymbol{P}\right)$,
and in the heavy quark mass limit the dominant contribution comes
from the central region $\zeta\sim1-\alpha_{\chi_{c}}/2$ ($z_{\chi_{c}}=1/2$).
If we construct a Taylor expansion near the point $z_{\chi_{c}}=1/2,\,\boldsymbol{P}=0$,
the first non-vanishing term in the latter can be reduced to NRQCD
LDMEs of $\chi_{c}$ mesons using the relations~(\ref{eq:LDME},~\ref{eq:Rel})
from Section~\ref{subsec:WFs}. The explicit expressions for the
coefficients in front of LDMEs are extremely lengthy (require taking
derivatives of the derivatives of the integrands) and won't be provided
explicitly due to space limitations. The evaluation of the cross-section
was performed without making such expansion, taking explicitly the
integrals over $d^{2}\ell\,d\zeta$ in momentum space.

\section{Numerical estimates}

\label{sec:Numer}

\subsection{Differential cross-sections}

\label{subsec:diff}For the sake of definiteness, we will use in what
follows the bCGC parametrization of the forward dipole scattering
amplitude, whose parameters were fixed from fits of HERA data~\cite{RESH},
\begin{align}
N\left(x,\,\boldsymbol{r},\,\boldsymbol{b}\right) & =\left\{ \begin{array}{cc}
N_{0}\,\left(\frac{r\,Q_{s}(x)}{2}\right)^{2\gamma_{{\rm eff}}(r)}, & r\,\le\frac{2}{Q_{s}(x)}\\
1-\exp\left(-\mathcal{A}\,\ln^{2}\left(\mathcal{B}r\,Q_{s}\right)\right), & r\,>\frac{2}{Q_{s}(x)}
\end{array}\right.~,\label{eq:CGCDipoleParametrization}\\
 & \mathcal{A}=-\frac{N_{0}^{2}\gamma_{s}^{2}}{\left(1-N_{0}\right)^{2}\ln\left(1-N_{0}\right)},\quad\mathcal{B}=\frac{1}{2}\left(1-N_{0}\right)^{-\frac{1-N_{0}}{N_{0}\gamma_{s}}},\\
 & Q_{s}(x,\,\boldsymbol{b})=\left(\frac{x_{0}}{x}\right)^{\lambda/2}T_{G}(b),\,\,\gamma_{{\rm eff}}(r)=\gamma_{s}+\frac{1}{\kappa\lambda\ln(1/x)}\ln\left(\frac{2}{r\,Q_{s}(x)}\right),\,\,T_{G}(b)=\exp\left(-\frac{b^{2}}{4\gamma_{s}B_{{\rm CGC}}}\right)\label{eq:gamma_eff}\\
\gamma_{s} & =0.6492,\quad\lambda=0.2023,\quad x_{0}=6.9\times10^{-4},\quad B_{{\rm CGC}}=5.5\,{\rm GeV}^{-2}\approx\left(0.463\,{\rm fm}\right)^{2}.
\end{align}
In the heavy quark mass limit, the binding energy of the quarkonium
formally is a small parameter, $\sim\mathcal{O}\left(v^{2}\right)\sim\mathcal{O}\left(\alpha_{s}^{2}\left(m_{c}\right)\right)$.
In order to have consistency with this expectation, we will assume
that the charm mass $m_{c}\approx M_{\chi_{c}}/2\approx1.8\,{\rm GeV}$.
We also will implement the corrections due to the real part of the
amplitude (\ref{eq:reA}) and skewedness (\ref{eq:Rg}) in exclusive
channels, as discussed in Section~\ref{subsec:Derivation}.

\begin{figure}
\includegraphics[width=6cm]{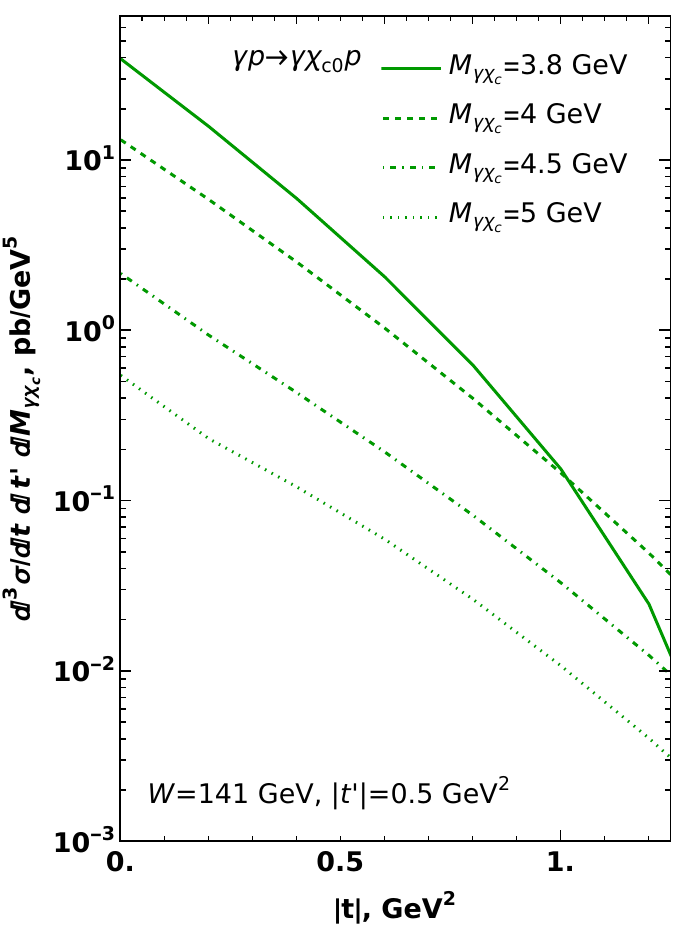}\includegraphics[width=6cm]{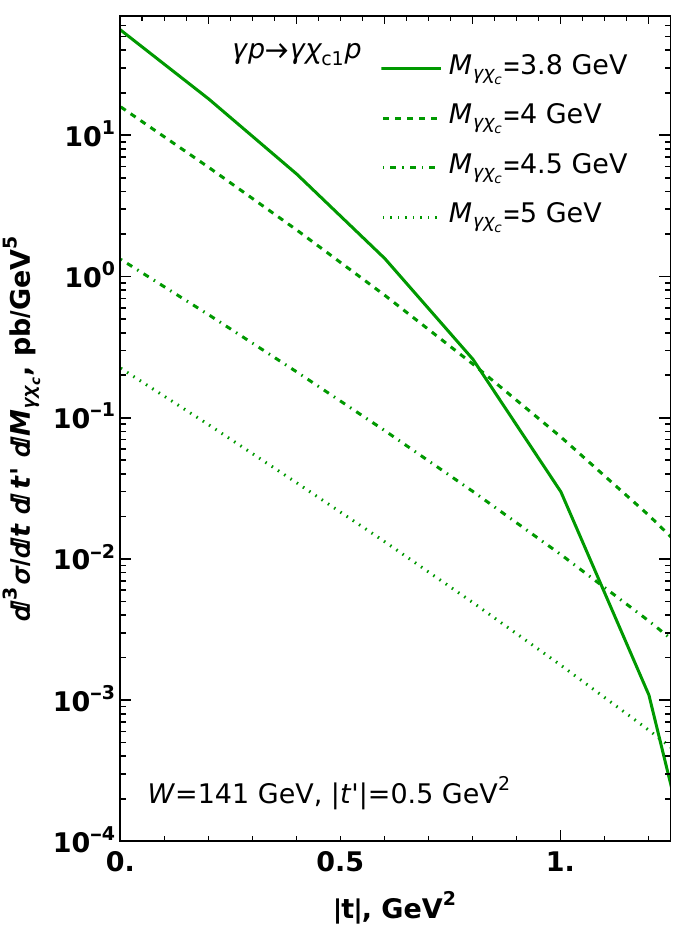}\includegraphics[width=6cm]{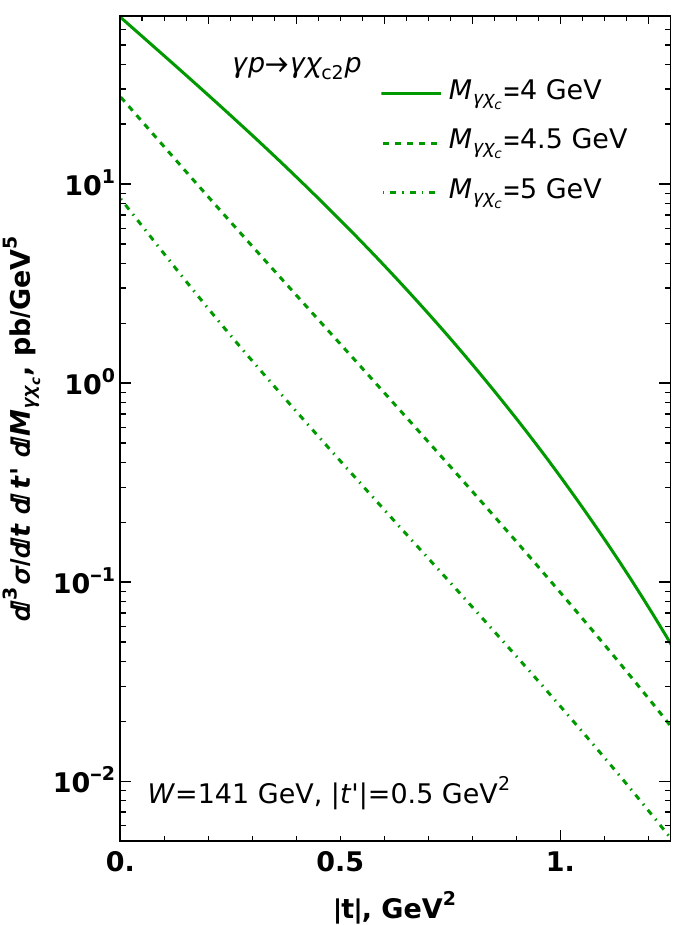}

\caption{The dependence of the photoproduction cross-section~(\ref{eq:Photo})
on the invariant momentum transfer to the target $t$. The left, central
and right columns correspond to the $\chi_{c0}$, $\chi_{c1}$ and
$\chi_{c2}$ mesons, respectively. The $t$-dependence at $M_{\gamma\chi_{c}}=3.8\,{\rm GeV}$
reflects the kinematic constraint $|t|\le M_{\gamma\chi_{c}}^{2}-M_{\chi_{c}}^{2}-\left|t'\right|$. }
\label{fig:tDep}
\end{figure}

We would like to start discussion from the threefold differential
unpolarized cross-section~(\ref{eq:Photo}). In the Figure~\ref{fig:tDep}
we have shown this observable as a function of the invariant momentum
transfer $t$ to the target. Since the parameter $\xi\sim M_{\gamma\chi_{c}}^{2}/W^{2}$
is very small in the chosen kinematics, the variable $t$ may be approximated
as $t\approx-\boldsymbol{\Delta}_{\perp}^{2}$. The pronounced dependence
on this variables largely stems from the explicit $\Delta_{\perp}$-dependence
which appears in the arguments of the functions $n,n_{\Phi}$-~(\ref{eq:nPhi}-\ref{eq:nPhi2c},\ref{eq:n0},\ref{eq:n1})
in the momentum space amplitudes (\ref{eq:Amplitude_PP-1}-\ref{eq:Amplitude_PP--1},~\ref{eq:Amplitude_PP-1-1}-\ref{eq:Amplitude_PP-1--1}).
The dependence on $\Delta_{\perp}$ which appears in prefactors in
front of these functions leads only to minor $\mathcal{O}\left(t/M_{\chi_{c}}^{2}\right)$
corrections. Indeed, we may observe that in all these prefactors the
external vectors $\boldsymbol{p}_{\perp},\boldsymbol{\Delta}_{\perp}$
always contribute only on the linear combination 
\begin{equation}
\boldsymbol{L}=\left(\alpha_{\chi_{c}}\boldsymbol{k}_{\perp}^{\gamma}-\bar{\alpha}_{\chi_{c}}\boldsymbol{p}_{\perp}^{\chi_{c}}\right)=\boldsymbol{p}_{\perp}+\boldsymbol{\Delta}_{\perp}\left(\frac{1}{2}-\alpha_{\chi_{c}}\right).\label{eq:LVec}
\end{equation}
After integration over the dummy momentum $\boldsymbol{\ell}$, the
absolute value of the amplitude may depend only on the square $\boldsymbol{L}^{2}$.
Since the vectors $\boldsymbol{p}_{\perp}$ and $\boldsymbol{\Delta}_{\perp}$
are mutually orthogonal in the high-energy kinematics (see Eq~(\ref{eq:tDep-1-1})),
and $\boldsymbol{p}_{\perp}$ is parametrically large, $\boldsymbol{p}_{\perp}^{2}\sim M_{\chi_{c}}^{2}$,
we may obtain
\begin{equation}
\,\,\,\boldsymbol{L}^{2}\approx\bar{\alpha}_{\chi_{c}}\left[\alpha_{\chi_{c}}M_{\gamma\chi_{c}}^{2}-M_{\chi_{c}}^{2}\right]-t\left(\frac{1}{2}-\alpha_{\chi_{c}}\right)^{2}\approx\bar{\alpha}_{\chi_{c}}\left[\alpha_{\chi_{c}}M_{\gamma\chi_{c}}^{2}-M_{\chi_{c}}^{2}\right]\left(1+\mathcal{O}\left(t/M_{\chi_{c}}^{2}\right)\right)\label{eq:LSq}
\end{equation}
Besides, the factor $\left(\frac{1}{2}-\alpha_{\chi_{c}}\right)^{2}$
in front of $t$ in~(\ref{eq:LSq}) provides additional numerical
suppression that justifies omission of this $t$-dependence. For comparison,
the $t$-dependence that originates from the functions $n,n_{\Phi}$
is related to the implemented impact parameter dependence of the dipole
amplitude and is controlled by the (inverse) proton size. In all the
phenomenological parametrizations fitted to exclusive processes, this
dependence is rapidly decreasing, in agreement with theoretical expectations~\cite{Lepage:1980fj}.
Technically, this implies that the $\chi_{c}\gamma$ pairs predominantly
have oppositely directed transverse momenta $\boldsymbol{p}_{\perp}^{\chi_{c}},\,\boldsymbol{k}_{\perp}^{\gamma}$.
For this reason, in what follows we will focus on the kinematics $|t|=|t_{{\rm min}}|$,
or the $t$-integrated observables that get the largest contribution
from small-$t$ kinematics.

\begin{figure}
\includegraphics[width=6cm]{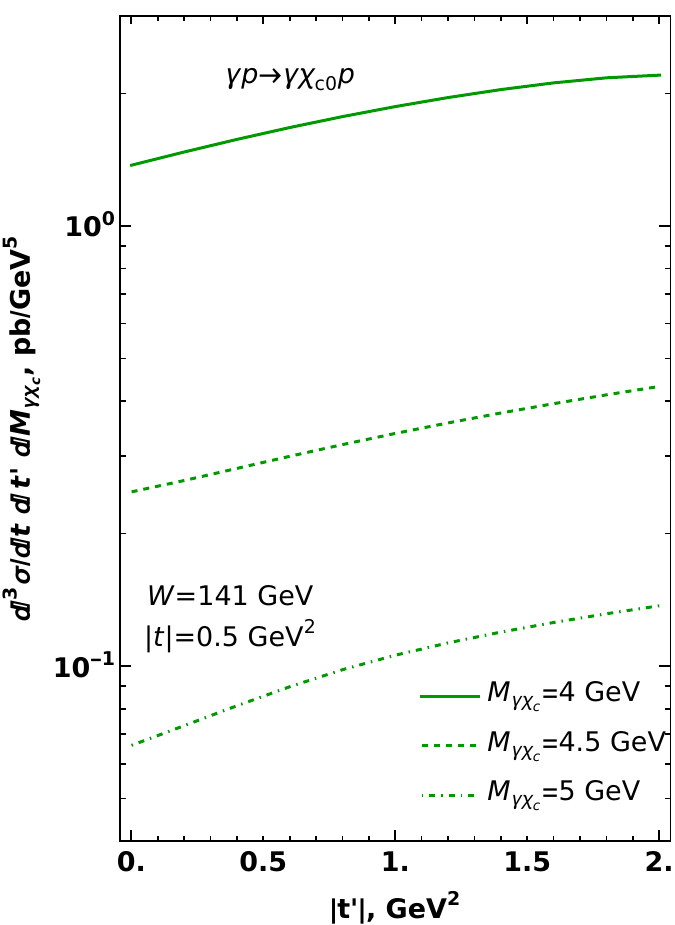}\includegraphics[width=6cm]{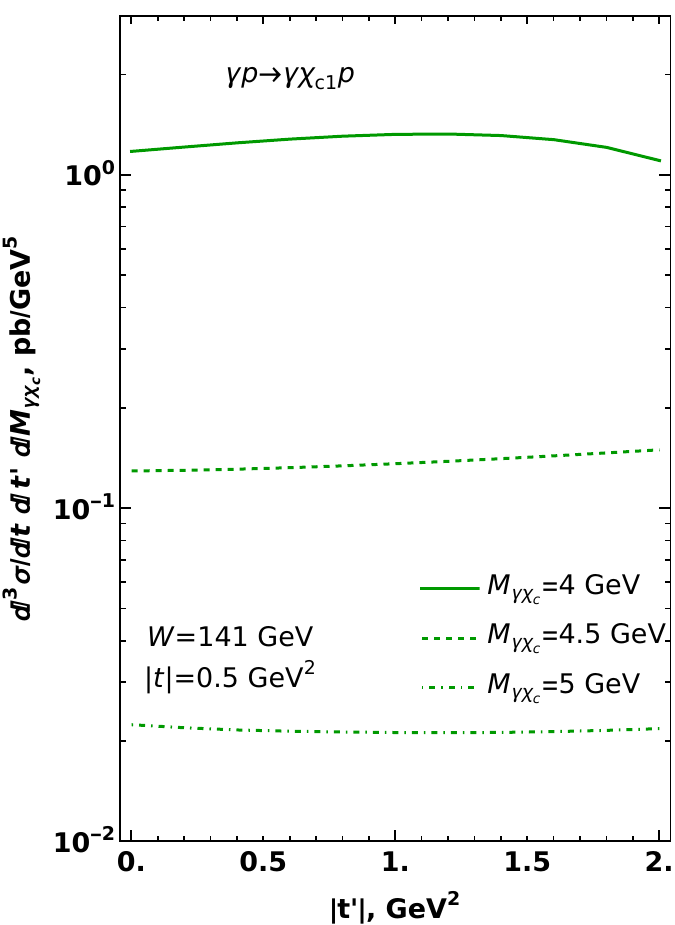}\includegraphics[width=6cm]{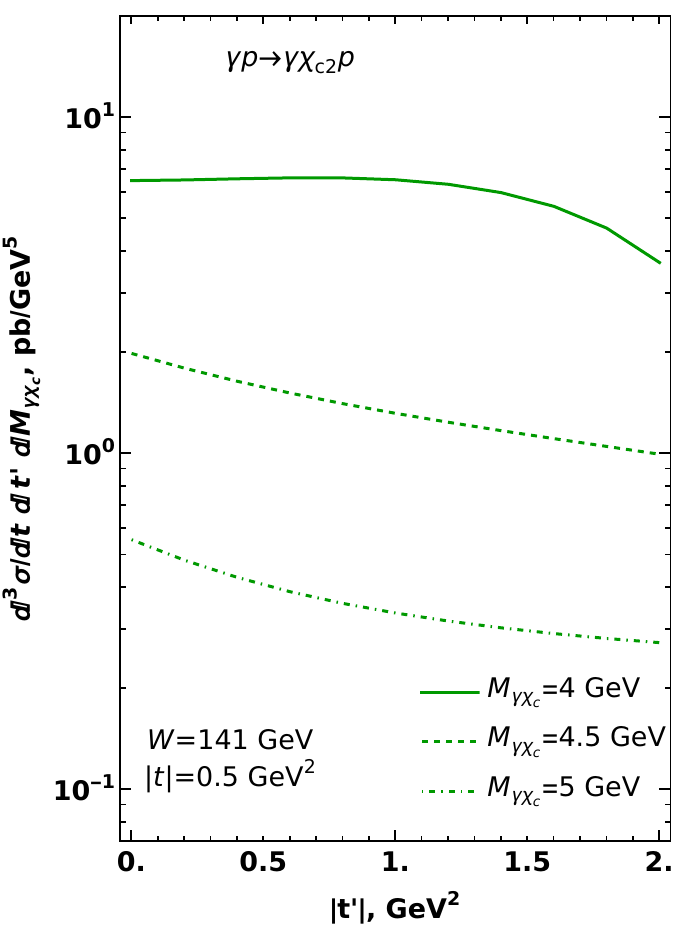}

\caption{The dependence of the photoproduction cross-section~(\ref{eq:Photo})
on the invariant momentum transfer to the photon (variable $t'=\left(q-k\right)^{2}$
defined in~(\ref{eq:uPrimetPrime})). The left, central and right
columns correspond to the $\chi_{c0}$, $\chi_{c1}$ and $\chi_{c2}$
mesons, respectively.}
\label{fig:tPrimeDep}
\end{figure}

\begin{figure}
\includegraphics[width=6cm]{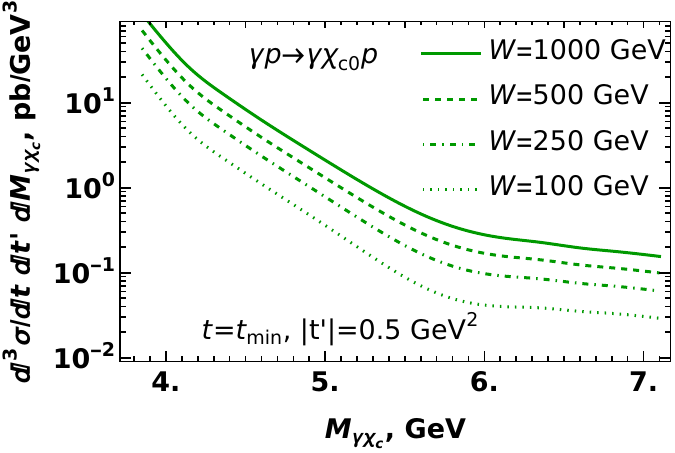}\includegraphics[width=6cm]{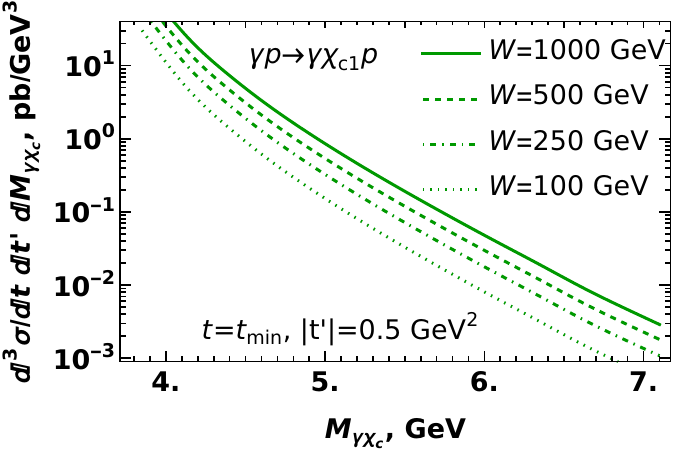}\includegraphics[width=6cm]{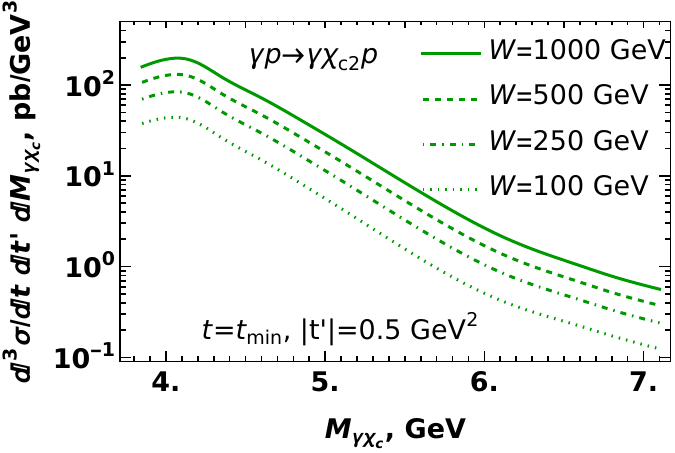}

\caption{The dependence of the photoproduction cross-section~(\ref{eq:Photo})
on the invariant mass $M_{\gamma\chi_{c}}$of the quakonium-photon
pair. The left, central and right columns correspond to the $\chi_{c0}$,
$\chi_{c1}$ and $\chi_{c2}$ mesons, respectively.}
\label{fig:M12Dep}
\end{figure}

 The dependence on $t'$  shown in the Figure~\ref{fig:tPrimeDep}
is very mild in the kinematics of small $|t'|$ because in the impact
factors this variable contributes in linear combination with parametrically
large scales $\sim M_{\gamma\chi_{c}}^{2},M_{\chi_{c}}^{2}$, and,
unless we consider the near-threshold kinematics $M_{\gamma\chi_{c}}^{2}\sim M_{\chi_{c}}^{2}$,
may be disregarded altogether. This variables also may be related
to the transverse momenta $\pm\boldsymbol{p}_{\perp}$ of the final
state $\chi_{c}$ and $\gamma$, and thus is correlated with the angle
between these particles. As we will explain below, the latter observation
allows to understand the completely different $t'$-dependence of
various helicity components. 
In the Figure~\ref{fig:M12Dep} we provide predictions for the dependence
on the invariant mass $M_{\gamma\chi_{c}}$. The latter variable plays
the role of the hard scale that controls the scale of all the transverse
momenta, and shows up explicitly in the impact factors provided in
previous section. This variable also determines the effective value
of the parameter $x\sim M_{\gamma\chi_{c}}^{2}/W^{2}$ in the dipole
amplitude~(\ref{eq:CGCDipoleParametrization}). Due to suppression
of the latter at large $x$, the cross-sections of all quarkonia are
strongly suppressed at large $M_{\gamma\chi_{c}}$. We also may observe
that unless we consider a near-threshold kinematics $M_{\gamma\chi_{c}}\sim M_{\chi_{c}}$,
the dependence on all variables largely factorizes. 

The cross-sections shown in the Figures~\ref{fig:tDep}-\ref{fig:M12Dep}
are larger than similar cross-sections found in the collinear factorization
framework~\cite{Siddikov:2025kah} by up to an order of magnitude.
Such enhanced sensitivity to the framework is not surprising: as could
be seen from explicit expressions in Section~\ref{subsec:WFs}, the
wave functions of $P$-wave $\chi_{c}$ mesons change sign under permutation
of quark and antiquark coordinates, and this property leads to large
cancellations during integration over the variables $\zeta,\,\boldsymbol{\ell}_{\perp}$
in the amplitudes (\ref{eq:Amplitude_PP-1}-\ref{eq:Amplitude_PP--1},~\ref{eq:Amplitude_PP-1-1}-\ref{eq:Amplitude_PP-1--1}).
Due to this property, the amplitudes of the CGC and collinear frameworks
differ by up to a factor of 2-3, and after squaring of the amplitude,
this gives to an order of magnitude difference in the cross-section.

\begin{figure}
\includegraphics[width=6cm]{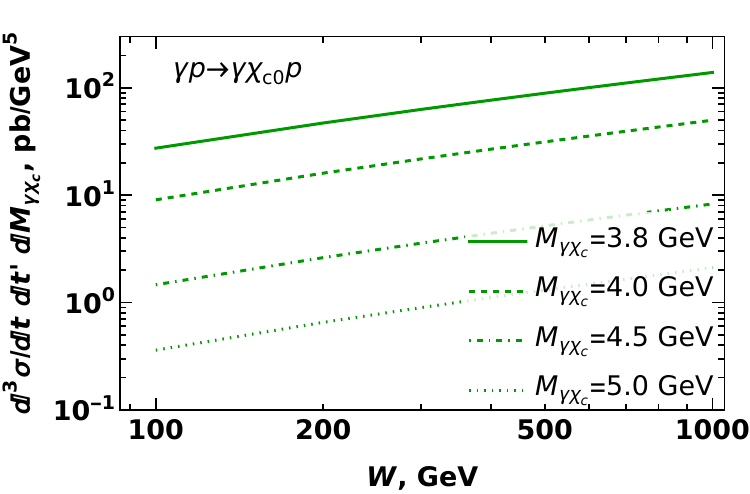}\includegraphics[width=6cm]{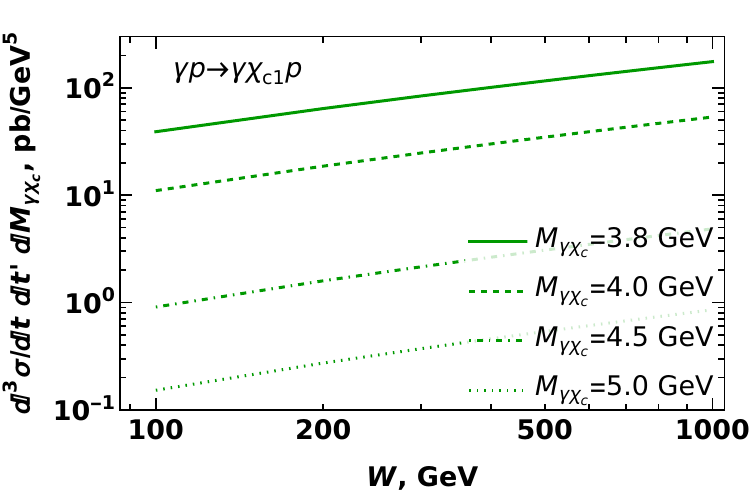}\includegraphics[width=6cm]{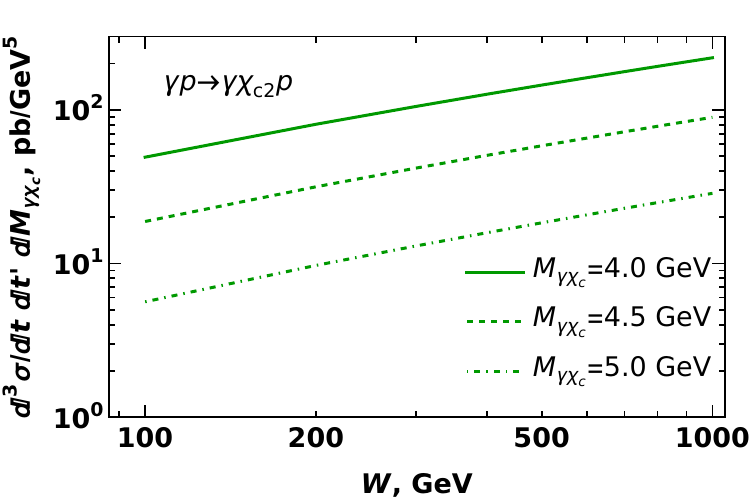}\caption{The energy dependence of the $\gamma p\to\chi_{c}\gamma p$ cross-section
for different spins and invariant energies. The nearly linear dependence
in double logarithmic coordinates suggests that the dependence may
be described by power law $\sim W^{\delta}$, where $\delta$ has
a very mild dependence on other kinematic variables, with typical
values $\delta\sim0.67-0.73$.}
\label{fig:WDep}
\end{figure}

The dependence on the invariant energy shown the Figure~\ref{fig:WDep}
can be described by the power law, $d\sigma(W)\sim W^{2\alpha}$,
where the constant $\alpha$ has a very mild dependence on other variables.
This behaviour follows from a mild $x$-dependence of the forward
dipole amplitude~(\ref{eq:CGCDipoleParametrization}) and reflects
the increase of the gluonic density at higher energies. Numerically,
the values of the parameter $\alpha$ are consistent with the energy
dependence implemented in the small-$r$ branch of~(\ref{eq:CGCDipoleParametrization}),
namely $\alpha\approx2\lambda\langle\gamma_{{\rm eff}}\rangle\approx(0.32,~0.37)$.
The early onset of the saturation effects would reveal itself as a
deviation from the power law behaviour.

\begin{figure}
\includegraphics[width=6cm]{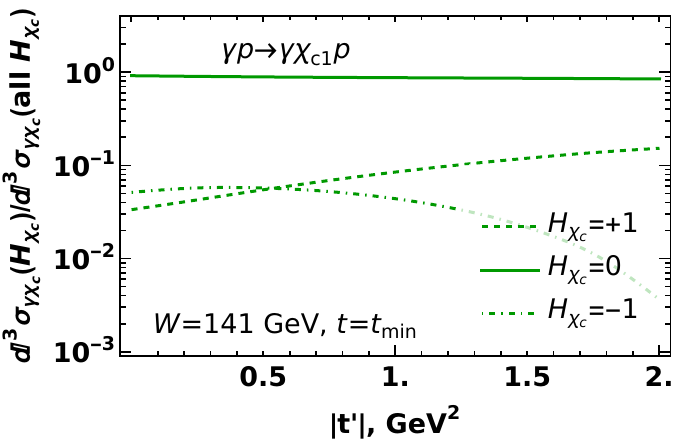}\includegraphics[width=6cm]{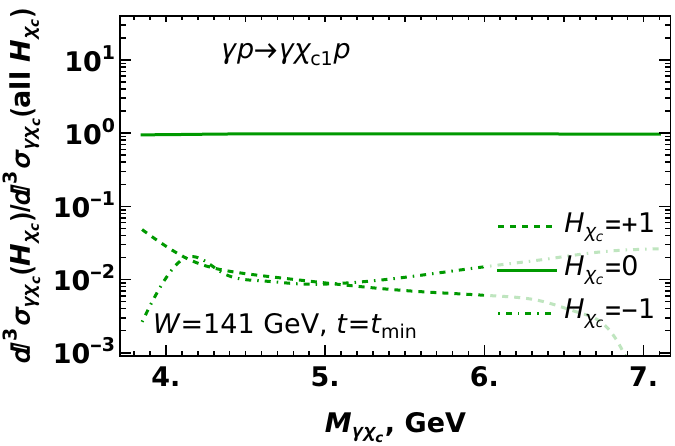}\includegraphics[width=6cm]{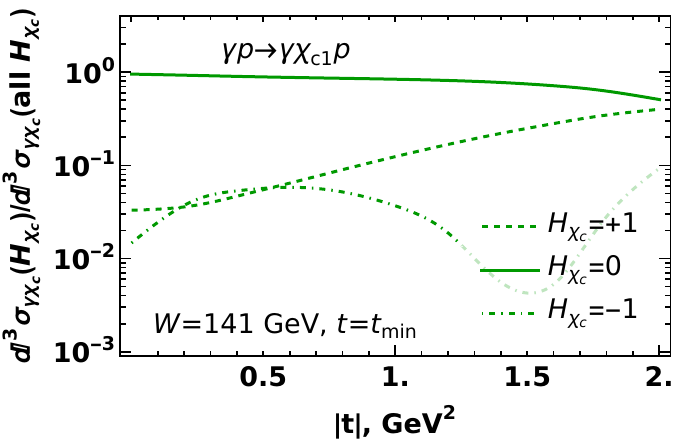}

\includegraphics[width=6cm]{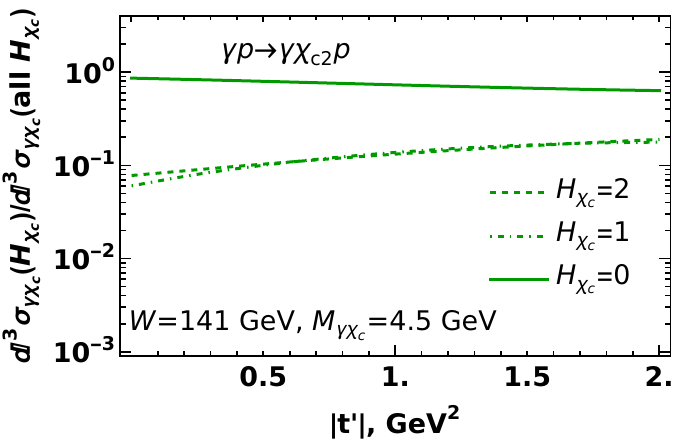}\includegraphics[width=6cm]{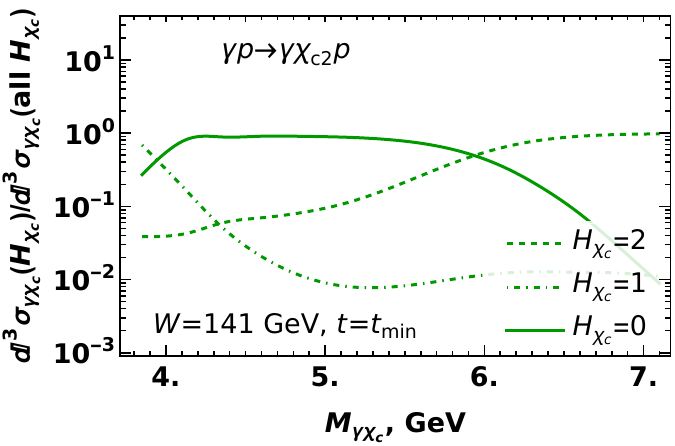}\includegraphics[width=6cm]{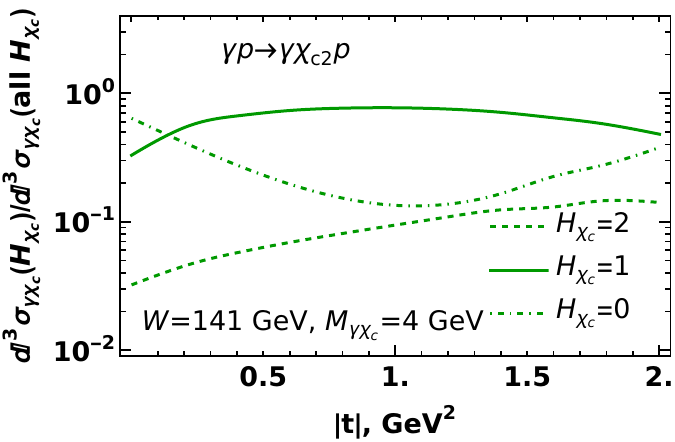}\caption{The relative contributions of different helicity components of $\chi_{c}$
in the total cross-section, as a function of kinematic variables $t,t',M_{\gamma_{\chi_{c}}}$.
The upper and lower rows correspond to $\chi_{c1}$ and $\chi_{c2}$,
respectively. For the sake of legibility, in the lower row (for $\chi_{c2}$)
we did not show the negligibly small contributions of the negative
helicity $H_{\chi_{c}}$.}
\label{fig:Polarized}
\end{figure}

\begin{figure}
\includegraphics[width=9cm]{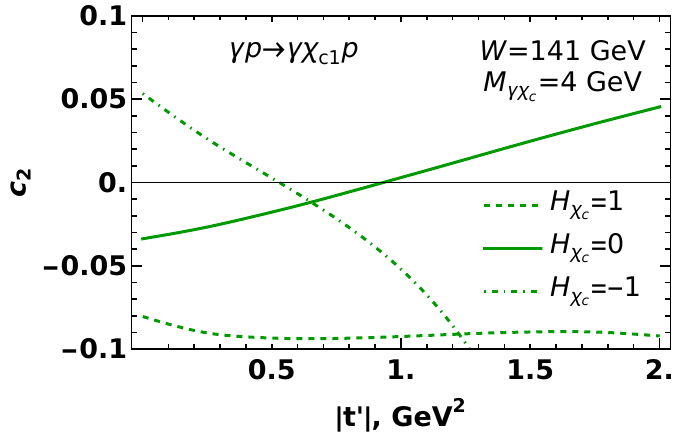}\includegraphics[width=9cm]{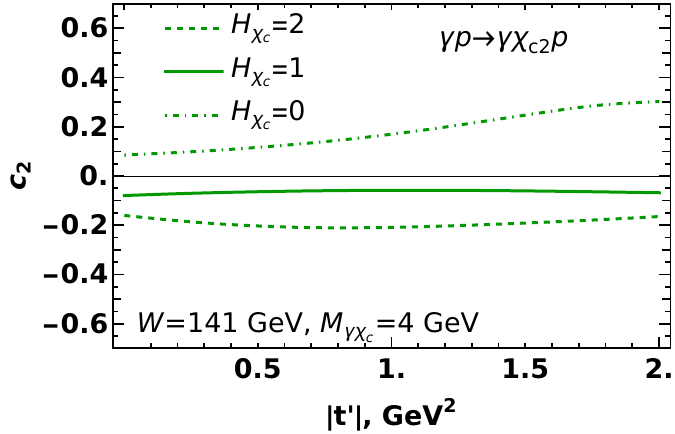}\caption{The angular harmonics  $c_{2}$ of electroproduction cross-section~(\ref{eq:c2s2}),
for different helicity components of $\chi_{c1}$ and $\chi_{c2}$
mesons, respectively. For the sake of legibility, in the right plot
(for $\chi_{c2}$) we did not show the asymmetries of the contributions
with negative helicity $H_{\chi_{c}}$, since the latter are negligibly
small.}
\label{fig:Polarized-1}
\end{figure}

Finally, in the Figure~\ref{fig:Polarized} we have shown the relative
contributions of different quarkonia helicities $H_{\chi_{c}}$ to
the unpolarized cross-section. In the kinematics of relatively small
$t'$, the quarkonia and photons have small transverse momenta and
thus are produced at relatively small angles with respect to the direction
of incoming photon. The helicities of all particles in this kinematics
nearly coincide with projection of the angular momentum onto the collision
axis, and for this reason are subject to the helicity conservation
rule,
\begin{equation}
H_{\gamma,{\rm in}}\approx H_{\gamma,{\rm out}}+H_{\chi_{c}},\qquad\qquad\qquad{\rm small}\,\,t'\approx0.\label{eq:HPlus}
\end{equation}
This selection rule allows to understand the dominance of the $H_{\chi_{c}}=0$
helicity component for $\chi_{c0},\chi_{c1}$: since the helicities
of the onshell photons only take values $\pm1$, and $\left|H_{\chi_{c}}\right|\not=2$
for these mesons, only the helicity state $H_{\chi_{c}}=0$ allows
to satisfy~(\ref{eq:HPlus}), provided that the photon helicity is
not flipped during the process, $H_{\gamma,{\rm in}}=H_{\gamma,{\rm out}}$.
We checked numerically that the contribution of the component with
helicity flip of the photon, $H_{\gamma,{\rm in}}=-H_{\gamma,{\rm out}}$
is indeed strongly suppressed. At larger values of the variables $|t'|,\,M_{\chi_{c}}$,
the angles between final-state particles and the collision axis also
increases rapidly, so the helicities are no longer are related to
projections of the angular momentum onto the same axis, and thus the
selection rule~(\ref{eq:HPlus}) becomes void. In order to understand
the relative size of the contributions with helicity flip of the photon,
in the Figure~\ref{fig:Polarized-1} we have shown the angular harmonics
$c_{2}$ defined in~(\ref{eq:c2s2}). We can see that for the harmonics
$H_{\chi_{c}}=0$ and $H_{\chi_{c}}=1$, which give the dominant contributions
in the cross-section, the asymmetry $c_{2}$ remains moderate. It
increases as a function of $|t'|$, in agreement with the above-mentioned
picture based on increase of the angles, however it does not exceed
$\sim$30 per cent in the kinematics of interest. Since the complex
phases of the amplitudes $\mathcal{A}_{\gamma p\to\chi_{c}\gamma p}^{(+,+)}$
and $\mathcal{A}_{\gamma p\to\chi_{c}\gamma p}^{(+,-)}$ are nearly
identical, the harmonics $s_{2}$ is vanishingly small. An experimental
confirmation of these theoretical expectations would constitute a
strong evidence in favor of the expected dominance of the contribution
without photon helicity flip.~

\subsection{Integrated cross-sections and counting rates}

\begin{figure}
\includegraphics[width=9cm]{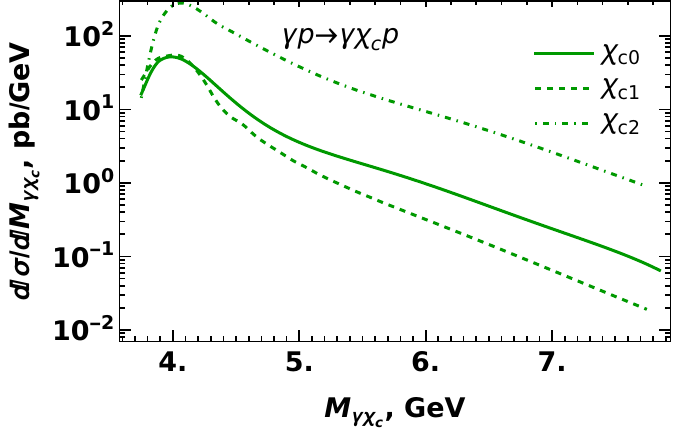}\includegraphics[width=9cm]{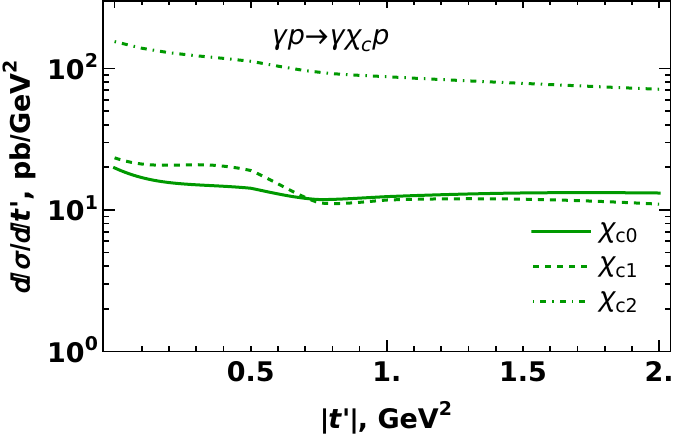}

\caption{The single-differential cross-sections $d\sigma/dt\,dM_{\gamma\chi_{c}}$,
$d\sigma/dM_{\gamma\chi_{c}}$ for different charmonia $\chi_{cJ}$.
Both plots correspond to the invariant energy $W=141$ GeV.}
\label{fig:M12}
\end{figure}

While the threefold differential cross-section are well-suited for
the phenomenological studies, experimentally it may be easier to study
the partially integrated (single-differential) cross-sections. In
the Figure~\ref{fig:M12} we have shown the cross-sections $d\sigma/\,dM_{\gamma\chi_{c}}$
and $d\sigma/dt'$ for different charmonia. As we discussed in the
previous section, for the threefold cross-section the dependence on
$W,\,t,\,M_{\gamma\chi_{c}}$ largely factorizes; for this reason
the dependence of the integrated cross-sections on these variables
is similar to that of the unintegrated cross-section.

The total (integrated) cross-sections of the photoproduction process
$\gamma p\to\chi_{cJ}\gamma p$ is given by
\begin{align}
\sigma_{{\rm tot}}^{\chi_{c0}}\left(W\approx100\,{\rm GeV},\,M_{\gamma\chi_{c}}\ge3.7\,{\rm GeV}\right) & \approx25\,{\rm pb},\\
\sigma_{{\rm tot}}^{\chi_{c1}}\left(W\approx100\,{\rm GeV},\,M_{\gamma\chi_{c}}\ge3.7\,{\rm GeV}\right) & \approx23\,{\rm pb},\\
\sigma_{{\rm tot}}^{\chi_{c2}}\left(W\approx100\,{\rm GeV},\,M_{\gamma\chi_{c}}\ge3.7\,{\rm GeV}\right) & \approx149\,{\rm pb},
\end{align}
and scales with energy approximately as $\sim W^{0.7}$. In our estimates
we introduced a cutoff on minimal invariant mass $M_{\gamma\chi_{c}}$
in order to avoid (huge) background from the radiative decays of $\psi(2S)\to\chi_{c}\gamma$~\footnote{Combining the experimental $\psi(2S)$ photoproduction cross-section~\cite{LHCb:2018rcm}
and the branching fractions of $\psi(2S)\to\chi_{cJ}\gamma$~\cite{Navas:2024X},
we estimate that the background contribution from $\psi(2S)$ radiative
decay is $\sigma_{{\rm tot}}^{({\rm rad})}\approx1.1\,{\rm nb}$. }. In the Tables~\ref{tab:XSections} and \ref{tab:XSections-1} we
provide tentative estimates for the total cross-sections of the electroproduction
$ep\to e\gamma\chi_{c}p$ and the ultraperipheral production $pp\to pp\gamma\chi_{c}$
in LHC kinematics. Since the spectrum of virtual photons is dominated
by quasireal photons, for the sake of simplicity we disregarded the
contributions of photons with large virtuality $Q^{2}\gtrsim1\,{\rm GeV^{2}}$,
and assumed that the photoproduction cross-section does not depend
on $Q^{2}$ at $Q^{2}\lesssim1\,{\rm GeV^{2}}$ in view of the smallness
of $\mathcal{O}\left(Q^{2}/M_{\chi_{c}}^{2}\right)$ corrections in
the heavy quark mass limit. We found that varying the upper cutoff
$Q_{{\rm max}}^{2}$ between 1 and $2\,{\rm GeV^{2}}$ changes the
cross-section within 10 per cent.

{\renewcommand{\baselinestretch}{1.5}
\begin{table}
\begin{tabular}{|c|c|c|c|c|c|c|c|}
\hline 
\multirow{2}{*}{} & \multirow{2}{*}{$\sigma_{{\rm tot}}^{(ep)}$} & \multicolumn{2}{c|}{\textbf{Production rates}} & \textbf{Decay} & \textbf{Combined} & \multicolumn{2}{c|}{\textbf{Counting rates}}\tabularnewline
\cline{3-4}\cline{7-8}
 &  & $N$ & $dN/dt$ & \textbf{channel} & \textbf{branching} & $N_{d}$ & $dN_{d}/dt$\tabularnewline
\hline 
$\chi_{c0}$ & 0.45~pb & 4.5$\times10^{4}$ & 380/day & $\chi_{c}\to J/\psi\,\gamma\,$ & 0.08~\% & 36 & 9.1/month\tabularnewline
\cline{1-4}\cline{6-8}
$\chi_{c1}$ & 0.41~pb & 4.1$\times10^{4}$ & 353/day & $J/\psi\to\mu^{+}\mu^{-}$ & 2~\% & 816 & 211/month\tabularnewline
\cline{1-4}\cline{6-8}
$\chi_{c2}$ & 2.5~pb & 2.5$\times10^{5}$ & 2100/day &  & 1.1~\% & 2750 & 695/month\tabularnewline
\hline 
\end{tabular}

\caption{The total electroproduction cross-sections, production and counting
rates for different $\chi_{c}$ mesons in EIC kinematics at electron-proton
energy $\sqrt{s_{ep}}=141\,{\rm GeV}$. The cross-section roughly
scales with energy as $\sim W^{0.7}$. For estimates of the production
and counting rates, we used the instantaneous luminosity $\mathcal{L}=10^{34}\,{\rm cm^{-2}s^{-1}=10^{-5}}{\rm fb}^{-1}$
and the integrated luminosity $\protect\int dt\,\mathcal{L}=100\,{\rm fb}^{-1}$.
The column \textquotedblleft combined branching\textquotedblright{}
corresponds to the product of branching fractions of $\chi_{c}\to J/\psi\,\gamma\,$
and $J/\psi\to\mu^{+}\mu^{-}$.}\label{tab:XSections}
\end{table}

\begin{table}
\begin{tabular}{|c|c|c|c|c|c|c|c|}
\hline 
\multirow{2}{*}{} & \multirow{2}{*}{$\sigma_{{\rm tot}}^{(pp)}$} & \multicolumn{2}{c|}{\textbf{Production rates}} & \textbf{Decay} & \textbf{Combined} & \multicolumn{2}{c|}{\textbf{Counting rates}}\tabularnewline
\cline{3-4}\cline{7-8}
 &  & $N$ & $dN/dt$ & \textbf{channel} & \textbf{branching} & $N_{d}$ & $dN_{d}/dt$\tabularnewline
\hline 
$\chi_{c0}$ & 0.3~pb & 3$\times10^{4}$ & 253/day & $\chi_{c}\to J/\psi\,\gamma\,$ & 0.08~\% & 24 & 6.1/month\tabularnewline
\cline{1-4}\cline{6-8}
$\chi_{c1}$ & 0.25~pb & 2.5$\times10^{4}$ & 215/day & $J/\psi\to\mu^{+}\mu^{-}$ & 2~\% & 500 & 129/month\tabularnewline
\cline{1-4}\cline{6-8}
$\chi_{c2}$ & 1.7~pb & 1.7$\times10^{5}$ & 1088/day &  & 1.1~\% & 1421 & 359/month\tabularnewline
\hline 
\end{tabular}

\caption{The total production cross-sections, production and counting rates
for different $\chi_{c}$ mesons, in ultraperipheral LHC kinematics
at proton-proton energy $\sqrt{s_{pp}}=13\,{\rm GeV}$. For estimates
of the production and counting rates, we used the instantaneous luminosity
$\mathcal{L}=10^{34}\,{\rm cm^{-2}s^{-1}}=10^{-5}{\rm fb}^{-1}s^{-1}$
and the integrated luminosity $\protect\int dt\,\mathcal{L}=100\,{\rm fb}^{-1}$.
The column \textquotedblleft combined branching\textquotedblright{}
corresponds to the product of branching fractions of $\chi_{c}\to J/\psi\,\gamma\,$
and $J/\psi\to\mu^{+}\mu^{-}$.}\label{tab:XSections-1}
\end{table}

}

\subsection{Comparison with odderon-mediated $\chi_{c}$ photoproduction}

\label{subsec:odderon} The odderons, or $C$-odd gluon exchanges
in $t$-channel, remain one of the least understood components of
the nonperturbative QCD~\cite{Odd1,Bartels:2001hw,Odd4,Odd5,Odd6,Odd7},
and the the contribution of the odderons to the dipole scattering
amplitude at present is largely unknown. A recent discovery of the
odderons from comparison of $pp$ and $p\bar{p}$ elastic cross-sections
measured at LHC and Tevatron~\cite{OddTotem1,OddTotem2} reinvigorated
the interest in this topic. However, the statistical significance
of the observed signal still remains under discussion due to sizable
uncertainties and possible background contributions. It could be extremely
difficult to study in detail the odderon amplitude from the channels
where odderons contribute as minor corrections. For this reason, the
processes which proceed via the $C$-odd $t$-channel exchanges got
into focus of odderon searches, and it is expected that such processes
will be studied experimentally both at HL-LHC and at the future EIC.

The exclusive photoproduction of $C$-even quarkonia, as for example
$\gamma p\to\eta_{c}p$, historically attracted a lot of interest,
since the heavy quark mass plays the role of the hard scale and justifies
at least partial description in perturbative QCD (see~\cite{Odd7,Benic:2023}
for a short overview). However, that process obtains a sizable contribution
from the photon-photon fusion (Primakoff mechanism), and for this
reason, recently the exclusive $\chi_{c}$ photoproduction $\gamma p\to\chi_{c}p$
has been recently suggested in~\cite{Jia:2022oyl,Benic:2024} as
a potential cleaner alternative. The cross-section of the latter process
is comparable to that of $\eta_{c}$ photoproduction; furthermore
it may be easier to study experimentally due to large branching fraction
of radiative decays of $\chi_{c}$ to $J/\psi$.

In this context, the photoproduction of $\chi_{c}\gamma$ pairs deserves
attention as a potential background to odderon-mediated $\chi_{c}$-photoproduction:
while the former is formally suppressed as $\mathcal{O}\left(\alpha_{{\rm em}}\right)$,
the latter is suppressed by smallness of the odderon amplitude, so
numerically their cross-sections are comparable. If the final-state
photon is not detected, potentially the $\chi_{c}\gamma$ photoproduction
can be misinterpreted as $\gamma p\to\chi_{c}p$ subprocess. Furthermore,
if there is no constraints (cuts) on the invariant mass of the produced
hadronic state, a sizable contribution could be obtained from the
radiative decays of heavier charmonia. In order to estimate accurately
these backgrounds, in general it is required to know in detail the
geometry and acceptance of the detector. For simplicity, now we will
consider that \emph{all} the final photons are not detected, and will
consider the cross-section $d\sigma/dt$ integrated over the phase
space of the produced photon. This approach provides an upper estimate
for the background. For the sake of definiteness, we will use for
comparison only the result of~\cite{Benic:2024} because the predictions
of~\cite{Jia:2022oyl} are provided in the kinematics of relatively
low energies, where the CGC approach might be not very reliable.

\begin{figure}
\includegraphics[width=18cm]{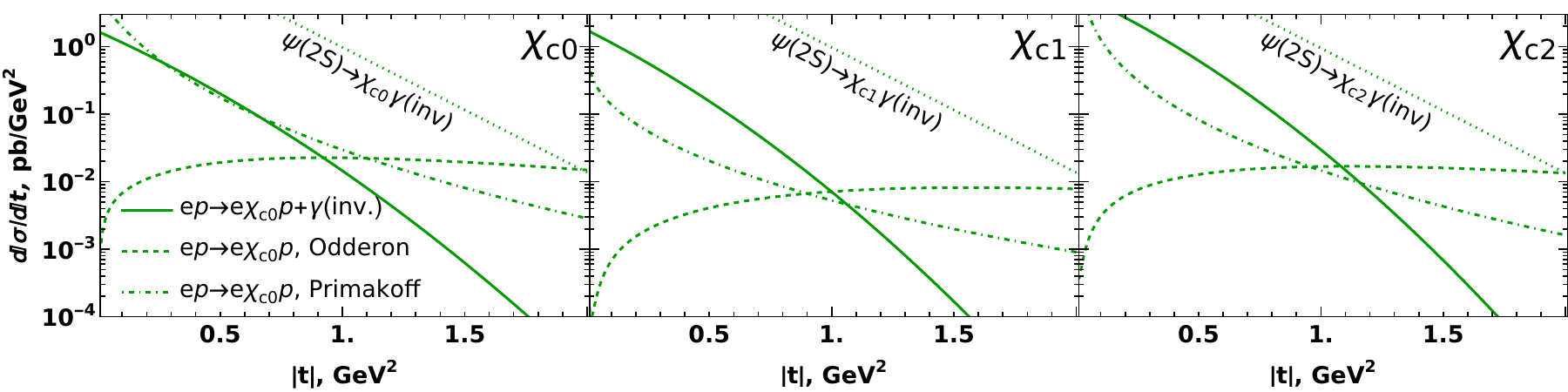}\caption{Comparison of different mechanisms which contribute to electroproduction
of mesons. The solid curve corresponds to contribution of $ep\to e\gamma\chi_{c}p$
process discussed in this paper, with invisible (integrated out) final
state photon. The upper dotted curve corresponds to the contribution
of $\chi_{c}\gamma$ pairs which stem from the radiative decays of
$\psi(2S)$ charmonia (the $\chi_{c}\gamma$ pairs in this mechanism
have invariant mass $M_{\gamma\chi_{c}}\sim M_{\psi(2S)}$). The dashed
and dot-dashed lines correspond to exclusive electroproduction $ep\to e\chi_{c}p$
mediated by odderons and the Primakoff mechanism, as provided in ~\cite{Benic:2024}
(see Figure 5).}
\label{fig:Odd}
\end{figure}

In the Figure~(\ref{fig:Odd}) we show the cross-sections of the
$\chi_{c}$ and $\chi_{c}\gamma$ production via different mechanisms.
The largest contribution comes from the $\psi(2S)$ photoproduction
with subsequent radiative decay $\psi(2S)\to\chi_{c}\gamma$, since
$\psi(2S)$ photoproduction does not require exchange of quantum numbers
in $t$-channel, and $\psi(2S)$ has remarkably large branching ratios
of radiative decays, ${\rm Br}\left(\psi(2S)\to\chi_{c0}\gamma\right)=9.79\pm0.2\,\%$,
${\rm Br}\left(\psi(2S)\to\chi_{c1}\gamma\right)=9.75\pm0.24\,\%$,
${\rm Br}\left(\psi(2S)\to\chi_{c2}\gamma\right)=9.52\pm0.2\,\%$
\cite{Navas:2024X}. We used for estimates the cross-section of $\psi(2S)$
photoproduction found in CGC framework in~\cite{Armesto:2014sma},
and checked that it can reproduce the experimentally measured total
($t$-integrated) cross-section~\cite{LHCb:2018rcm} and the diffractive
slope~\cite{H1:2002yab}. While the radiative decays of other excited
quarkonia potentially could also contribute to the observed backgrounds,
at present it is not possible to estimate accurately their contribution
due to a sizable uncertainty in their wave functions.

The exclusive (non-resonant) production of $\chi_{c}\gamma$ pairs
suggested in this paper exceeds significantly the odderon-mediated
$\chi_{c}$ photoproduction at small-$t$ ($|t|\lesssim1\,{\rm GeV}^{2}$),
although is less relevant in the large-$t$ kinematics. The cross-sections
of both mechanisms of $\chi_{c}\gamma$ production (direct and via
$\psi(2S)$ radiative decays) increase with energy, whereas the contribution
of odderons should decrease due to different odderon intercept.

For the sake of comparison we also have shown the contribution of
the photon-photon fusion (so-called Primakoff mechanism) which exceeds
the odderon contribution in the small-$t$ kinematics, as discussed
in~\cite{Benic:2024}. While the Primakoff mechanism is less relevant
for odderon searches on neutrons, namely for $\gamma n\to\chi_{c}n$
subprocess, the backgrounds due to the $\chi_{c}\gamma$ production
with undetected photon are the same for proton and neutron targets. 

The contributions of different mechanisms discussed in this section
can be separated, imposing the kinematic cuts on the variable $\left(q-\Delta\right)^{2}\approx t-Q^{2}-2q\cdot\Delta$.
For exclusive $\chi_{c}$ production via odderon-mediated and Primakoff
mechanisms, this variable corresponds to the square of charmonium
mass, $M_{\chi_{c}}^{2}$. For production via radiative decays of
$\psi(2S)$, this variable equals $M_{\psi(2S)}^{2}$. Finally, for
the direct (non-resonant) production of $\chi_{c}\gamma$ pairs, this
variable coincides with invariant mass $M_{\gamma\chi_{c}}^{2}$ of
$\chi_{c}\gamma$ pair.

\section{Conclusions}

\label{sec:Conclusions}Using the CGC framework, we analyzed the exclusive
$\chi_{c}\gamma$ photoproduction, including its helicity dependence,
for all $\chi_{cJ}$ states. We found that the amplitude of the process
may be represented as a convolution of the process-dependent impact
factors and forward dipole scattering amplitude. In the heavy quark
mass limit, the kinematic distribution of produced $\chi_{c}\gamma$
pairs may be related to the dependence of the forward dipole scattering
amplitude on dipole size, impact parameter and rapidity dependence.
We also estimated numerically the cross-sections in the ultraperipheral
kinematics at LHC and the future EIC. The cross-sections of all $\chi_{c}$
mesons are comparable to each other (within a factor of two), and
of the same order of magnitude as the cross-section of $\eta_{c}$-production
found earlier in~\cite{Siddikov:2024bre}.  The expected production
rates of $\chi_{c}\gamma$ pairs constitute $\sim10^{4}-10^{5}$ events
per each 100 ${\rm fb}^{-1}$ of integrated luminosity, both at EIC
and LHC. The expected detection (counting) rates for $\chi_{c1}$
and $\chi_{c2}$ mesons constitute $\sim10^{2}-10^{3}$ events per
each 100 ${\rm fb}^{-1}$ of integrated luminosity, if the $\chi_{c}$
mesons are detected via their $\chi_{c}\to J/\psi\,\gamma$ decays.
These numbers suggest that the cross-section may be measured with
reasonable precision. We also found that the production of $\chi_{c}\gamma$
pairs with invisible photon (either direct or via radiative decay
of $\psi(2S)\,\to\chi_{c}\gamma$) gives a sizable background to exclusive
photoproduction of $\chi_{c}$ mesons, which was suggested recently
in~~\cite{Jia:2022oyl,Benic:2024} as an alternative tool for study
of odderons. While the cross-sections of $\chi_{c}\gamma$ production
decrease rapidly as a function of the momentum transfer $|t|$, the
radiative decay of $\psi(2S)$ remains the dominant mechanism up to
$|t|\sim2\,{\rm GeV^{2}}$. These findings are in line with our previous analysis~\cite{Siddikov:2024bre}, which concluded that odderon-mediated quarkonia production can get sizeable backgrounds from quarkonia-photon production.

\section*{Acknowledgments}

We thank our colleagues at UTFSM university for encouraging discussions.
This research was partially supported by Proyecto ANID PIA/APOYO AFB220004
(Chile), and ANID grants Fondecyt Regular \textnumero 1251975 and
Fondecyt Postdoctoral \textnumero 3230699. I. Z. also expresses his
gratitude to the Institute of Physics of PUCV. Powered@NLHPC: This
research was partially supported by the supercomputing infrastructure
of the NLHPC (ECM-02).

\appendix

 \end{document}